\documentclass[twocolumn, twocolappendix]{aastex63}

\usepackage{amsmath}
\usepackage{mathrsfs}
\usepackage{afterpage}
\usepackage{scalerel}
\usepackage{natbib}
\bibpunct[; ]{(}{)}{;}{a}{}{,}
\usepackage{multirow}
\usepackage{graphics}
\usepackage{threeparttable}
\usepackage[varg]{txfonts}
\usepackage{xcolor}
\usepackage{bm}
\usepackage{wrapfig}
\hypersetup{colorlinks,linkcolor={blue},citecolor={blue},urlcolor={blue}}

\begin{document}
\title{Spectral Energy Distributions in Three Deep-Drilling Fields of the Vera C. Rubin Observatory Legacy Survey of Space and Time: Source Classification and Galaxy Properties}

\author[0000-0002-4436-6923]{Fan Zou}
\affiliation{Department of Astronomy and Astrophysics, 525 Davey Lab, The Pennsylvania State University, University Park, PA 16802, USA}
\affiliation{Institute for Gravitation and the Cosmos, The Pennsylvania State University, University Park, PA 16802, USA}

\author[0000-0002-0167-2453]{W. N. Brandt}
\affiliation{Department of Astronomy and Astrophysics, 525 Davey Lab, The Pennsylvania State University, University Park, PA 16802, USA}
\affiliation{Institute for Gravitation and the Cosmos, The Pennsylvania State University, University Park, PA 16802, USA}
\affiliation{Department of Physics, 104 Davey Laboratory, The Pennsylvania State University, University Park, PA 16802, USA}

\author[0000-0002-4945-5079]{Chien-Ting Chen}
\affiliation{Marshall Space Flight Center, Huntsville, AL 35811, USA}
\affiliation{Science and Technology Institute, Universities Space Research Association, Huntsville, AL 35805, USA}

\author[0000-0001-6755-1315]{Joel Leja}
\affiliation{Department of Astronomy and Astrophysics, 525 Davey Lab, The Pennsylvania State University, University Park, PA 16802, USA}
\affiliation{Institute for Gravitation and the Cosmos, The Pennsylvania State University, University Park, PA 16802, USA}
\affiliation{Institute for Computational and Data Sciences, The Pennsylvania State University, University Park, PA 16802, USA}

\author[0000-0002-8577-2717]{Qingling Ni}
\affiliation{Institute for Astronomy, University of Edinburgh, Royal Observatory, Edinburgh, EH9 3HJ, UK}

\author[0000-0001-9519-1812]{Wei Yan}
\affiliation{Department of Astronomy and Astrophysics, 525 Davey Lab, The Pennsylvania State University, University Park, PA 16802, USA}
\affiliation{Institute for Gravitation and the Cosmos, The Pennsylvania State University, University Park, PA 16802, USA}

\author[0000-0001-8835-7722]{Guang Yang}
\affiliation{Department of Physics and Astronomy, Texas A$\&$M University, College Station, TX, 77843-4242 USA}
\affiliation{George P. and Cynthia Woods Mitchell Institute for Fundamental Physics and Astronomy, Texas A$\&$M University, College Station, TX, 77843-4242 USA}

\author[0000-0002-1653-4969]{Shifu Zhu}
\affiliation{Department of Astronomy and Astrophysics, 525 Davey Lab, The Pennsylvania State University, University Park, PA 16802, USA}
\affiliation{Institute for Gravitation and the Cosmos, The Pennsylvania State University, University Park, PA 16802, USA}

\author[0000-0002-9036-0063]{Bin Luo}
\affiliation{School of Astronomy and Space Science, Nanjing University, Nanjing, Jiangsu 210093, China}
\affiliation{Key Laboratory of Modern Astronomy and Astrophysics (Nanjing University), Ministry of Education, Nanjing 210093, China}

\author[0000-0003-1991-370X]{Kristina Nyland}
\affiliation{U.S. Naval Research Laboratory, 4555 Overlook Ave SW, Washington, DC 20375, USA}

\author[0000-0003-0680-9305]{Fabio Vito}
\affiliation{Scuola Normale Superiore, Piazza dei Cavalieri 7, 56126 Pisa, Italy}
\affiliation{INAF -- Osservatorio di Astrofisica e Scienza dello Spazio di Bologna, Via Gobetti 93/3, I-40129 Bologna, Italy}

\author[0000-0002-1935-8104]{Yongquan Xue}
\affiliation{CAS Key Laboratory for Research in Galaxies and Cosmology, Department of Astronomy, University of Science and Technology of China, Hefei 230026, China}
\affiliation{School of Astronomy and Space Sciences, University of Science and Technology of China, Hefei 230026, China}

\email{E-mail: fuz64@psu.edu}

\begin{abstract}
\mbox{W-CDF-S}, ELAIS-S1, and XMM-LSS will be three Deep-Drilling Fields (DDFs) of the Vera C. Rubin Observatory Legacy Survey of Space and Time (LSST), but their extensive multi-wavelength data have not been fully utilized as done in the COSMOS field, another LSST DDF. To prepare for future science, we fit source spectral energy distributions (SEDs) from \mbox{X-ray} to far-infrared in these three fields mainly to derive galaxy stellar masses and star-formation rates. We use \texttt{CIGALE} v2022.0, a code that has been regularly developed and evaluated, for the SED fitting. Our catalog includes 0.8 million sources covering $4.9~\mathrm{deg^2}$ in \mbox{W-CDF-S}, 0.8 million sources covering $3.4~\mathrm{deg^2}$ in ELAIS-S1, and 1.2 million sources covering $4.9~\mathrm{deg^2}$ in XMM-LSS. Besides fitting normal galaxies, we also select candidates that may host active galactic nuclei (AGNs) or are experiencing recent star-formation variations and use models specifically designed for these sources to fit their SEDs; this increases the utility of our catalog for various projects in the future. We calibrate our measurements by comparison with those in well-studied smaller regions and briefly discuss the implications of our results. We also perform detailed tests of the completeness and purity of SED-selected AGNs. Our data can be retrieved from a public website.
\end{abstract}
\keywords{Sky surveys, Celestial objects catalogs, Galaxies, Active galactic nuclei}

\section{Introduction}
The Vera C. Rubin Observatory Legacy Survey of Space and Time (LSST; \citealt{Ivezic19}) will be one of the most ambitious surveys in the coming years. It will survey the southern sky repeatedly in six optical bands ($ugrizy$) for ten years and observe billions of galaxies. Currently, five Deep-Drilling Fields (DDFs; e.g., \citealt{Brandt18, Scolnic18}) have been selected: COSMOS (Cosmic Evolution Survey), \mbox{W-CDF-S} (Wide Chandra Deep Field-South), ELAIS-S1 (European Large-Area ISO Survey-S1), XMM-LSS (XMM-Large Scale Structure), and EDF-S (Euclid Deep Field-South). Rubin will observe them with a higher cadence and greater sensitivity than those characterizing the wide survey. Rich multi-wavelength datasets (archival or planned) are available in all the DDFs. To name just a few, these include the XMM-Spitzer Extragalactic Representative Volume Survey (XMM-SERVS)\footnote{\url{http://personal.psu.edu/wnb3/xmmservs/xmmservs.html}} in \mbox{X-rays} \citep{Chen18, Ni21}, the Spitzer DeepDrill survey in the infrared \citep{Lacy21}, and the MeerKAT International GHz Tiered Extragalactic Exploration (MIGHTEE) survey in the radio \citep{Jarvis16, Heywood21}. The DDFs will be valuable for many kinds of studies involving time-domain astronomy, ultra-deep imaging, or multi-wavelength investigations. The selection of the EDF-S field as the fifth DDF was finalized only recently (in March 2022), during the review process of this article, and EDF-S currently has poorer multi-wavelength data than the other DDFs. We leave the corresponding analyses and discussion of this field to the future and largely focus on the four original DDFs (i.e., COSMOS, \mbox{W-CDF-S}, ELAIS-S1, and XMM-LSS) in the following text.\par
The COSMOS field has been extensively studied, with source properties cataloged carefully. Especially, the COSMOS2015 \citep{Laigle16} and COSMOS2020 \citep{Weaver22} catalogs contain refined photometry, photometric redshifts (photo-$z$s), and physical properties of sources derived from their spectral energy distributions (SEDs). The remaining three original DDFs (i.e., \mbox{W-CDF-S}, ELAIS-S1, and XMM-LSS), on the other hand, have not been fully investigated. To prepare for the upcoming LSST era, we have derived forced photometry and photo-$z$s for these fields in our previous works (\citealt{Nyland17, Chen18, Zou21a, Zou21b}; Nyland et al., in preparation) and present detailed SED fitting in this work.\par
Multiwavelength SEDs contain the imprints of all the physical processes in galaxies, and different parts of SEDs are generally dominated by different processes -- \mbox{X-rays} mainly trace the emission from active galactic nuclei (AGNs), UV-to-optical light is from young stars (and/or AGNs) and is absorbed by dust (if present), near-infrared (NIR) emission is mainly from intermediate-age and old stars, and mid-infrared (MIR) to far-infrared (FIR) emission is from dust. Therefore, SEDs can provide many insights about source properties, which is particularly important for large extragalactic photometric surveys. A notable example is deriving redshifts from SEDs. Spectroscopic redshifts are expensive and generally limited to bright sources; on the other hand, multiwavelength photometry is usually much easier to obtain, and fitting the resulting SEDs can provide photo-$z$s. This is also true for many other parameters, whose single tracers are often expensive to obtain (e.g., H$\alpha$ traces star formation well but is hard to measure), and thus SED fitting becomes vital for these cases.\par
By fitting SEDs with pre-constructed models, all the model parameters can be estimated, but their reliability is often not guaranteed. SED models are built upon all the detailed physical processes or empirical ones, and the internal uncertainties of these models themselves may cause biases in SED-fitting results; even if the models are correct in an average sense, they are often unable to span all the possible variations for individual galaxies, due to the limitations of both model flexibility and computational requirements, and hence many simplified assumptions are often inevitable. More importantly, SED fitting usually involves many parameters describing many physical processes, which couple together and form a complicated, nonlinear system. All of these issues have presented strong challenges to both the SED-fitting algorithms and interpretation of their results. Great efforts have been devoted to both of these. \citet{Walcher11} and \citet{Conroy13} are two useful reviews for fitting galaxy SEDs. There are many additional valuable related works that have appeared after these two reviews. Fig.~1 in \citet{Thorne21} summarizes the main features of the currently most popular SED-fitting codes, and \citet{Baes20} is a more recent review.\par
SED fitting can provide direct information for LSST sources and thus can serve as a basis for a variety of works in the future. For instance, AGNs are important because supermassive black holes (SMBHs) coevolve with their host galaxies over cosmic time (e.g., \citealt{Kormendy13, Brandt21}; and references therein). In fact, AGN studies will be a pillar of the science that will be performed in the LSST DDFs (e.g., \citealt{Brandt18}). We thus need to derive AGN host-galaxy properties via SED fitting, exploiting the rich multi-wavelength coverage of the DDFs, which has currently not been fully utilized. Furthermore, galaxies that are experiencing rapid bursting or quenching (BQ for short) of star formation are also scientifically important for both galaxy and SMBH studies. For example, they are good candidates for experiencing nonsecular galaxy evolution (e.g., \citealt{Smethurst15}) and can also help us understand the evolution of galaxies across the main sequence (MS; e.g., \citealt{Ciesla18}). The driving physics of the BQ phases is still unclear, and one possible cause is AGN activity (though it may not be the dominant one). There is indeed observational evidence showing that the quenching and AGN activities are correlated (e.g., \citealt{Smethurst16, Alatalo17, Greene20}). Additionally, tidal disruption events (TDEs) significantly prefer (post-)starburst galaxies, with the fraction of post-starburst galaxies among TDE hosts enhanced by a factor of $\sim20-200$ compared to general galaxy populations (e.g., \citealt{French20} and references therein), and such TDE hosts are undergoing rapid instead of slow quenching \citep{French17}. Selecting BQ galaxies in advance can thus help the identification and follow-up observations of TDE candidates in the LSST era \citep{French18}. Additionally, the SMBH masses of TDEs cannot exceed the Hills mass \citep{Hills75}, which provides a soft constraint for the TDE host-galaxy stellar masses ($M_\star$) given the correlation between the SMBH mass and $M_\star$; therefore, measuring $M_\star$ can also help TDE searches. These two types of sources, AGNs and BQ galaxies, have distinct SED features, and thus SEDs can be used to select them and gain insights. Especially, the XMM-SERVS survey provides medium-deep \mbox{X-ray} coverage and can thus significantly help the AGN selection and modeling. Their different SEDs from those of normal galaxies also make it necessary to model their SEDs appropriately in a different way from those of normal galaxies. It has been shown that using normal-galaxy templates to fit AGN host galaxies or BQ galaxies leads to inaccurate results (e.g., \citealt{Ciesla15, Ciesla17, Salvato19}) because for AGNs, the AGN emission is wrongly attributed to the galaxy emission, and for BQ galaxies, normal parametric star-formation histories (SFHs) do not have the flexibility to sample BQ-galaxy SEDs well.\par
There has been much work investigating AGNs and BQ galaxies in the COSMOS field (e.g., \citealt{Aufort20, Ni20}), separately from the general COSMOS catalogs, but as far as we know, no works are available that have systematically analyzed AGNs and BQ galaxies in our fields. To increase the utility of our work and prepare for broader investigations in the future, we decided to select AGNs and BQ galaxies and use models designed for them to fit their SEDs aside from the fitting of normal galaxies.\par
This work mainly provides catalogs recording source classifications, $M_\star$, star-formation rates (SFR), and other related properties. Throughout this paper, we focus on the \mbox{W-CDF-S} field in the main text and put the results for ELAIS-S1 and XMM-LSS into two appendices to keep the narrative flow clear. \mbox{W-CDF-S} is chosen as the representative example because it has the most complete previous literature for comparison. For example, \mbox{CDF-S}, which is embedded in \mbox{W-CDF-S}, has the deepest X-ray observations ever obtained \citep{Luo17, Xue17, Brandt21} and thus can provide a largely complete pure AGN sample for calibration of AGN selection.\par
This paper is structured as follows. Section~\ref{sec: data} describes the data. In Section~\ref{sec: sed_step1}, we run the SED fitting and classify our sources into four categories -- star, AGN candidate, BQ-galaxy candidate, or normal galaxy. Section~\ref{sec: sedfitting} presents the analyses of our results and relevant discussions. Section~\ref{sec: summary} summarizes this work. Appendices~\ref{append: es1} and \ref{append: xmmlss} present the SED analyses in ELAIS-S1 and XMM-LSS, respectively. We adopt a flat $\Lambda\mathrm{CDM}$ cosmology with $H_0=70~\mathrm{km~s^{-1}~Mpc^{-1}}$, $\Omega_\Lambda=0.7$, and $\Omega_M=0.3$.

\section{Data}
\label{sec: data}
\subsection{Overview}
Our full sample includes 0.8 million sources covering $4.9~\mathrm{deg^2}$ in \mbox{W-CDF-S}, 0.8 million sources covering $3.4~\mathrm{deg^2}$ in ELAIS-S1, and 1.2 million sources covering $4.9~\mathrm{deg^2}$ in XMM-LSS. All sources are required to be detected in the VISTA Deep Extragalactic Observations (VIDEO) survey (i.e., detected in any one of the VIDEO $ZYJHK_s$ bands; \citealt{Jarvis13}) because the VIDEO data are necessary for obtaining quality forced photometry (Section~\ref{sec: tractorphot}) and sufficient SED coverage in wavelength. The relatively smaller source surface-number density on $\mathrm{deg^2}$ scales in \mbox{W-CDF-S} is caused by the fact that the currently available VIDEO data are shallower in some parts of \mbox{W-CDF-S}. For example, the $Z$ band only covers $1.8~\mathrm{deg^2}$ in \mbox{W-CDF-S}. Due to this reason, the source surface number density shows a global variation across \mbox{W-CDF-S}, and one should not analyze, e.g., the demographics or spatial clustering of sources without accounting for this factor. ELAIS-S1 and XMM-LSS do not have this issue -- they are covered by all the VIDEO bands. The basic information for the three fields is listed in Table~\ref{tbl_fieldinfo}. We also refer readers to Tables~1 in \citet{Chen18} and \citet{Ni21} for similar summaries.\par

\begin{table*}
\caption{Basic information for the three fields}
\label{tbl_fieldinfo}
\centering
\resizebox{\hsize}{!}{
\begin{threeparttable}
\begin{tabular}{cccc}
\hline
\hline
& \mbox{W-CDF-S} & ELAIS-S1 & XMM-LSS\\
\hline
Center (J2000) & $\mathrm{RA=03^h32^m09^s, Dec=-28^\circ08'32''}$ & $\mathrm{RA=00^h37^m47^s, Dec=-44^\circ00'07''}$ & $\mathrm{RA=02^h22^m10^s, Dec=35^\circ32'30''}$\\
\hline
Area & $4.9~\mathrm{deg^2}$ & $3.4~\mathrm{deg^2}$ & $4.9~\mathrm{deg^2}$\\
\hline
Source number & 799607 & 826242 & 1247954\\
Star$^\star$ & 42628 & 56850 & 49230\\
AGN$^\star$ & 19612 & 18454 & 41568\\
BQ galaxy$^\star$ & 3624 & 4304 & 20852\\
Normal galaxy$^\star$ & 733743 & 746634 & 1136304\\
Reliable SED AGNs$^\sharp$ & 2652 & 2507 & 3658\\
\hline
\mbox{X-ray} survey & XMM-SERV$\mathrm{S^a}$: \mbox{X-ray} & XMM-SERV$\mathrm{S^b}$: \mbox{X-ray} & XMM-SERV$\mathrm{S^b}$: \mbox{X-ray}\\
\hline
UV survey & GALE$\mathrm{X^c}$: FUV and NUV & GALE$\mathrm{X^c}$: FUV and NUV & GALE$\mathrm{X^c}$: FUV and NUV\\
\hline
\multirow{3}{*}{Optical surveys} & \multirow{2}{*}{VOIC$\mathrm{E^d}$: $ugri$} & VOIC$\mathrm{E^d}$: $u$ & \multirow{2}{*}{CFHTL$\mathrm{S^h}$: $ugriz$}\\
& \multirow{2}{*}{HS$\mathrm{C^e}$: $griz$} & ESI$\mathrm{S^f}$: $BVR$ & \multirow{2}{*}{HS$\mathrm{C^i}$: $grizy$}\\
& & DE$\mathrm{S^g}$: $grizY$\\
\hline
\multirow{4}{*}{IR surveys} & VIDE$\mathrm{O^j}$: $ZYJHK_s$ & VIDE$\mathrm{O^j}$: $ZYJHK_s$ & VIDE$\mathrm{O^j}$: $ZYJHK_s$\\
& DeepDril$\mathrm{l^k}$: 3.6 and 4.5~$\mu\mathrm{m}$ & DeepDril$\mathrm{l^k}$: 3.6 and 4.5~$\mu\mathrm{m}$ & DeepDril$\mathrm{l^k}$: 3.6 and 4.5~$\mu\mathrm{m}$\\
& SWIR$\mathrm{E^l}$: 5.8, 8, 24, 70, and 160~$\mu\mathrm{m}$ & SWIR$\mathrm{E^l}$: 5.8, 8, 24, 70, and 160~$\mu\mathrm{m}$ & SWIR$\mathrm{E^l}$: 5.8, 8, 24, 70, and 160~$\mu\mathrm{m}$\\
& HerME$\mathrm{S^m}$: 100, 160, 250, 350, and 500~$\mu\mathrm{m}$ & HerME$\mathrm{S^m}$: 100, 160, 250, 350, and 500~$\mu\mathrm{m}$ & HerME$\mathrm{S^m}$: 100, 160, 250, 350, and 500~$\mu\mathrm{m}$\\
\hline
\hline
\end{tabular}
\begin{tablenotes}
\item
\textit{Notes.} $^\star$ These rows are the numbers of sources whose ``best'' results are from the corresponding categories; see Section~\ref{sec: bestsedfittingresults} for more details. $^\sharp$ These are the numbers of calibrated, reliable SED AGNs; see Section~\ref{sec: sedagn} and Appendices~\ref{append: es1} and \ref{append: xmmlss} for more details. The full names of the survey or mission acronyms are listed as the following. XMM-SERVS is The XMM-Spitzer Extragalactic Representative Volume Survey, GALEX is The Galaxy Evolution Explorer, VOICE is The VST Optical Imaging of the CDF-S and ELAIS-S1 Fields, HSC is The Hyper Suprime-Cam, ESIS is The ESO-Spitzer Imaging extragalactic Survey, DES is The Dark Energy Survey, CFHTLS is The Canada-France-Hawaii Telescope Legacy Survey, VIDEO is The VISTA Deep Extragalactic Observations, DeepDrill is The Spitzer Survey of Deep-Drilling Fields, SWIRE is The Spitzer Wide-area Infrared Extragalactic survey, and HerMES is The Herschel Multi-tiered Extragalactic Survey. Example references: [a] \citet{Chen18}; [b] \citet{Ni21}; [c] \citet{Martin05}; [d] \citet{Vaccari16}; [e] \citet{Ni19}; [f] \citet{Berta06}; [g] \citet{Abbott21}; [h] \citet{Hudelot12}; [i] \citet{Aihara18}; [j] \citet{Jarvis13}; [k] \citet{Lacy21}; [l] \citet{Lonsdale03}; [m] \citet{Oliver12}.
\end{tablenotes}
\end{threeparttable}
}
\end{table*}

Our photometry has been collected in a non-simultaneous manner, which not only applies to different bands, but also for single bands because the single-band images were merged from observations that often span several years. Possible photometric variability is not expected to influence our general results because sources with strong photometric variations are mainly bright type~1 AGNs that outshine their host galaxies, which are rare and also need extra caution that is beyond our general analyses (see Section~\ref{sec: qsohost} for further discussion). The fact that our single-band images are often from several observations also suppresses the impact of possible variability. Nevertheless, multi-epoch SED analyses can help investigate some rare but interesting non-galaxy sources (e.g., \citealt{Senarath21}) in the future, and LSST will provide high-cadence light curves for time-domain science.\par
We will mainly rely upon \texttt{CIGALE} (Code Investigating GALaxy Emission) v2022.0\footnote{\url{https://cigale.lam.fr}} \citep{Boquien19, Yang20, Yang22} for the SED fitting. This code is based on an energy-balance principle and can decompose an SED into several user-defined components (including AGNs) from \mbox{X-ray} to radio. We choose \texttt{CIGALE} mainly for three reasons. First, its efficient parallel algorithm enables fast modeling for millions of sources. Second, its ability to fit AGN SEDs is the most advanced among current SED-fitting codes and has been well probed in the literature. For example, it allows modeling of the \mbox{X-ray} photometry and has state-of-the-art AGN templates \citep{Yang20, Yang22}. Previous literature has explored the best fitting strategies and justified its reliability for modeling AGN SEDs; see, e.g, \citet{Ciesla15, Buat21, Mountrichas20, Padilla21}. Third, dedicated studies of using \texttt{CIGALE} to fit BQ galaxies are also available; see, e.g., \citet{Boselli16, Aufort20, Ciesla21}. Due to its efficiency, accuracy, and flexibility, \texttt{CIGALE} has been widely used in other extragalactic survey studies (e.g., \citealt{Malek18, Zou19, Ni20}). However, \texttt{CIGALE} is not used to derive our photo-$z$s because this function has not been thoroughly evaluated, and its large number of parameters may lead to strong degeneracy in photo-$z$s. Our photo-$z$s were derived using other dedicated SED-fitting codes; see Section~\ref{sec: redshift} and references therein.\par
We utilize photometry from the \mbox{X-ray} to FIR to perform the SED fitting. The following subsections will present our compilation and reduction of the photometry and redshifts in \mbox{W-CDF-S}\footnote{We will not explicitly write ``\mbox{W-CDF-S}'' hereafter. Unless noted in the main text, we always refer to \mbox{W-CDF-S} instead of ELAIS-S1 or XMM-LSS.} as a representative example. Almost the whole field is covered by \mbox{X-ray} to FIR surveys, and the multi-wavelength coverage is presented in Fig.~1 of \citet{Ni21}.

\subsection{X-Ray Photometry}
\label{sec: xrayphot}
Our \mbox{X-ray} photometry is from the XMM-SERVS survey \citep{Chen18, Ni21}, which has observed the \mbox{W-CDF-S} field for 2.3 Ms, reaching a flux limit of $\approx1.0\times10^{-14}~\mathrm{erg~cm^{-2}~s^{-1}}$ in the $0.5-10$ keV band. \mbox{X-ray} sources have already been matched to \textit{The Tractor} catalog (Section~\ref{sec: tractorphot}; Nyland et al., in preparation) in \citet{Ni21}. Simple positional closest-radius matching is not suitable for matching these \mbox{X-ray} sources to their multi-wavelength counterparts because XMM-Newton has non-negligible positional uncertainties, and thus \citet{Ni21} used a Bayesian method that takes the offsets, magnitudes, and colors into consideration simultaneously to do the matching, as detailed in their Section~4. 3319 of our sources have reliable \mbox{X-ray} counterparts in \citet{Ni21}, and the others will be assigned \mbox{X-ray} upper limits in this section. The impacts of the \mbox{X-ray} data as well as the upper limits to our SED fitting will be discussed in detail in Section~\ref{sec: xraydatapoint}. There are 734 \mbox{X-ray} sources in \citet{Ni21} not included in our VIDEO-based sample. About one-third of them are not included because their VIDEO counterparts cannot be reliably assigned as there may be multiple possible counterparts corresponding to a single \mbox{X-ray} source, and the others are undetected in VIDEO (see Section~4 in \citealt{Ni21} for more details).\par
The \mbox{X-ray} point-source catalog in \citet{Ni21} only presents observed \mbox{X-ray} fluxes, i.e., uncorrected for intrinsic obscuration (but corrected for Galactic obscuration). However, \texttt{CIGALE} needs absorption-corrected \mbox{X-ray} fluxes, and we thus estimate the intrinsic \mbox{X-ray} luminosities directly based on the \mbox{X-ray} count maps using a Bayesian approach. Bright sources are generally not affected by the Bayesian approach, and the prior (see below) can regulate the posteriors of faint sources so that the Eddington bias can be corrected. Our following method is optimized for AGNs because the majority of the \mbox{X-ray} sources with longer-wavelength counterparts are AGNs. Pure galaxies may present low-level \mbox{X-ray} emission mainly from \mbox{X-ray} binaries. \citet{Lehmer16} presented scaling relations for the \mbox{X-ray} luminosity from \mbox{X-ray} binaries as functions of $M_\star$, SFR, and $z$. We estimated such galaxy-only \mbox{X-ray} luminosities using our $M_\star$ and SFR measurements (Section~\ref{sec: sedfitting}) and confirmed that they are generally orders of magnitude lower than our observed luminosities, and the excess \mbox{X-ray} emission is expected to arise from AGNs.\par
We take the column density, $N_\mathrm{H}$, and intrinsic $2-10$ keV luminosity, $L_\mathrm{X}$, as the free parameters. The model flux between the observed-frame energy range, $E_\mathrm{low}-E_\mathrm{high}$, is
\begin{align}
f_\mathrm{X}=
\begin{cases}
\frac{L_\mathrm{X}}{4\pi D_L^2}(1+z)^{2-\Gamma}\frac{E_\mathrm{high}^{2-\Gamma}-E_\mathrm{low}^{2-\Gamma}}{10^{2-\Gamma}-2^{2-\Gamma}}\eta,~\Gamma\neq2\\
\frac{L_\mathrm{X}}{4\pi D_L^2}\frac{\ln\frac{E_\mathrm{high}}{E_\mathrm{low}}}{\ln5}\eta,~\Gamma=2
\end{cases}
,\label{eq_def_fx}
\end{align}
where the full derivation is presented in Appendix~\ref{append: eq1}, $E_\mathrm{low}$ and $E_\mathrm{high}$ are in keV, $D_L$ is the luminosity distance at redshift $z$, $\Gamma$ is the power-law photon index of the source's intrinsic spectrum (assumed to be 1.8), and $\eta=\eta(N_\mathrm{H}, z; E_\mathrm{low}, E_\mathrm{high}, \Gamma)$ is the flux-reduction factor if the source emission (assumed to be a power-law) is absorbed by both the Galaxy and the source itself, where the column density of the Galaxy toward the \mbox{W-CDF-S} is taken to be $8.4\times10^{19}~\mathrm{cm^{-2}}$ \citep{Ni21}. $\eta$ is calculated using \texttt{XSPEC} \citep{Arnaud96}. The model flux is then converted to the predicted source counts within $5\times5$ pixels (i.e., $20''\times20''$) using the single-camera exposure ($t$) maps, encircled energy fraction (EEF) maps, and energy conversion factors (ECFs) in \citet{Ni21}, where EEF is the expected fraction of source photons falling within the given aperture centered at the position of the source, and ECF is the expected ratio between the source flux and the source counts. We follow Eq.~9 in \citet{Ruiz21}:
\begin{align}
M(k)=f_\mathrm{X}(k)\sum_{i=1}^3t_i(k)\mathrm{EEF}_i(k)\mathrm{ECF}_i(k),\label{eq: flux2counts}
\end{align}
where the subscript, $i$, denotes the cameras (EPIC PN, MOS1, and MOS2), and $k$ denotes soft (SB), hard (HB), and full (FB) bands. As done in \citet{Ni21}, $E_\mathrm{low}=0.5$~keV, $E_\mathrm{high}=2$~keV, and $M(\mathrm{SB})$ is calculated between $0.2-2$~keV for SB; $E_\mathrm{low}=2$~keV, $E_\mathrm{high}=10$~keV, and $M(\mathrm{HB})$ is calculated between $2-12$~keV for HB; $E_\mathrm{low}=0.5$~keV, $E_\mathrm{high}=10$~keV, and $M(\mathrm{FB})$ is calculated between $0.2-12$~keV for FB. In fact, the actual ECFs depend on the spectral shape, and our adopted values from \citet{Ni21} are only approximations. Based on the standard XMM-Newton response files,\footnote{\url{https://www.cosmos.esa.int/web/xmm-Newton/epic-response-files}} the HB ECF is estimated to vary within $\sim0.05~\mathrm{dex}$ around the value corresponding to $\Gamma=1.4$ and $N_\mathrm{H}=0$ for different $\Gamma$ and $N_\mathrm{H}$, and the SB ECF may deviate up to $\sim0.2~\mathrm{dex}$ when the SB counts are larger than the HB counts. Therefore, the variation of the ECF is only modest and unlikely to bias our results significantly.\par
To compare the model counts with the observed counts, we further assume that the expected background intensity can be accurately measured. This assumption, also adopted in \citet{Ruiz21}, is reasonable because the intensity is estimated based on many background counts.\footnote{For example, for a typical background aperture with a radius of $60''$, which is much larger than the source aperture, there are over 2000 background counts, much larger than the typical source counts of a few tens to a few hundreds. Therefore, the relative uncertainty of the background intensity is generally much smaller than that of the source intensity.} The likelihood is thus\par
\begin{align}
\mathcal{L}=\prod_{k\in\mathrm{\{SB, HB\}}}\frac{[M(k)+B(k)]^{S(k)}e^{-[M(k)+B(k)]}}{S(k)!},
\end{align}
where $B(k)$ is the estimated background intensity within the source region (defined as $5\times5$ pixels around the source) based on the background maps \citep{Ni21}, $S(k)$ is the observed counts within the source region, and $\mathcal{L}$ is essentially the Poisson probability of observing $S(k)$ photons when the expected counts are $M(k)+B(k)$.\par
The prior is adopted as the product of the \mbox{X-ray} luminosity function (XLF) in \citet{Ananna19},\footnote{We have also tried using the XLF in \citet{Ueda14} and obtained similar results.} which is a function of not only $L_\mathrm{X}$ and $z$, but also $N_\mathrm{H}$, and the probability that the source is detected (see the next paragraph). We further set $\mathrm{XLF}=0$ when $N_\mathrm{H}\geq10^{24}~\mathrm{cm^{-2}}$ for two reasons. First, other complex components besides the simple transmission are important when $N_\mathrm{H}\geq10^{24}~\mathrm{cm^{-2}}$ (e.g., \citealt{Li19}), and it is impossible to use only two data points (i.e., SB and HB counts) to constrain them; secondly, the $N_\mathrm{H}$ distribution itself is not well understood when $N_\mathrm{H}\geq10^{24}~\mathrm{cm^{-2}}$ (e.g., \citealt{Ueda14, Yang21a}). Compton-thick (CT) AGNs should be selected and analyzed individually. For example, \mbox{X-ray} spectral analyses should be optimized specifically for heavily obscured sources to select CT AGNs (\citealt{Lanzuisi18}; Yan et al., in preparation). Besides, Yan et al. (in preparation) searched for CT AGNs in XMM-SERVS and only found several dozen candidates, indicating that the assumption of $N_\mathrm{H}\leq10^{24}~\mathrm{cm^{-2}}$ is appropriate for most of our sources.\par
We adopt the Poisson likelihood to estimate the detection probability (i.e., $D$) of a source, which, in addition to the XLF, constitutes the adopted prior. The Poisson likelihood roughly follows a one-to-one relationship with the sophisticated PSF-fitting likelihood during the real detection process, though the scatter is large \citep{Liu20, Ni21}. First, we would expect a source to be detected in a band if its counts are strictly larger than a detection threshold (denoted as $N_\mathrm{thres}$) at a given significance (denoted as $P_\mathrm{Poisson}$), where $P_\mathrm{Poisson}(\mathrm{SB})=0.03$,  $P_\mathrm{Poisson}(\mathrm{HB})=7.5\times10^{-5}$, and $P_\mathrm{Poisson}(\mathrm{FB})=0.03$ (see \citealt{Ni21} for more details). $N_\mathrm{thres}$ is thus the minimum non-negative integer that satisfies
\begin{align}
\mathfrak{Prob}\{\mathbb{POI}(B)\leq N_\mathrm{thres}\}&\geq1-P_\mathrm{Poisson},\label{eq: def_Nthres_part1}\\
\Leftrightarrow\mathcal{P}_\mathrm{IG}(N_\mathrm{thres}+1, B)&\leq P_\mathrm{Poisson},
\end{align}
where $\mathfrak{Prob}$ means probability, $B$ is the expected background counts, $\mathbb{POI}(B)$ represents a Poisson random variable with rate $B$, and $\mathcal{P}_\mathrm{IG}(a, x)$ is the regularized lower incomplete gamma function.\footnote{The conventional notations of ``probability'', ``Poisson distribution'', and ``regularized lower incomplete gamma function'' are all ``$P$'', and thus we use different styles to distinguish them.} We denote $A(y, x)$ as the inversion of $\mathcal{P}_\mathrm{IG}(a, x)$ that takes $a$ as the independent variable and $x$ as the parameter, i.e., $\mathcal{P}_\mathrm{IG}(A(y, x), x)=y$. Then
\begin{align}
N_\mathrm{thres}=
\begin{cases}
\lceil A(P_\mathrm{Poisson}, B)\rceil-1,~B>0\\
0,~B=0
\end{cases},
\end{align}
where $\lceil x\rceil$ means the ceiling function of $x$. Based on these, we derive the band-merged detection probability, $D(N_\mathrm{H}, L_\mathrm{X}, z)$, in Appendix~\ref{append: xray_detection}.\par
\begin{figure}[h]
\centering
\resizebox{\hsize}{!}{\includegraphics{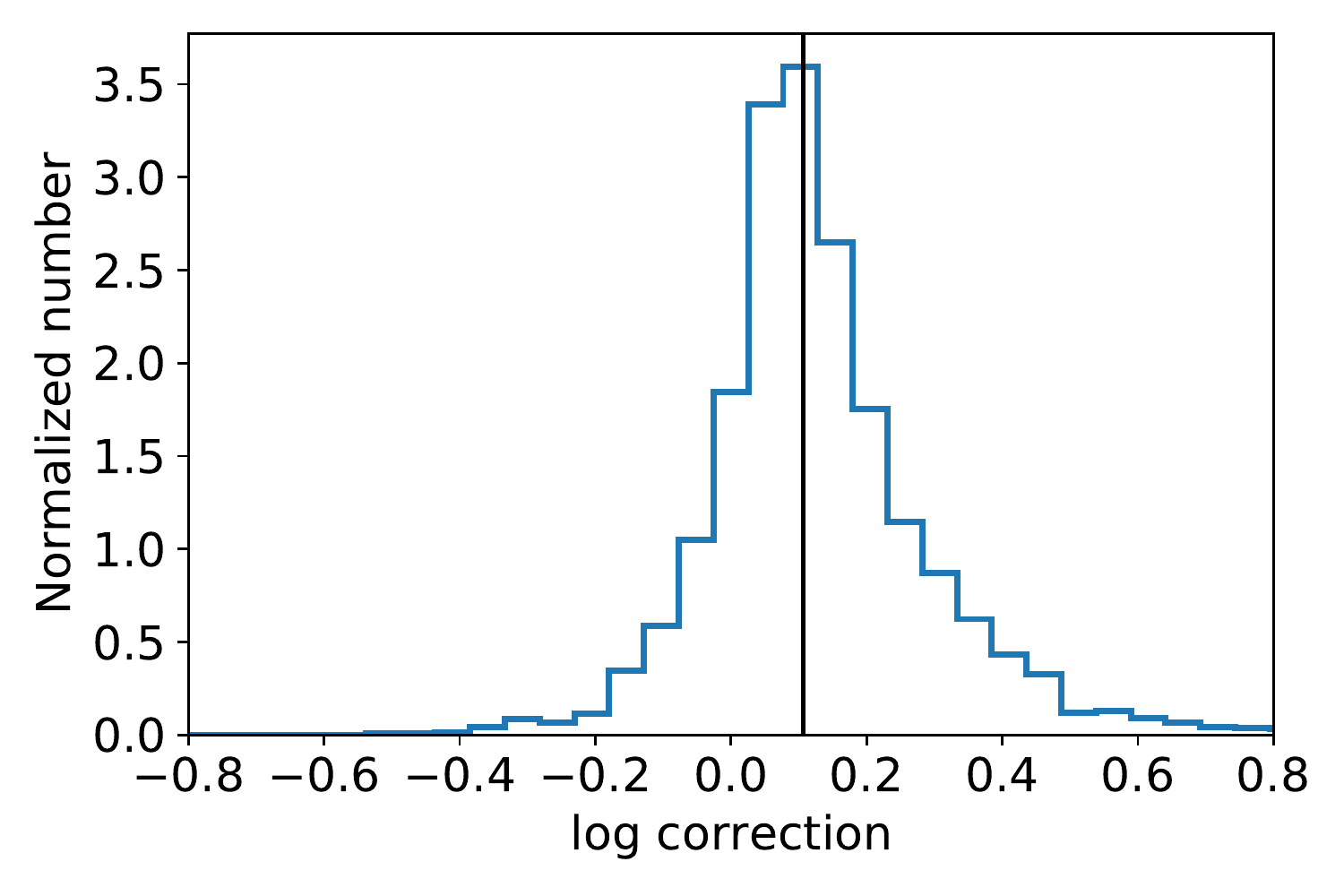}}
\caption{The distribution of the correction factor, defined as the ratio between $L_\mathrm{X}$ and the observed \mbox{X-ray} luminosities, for \mbox{X-ray}-detected sources. The vertical black line marks the median correction (0.1~dex).}
\label{fig_xraycorr}
\end{figure}

The posterior is thus $\mathfrak{Prob}(N_\mathrm{H}, L_\mathrm{X})\propto\mathcal{L}(N_\mathrm{H}, L_\mathrm{X})\times\mathrm{XLF}\times D$, and the expected $L_\mathrm{X}$ and its standard deviation are estimated by integrating the posterior using the \texttt{HCubature} module in \texttt{Julia} \citep{Julia17}. To prevent the XLF, which increases rapidly at low $L_\mathrm{X}$, from dominating the integration,\footnote{The large value given by the XLF cannot be fully counterbalanced by the small value of $D$ for very faint fluxes because the XLF increases roughly following a power-law while $D$ converges to a finite, non-zero constant when the source flux decreases.} we set the posterior to be 0 when $D(N_\mathrm{H}=0, L_\mathrm{X}, z)$ drops below 0.2,\footnote{Also note that when $D$ is small, the difference between the Poisson likelihood and the actual PSF-fitting likelihood shows large variations (see Fig.~5 in \citealt{Liu20}), and thus $D$ itself may deviate from the actual detection probability. The threshold, 0.2, only serves as an empirical value, and this value is also not too much larger than the smallest possible value of $D$ (i.e., $D(L_\mathrm{X}=0)=0.04$.)} and this threshold corresponds to fluxes small enough to be roughly several tens of times smaller than the sensitivities. $L_\mathrm{X}$ is then converted to the intrinsic $2-10$~keV flux, as the SED fitting requires. Fig.~\ref{fig_xraycorr} displays the distribution of the correction factor, defined as the ratio between $L_\mathrm{X}$ and the observed $2-10$~keV \mbox{X-ray} luminosities in \citet{Ni21}. The median correction is 0.1~dex, which is modest and indicates that absorption effects are unlikely to cause significant biases to our results. Such small corrections are also confirmed at similar \mbox{X-ray} fluxes through direct \mbox{X-ray} spectral fitting in \citet{Yang18b}.

\begin{figure}
\centering
\resizebox{1.1\hsize}{!}{\includegraphics{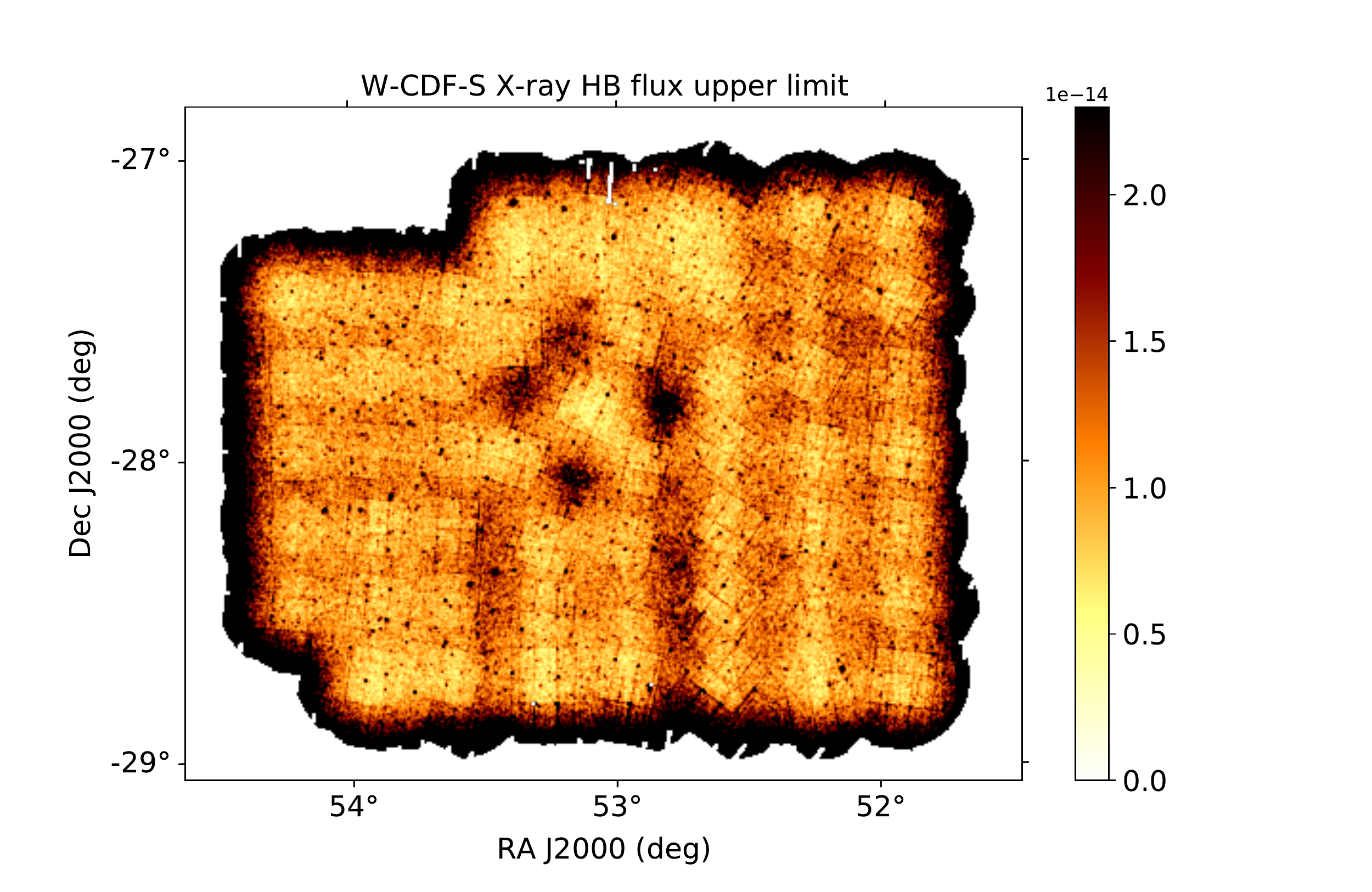}}
\caption{The HB flux upper-limit map with units of $\mathrm{erg~cm^{-2}~s^{-1}}$. We derive flux upper limits for \mbox{X-ray}-undetected sources based on this map.}
\label{fig_fupp_HB}
\end{figure}

\texttt{CIGALE} supports using flux upper limits to constrain the fitting (see \citealt{Boquien19} for more details), and thus we derive the $3\sigma$ observed HB flux upper-limit map following the method in \citet{Ruiz21} for the remaining sources undetected in any of the \mbox{X-ray} bands. The HB is adopted because it is less affected by absorption effects. The upper limit is
\begin{widetext}
\begin{align}
f_\mathrm{X, upp}=\frac{\mathcal{P}_\mathrm{IG}^{-1}(S(\mathrm{HB})+1, 0.9987\mathcal{Q}_\mathrm{IG}(S(\mathrm{HB})+1, B(\mathrm{HB}))+\mathcal{P}_\mathrm{IG}(S(\mathrm{HB})+1, B(\mathrm{HB})))-B(\mathrm{HB})}{\sum_{i=1}^3t_i(\mathrm{HB})\mathrm{EEF}_i(\mathrm{HB})\mathrm{ECF}_i(\mathrm{HB})},
\end{align}
\end{widetext}
where $\mathcal{P}_\mathrm{IG}^{-1}(a, y)$ is the inverse function of $\mathcal{P}_\mathrm{IG}(a, x)$ (i.e., $\mathcal{P}_\mathrm{IG}(a, \mathcal{P}_\mathrm{IG}^{-1}(a, y))=y$), $\mathcal{Q}_\mathrm{IG}(a, x)=1-\mathcal{P}_\mathrm{IG}(a, x)$ is the regularized upper incomplete gamma function, and 0.9987 is the one-sided $3\sigma$ confidence level. Fig.~\ref{fig_fupp_HB} shows the resulting map. \citet{Ni21} also provide sensitivity maps, but our flux upper limit is conceptually different from sensitivity. Their subtle differences are detailed in \citet{Kashyap10}, where ``upper limit'' in our article is referred to as ``upper bound'' in theirs. Briefly, the sensitivity in \citet{Ni21} is roughly the detection threshold and thus only depends on the background intensity, but our flux upper limit is the largest possible value that a source can have at a given confidence level and depends on both the background and the signals within the source region (no matter whether the source is detected or not). Moreover, the detection significance of the HB sensitivity map in \citet{Ni21} is $7.5\times10^{-5}$, much more conservative than our adopted upper-limit significance ($1.3\times10^{-3}$).\par

\begin{figure}
\centering
\resizebox{\hsize}{!}{\includegraphics{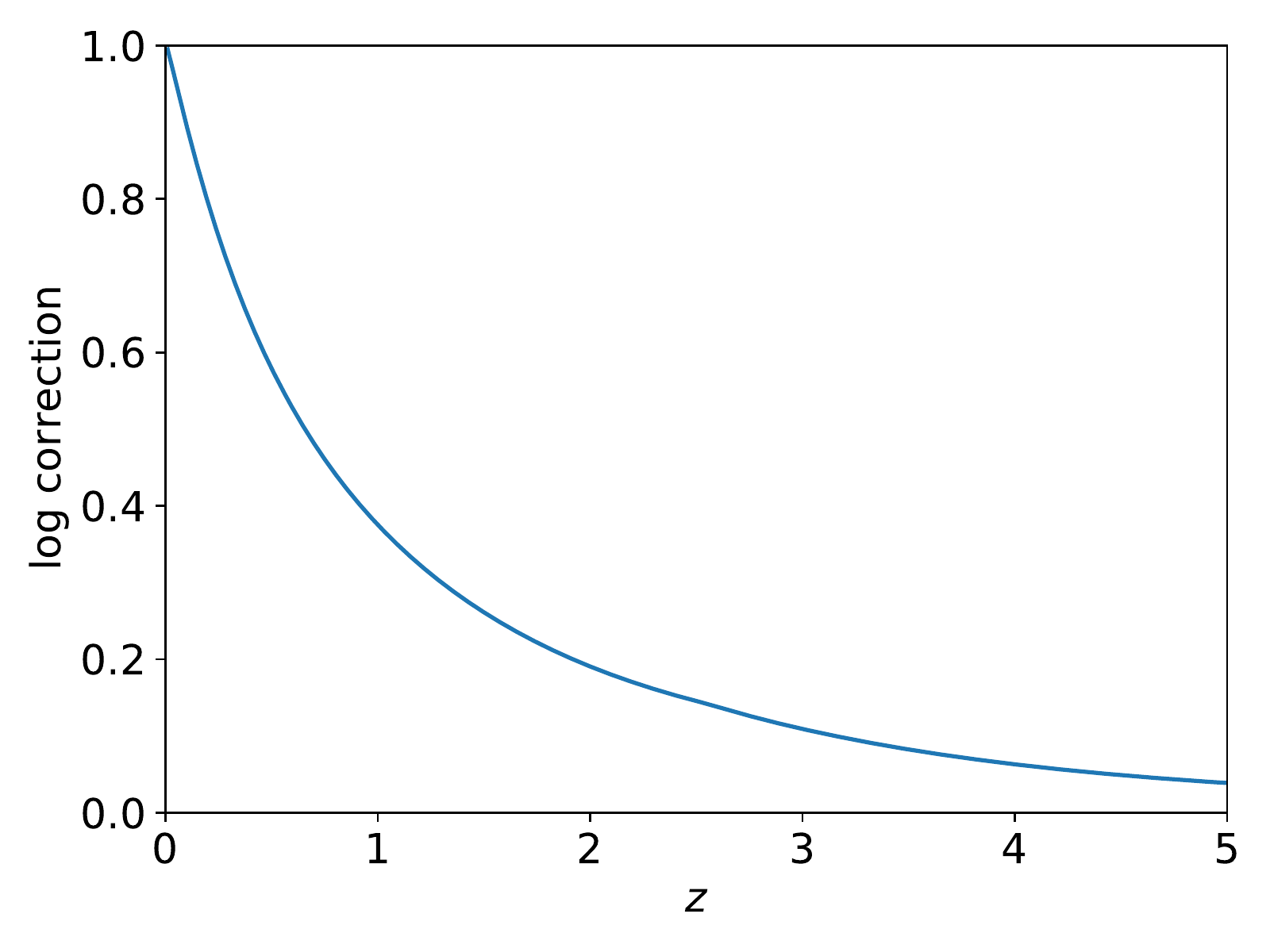}}
\caption{The correction factor of the HB flux upper limit for undetected sources, $C_\mathrm{undet}$, as a function of $z$. We multiply the observed HB flux upper limits by this factor to obtain the intrinsic HB flux upper limits.}
\label{fig_senscorr}
\end{figure}

We then convert the observed HB flux upper limits to the intrinsic HB flux upper limits for undetected sources using the $\eta_\mathrm{HB}$ function, i.e., the $\eta$ in Eq.~\ref{eq_def_fx} for HB. Since $\eta_\mathrm{HB}$ depends on $N_\mathrm{H}$, whose distribution further depends on $z$ and $L_\mathrm{X}$, and both $N_\mathrm{H}$ and $L_\mathrm{X}$ are unknown for undetected sources, we would like to derive typical correction factors independent of $N_\mathrm{H}$ and $L_\mathrm{X}$. The redshift dependence of the $N_\mathrm{H}$ distribution is addressed by the XLF. The undetected sources in which we are interested when deriving the corrections are those that may be detected if their $N_\mathrm{H}$ values were 0, and thus we add a weight of $D(N_\mathrm{H}=0, L_\mathrm{X}, z)$. The expected HB flux-correction factor for undetected sources is thus
\begin{widetext}
\begin{align}
C_\mathrm{undet}(z)=\frac{\int_{20}^{24}\int_{\log{L_\mathrm{X, low}}}^{50}D(N_\mathrm{H}=0, L_\mathrm{X}, z)\left[1-D(N_\mathrm{H}, L_\mathrm{X}, z)\right]\mathrm{XLF}(\log N_\mathrm{H}, \log L_\mathrm{X}, z)/\eta_\mathrm{HB}d\log L_\mathrm{X}d\log N_\mathrm{H}}{\int_{20}^{24}\int_{\log{L_\mathrm{X, low}}}^{50}D(N_\mathrm{H}=0, L_\mathrm{X}, z)\left[1-D(N_\mathrm{H}, L_\mathrm{X}, z)\right]\mathrm{XLF}(\log N_\mathrm{H}, \log L_\mathrm{X}, z)d\log L_\mathrm{X}d\log N_\mathrm{H}},\label{eq: def_Cundet}
\end{align}
\end{widetext}
where the upper integration bound of $L_\mathrm{X}$ ($10^{50}~\mathrm{erg~s^{-1}}$) is an arbitrary large number, and the lower integration bound of $L_\mathrm{X}$, $\log{L_\mathrm{X, low}}$, is set to prevent the rapidly increasing XLF from dominating the integration in the small $L_\mathrm{X}$ regime. As for the detected case, we define $\log{L_\mathrm{X, low}}$ as the value when $D(N_\mathrm{H}=0, L_\mathrm{X})$ drops down to 0.2. In principle, Eq.~\ref{eq: def_Cundet} is valid for every pixel and can thus lead to $C_\mathrm{undet}$ maps as a function of $z$, but this is too computationally demanding. Instead, we simply adopt the median values of the background maps as $B(k)$, i.e., $B(\mathrm{SB})=32.5$, $B(\mathrm{HB})=43.9$, and $B(\mathrm{FB})=76.6$ within $5\times5$ pixels. The corresponding $N_\mathrm{thres}$ values are 44, 71, and 93 counts for SB, HB, and FB, respectively. The conversion factor from flux to counts in each band is also adopted as the median value of the conversion-factor map (i.e., $\sum_{i=1}^3t_i(k)\mathrm{EEF}_i(k)\mathrm{ECF}_i(k)$; cf., Eq.~\ref{eq: flux2counts}) -- $2.1\times10^{16}$ and $2.5\times10^{15}$~$\mathrm{counts~erg^{-1}~cm^2~s}$ for SB and HB, respectively. The resulting $C_\mathrm{undet}(z)$ is shown in Fig.~\ref{fig_senscorr}. The intrinsic flux upper limit of a source in observed-frame $2-10~\mathrm{keV}$ is obtained by multiplying the value from the HB flux upper-limit map with $C_\mathrm{undet}$ at its redshift.\par

The analyses in this section are done for all the XMM-SERVS fields, i.e., also for ELAIS-S1 and XMM-LSS.

\subsection{UV Photometry}
We collect UV photometry from GALEX (\citealt{Martin05}). Sources in \mbox{W-CDF-S} usually have multiple measurements in the GALEX database, and we follow a similar method as \citet{Bianchi17} to select unique measurements for each source. We first rank all the GALEX sources based on the detection status, exposure time, and distance from the center of the observed field (\texttt{fov\_radius} in the GALEX catalog). The rank of detection is the following: sources detected in both NUV and FUV are ranked the highest, those only detected in NUV are the second, and those only detected in FUV are ranked the lowest. If two sources have the same detection status, the one with higher exposure time is ranked higher. If two sources further have the same exposure time, the one with smaller \texttt{fov\_radius} is ranked higher. Proceeding from the source with the highest rank to the one with the lowest rank, we link surrounding sources that are within $2.5''$ of the primary source and from different observations to the primary one and remove these surrounding sources in the catalog, and the remaining primary sources constitute our unique-source catalog. We confirmed that the fluxes of the removed sources are generally consistent with those of the primary sources, indicating that they are indeed from the same objects.\par
We then cross match the cleaned GALEX catalog to \textit{The Tractor} catalog (Section~\ref{sec: tractorphot}) with a matching radius of $2''$.

\subsection{\textit{The Tractor} Photometry}
\label{sec: tractorphot}
The photometry for \mbox{W-CDF-S} from $0.36-4.5~\mu\mathrm{m}$ is compiled in Nyland et al. (in preparation), including VOICE $ugri$ \citep{Vaccari16}, HSC $griz$ \citep{Ni19}, VIDEO $ZYJHK_s$ \citep{Jarvis13}, and DeepDrill IRAC 3.6 and 4.5 $\mu\mathrm{m}$ \citep{Lacy21}. They adopted the band with the longest wavelength among the VIDEO bands in which a given source is detected as the fiducial band to derive the forced photometry in other bands using \textit{The Tractor} code \citep{Lang16}. This technique provides self-consistent photometry across different bands, partly deblends low-resolution images, and extends photometric measurements to a fainter magnitude regime, and thus the resulting photometric catalog is expected to be suitable for our multi-wavelength study. More details of the forced-photometry measurements are presented in \citet{Nyland17}, \citet{Zou21a}, and Nyland et al. (in preparation).\par
We further found that residual atmospheric extinction may slightly affect the HSC $g$-band photometry for the \mbox{W-CDF-S}. This is because the airmass of HSC observations of \mbox{W-CDF-S} is generally high ($\sim1.5-2$), causing the fluxes of blue sources to be relatively more suppressed compared to redder sources in single broad bands, especially in the $g$ band. HSC uses bright stars to calibrate the photometry, but the intrinsic spectra of stars are different from those of galaxies; hence, the calibration may be slightly biased for galaxies, and the bias depends upon their colors. We empirically correct this issue by matching the HSC $g$-band photometry to the VOICE $g$-band photometry, and the following formula gives the correction:
\begin{align}
g_\mathrm{HSC}^\mathrm{new}=g_\mathrm{HSC}+0.0601(g_\mathrm{HSC}-i_\mathrm{HSC})-0.129.\label{eq: corr_g_hsc}
\end{align}
There is still a systematic $\sim0.02$~mag difference between $g_\mathrm{HSC}^\mathrm{new}$ and VOICE $g$, and thus we add an additional 0.02 mag error to the $g$-band error to account for the uncertainty of the calibration. For sources without $i_\mathrm{HSC}$, we increase the additional error term to 0.1 mag, which is the typical correction value from Eq.~\ref{eq: corr_g_hsc}. Note that this correction is only applied to \mbox{W-CDF-S} as \mbox{XMM-LSS} does not suffer from this issue and ELAIS-S1 lacks HSC data.\par
Duplicated bands (i.e., VOICE $gri$ and HSC $gri$) are all included to provide more information and also reduce the risk of missing some bands due to bad photometry in either survey.

\subsection{Photometry between $5.8-500~\mu m$}
We adopt photometric data at wavelengths longer than $5.8~\mu m$ from the HELP project \citep{Shirley19, Shirley21}, including IRAC 5.8~$\mu\mathrm{m}$, IRAC 8~$\mu\mathrm{m}$, MIPS 24~$\mu\mathrm{m}$, PACS 100~$\mu\mathrm{m}$, PACS 160~$\mu\mathrm{m}$, SPIRE 250~$\mu\mathrm{m}$, SPIRE 350~$\mu\mathrm{m}$, and SPIRE 500~$\mu\mathrm{m}$. The photometric data are deblended for sources detected in IRAC bands using the XID+ tool \citep{Hurley17}.\par
Given the importance of FIR data in constraining SFRs (e.g., \citealt{Ciesla15}), we further derive flux upper limits from 24~$\mu\mathrm{m}$ to 500 $\mu\mathrm{m}$, including MIPS 70 and 160~$\mu\mathrm{m}$ and the aforementioned FIR bands in HELP. These provide FIR constraints for $\approx50\%-70\%$ (the exact fraction varies across different bands) of our sources. Though the constraints are generally loose for the main population, they can help constrain galaxies with extreme SFRs -- we found that without the upper limits, 20\% of these sources with $\mathrm{SFR}>1000~M_\odot~\mathrm{yr^{-1}}$ will have SFR measurements overestimated by over 50\%.\par
Similar to the \mbox{X-ray} HB flux upper-limit map in Section~\ref{sec: xrayphot}, we will generate FIR upper-limit maps, in which each pixel value equals the flux upper limit if a source is located at the pixel. We conduct point-response-function (PRF) fitting for each pixel, assuming that a source is located at the center of this pixel. As given in \citet{Smith12}, the best-fit flux and error are
\begin{align}
f&=\frac{\sum_i\frac{d_ip_i}{\sigma_i^2}}{\sum_i\frac{p_i^2}{\sigma_i^2}},\label{eq: prffittingflux}\\
\sigma_\mathrm{inst}&=\frac{1}{\sqrt{\sum_i\frac{p_i^2}{\sigma_i^2}}},\label{eq: prffittingerr}
\end{align}
where $d_i$, $p_i$, $\sigma_i$ are the image, PRF, and error map values at pixel $i$, respectively. Note that Eq.~\ref{eq: prffittingerr} only describes the instrumental noise, which would vanish relative to $f$ if the exposure time increased to infinity and is valid only if all the pixel values are independent. Actual instrumental noise values are usually inflated by a factor (denoted as $C_\mathrm{corr}$) due to the correlations among pixels, and the variance of the sky itself due to unresolved sources may also contribute to the total noise, named the confusion noise (denoted as $\sigma_\mathrm{conf}$) and is often assumed to be constant across a field (e.g., \citealt{Nguyen10, Hurley17}). The total noise $\sigma_\mathrm{tot}$ is thus
\begin{align}
\sigma_\mathrm{tot}=\sqrt{(C_\mathrm{corr}\sigma_\mathrm{inst})^2+\sigma_\mathrm{conf}^2}.
\end{align}
We then define the flux upper limit as
\begin{align}
f_\mathrm{upp}=\max\{f+3\sigma_\mathrm{tot}, \sigma_\mathrm{tot}\},\label{eq: fupp}
\end{align}
where $f_\mathrm{upp}$ is truncated at $\sigma_\mathrm{tot}$ to prevent the upper limit from being too small to be reliable. Again, this upper limit should be distinguished from sensitivity (usually $5\sigma_\mathrm{tot}$), as explained in Section~\ref{sec: xrayphot} and \citet{Kashyap10}.\par
The FIR data are from the SWIRE survey (MIPS; \citealt{Lonsdale03, Surace05}) and HerMES survey (PACS and SPIRE; \citealt{Oliver12}), on which the HELP project is based. We calibrate the error following the procedures explained below and derive the flux upper limit based on Eqs.~\ref{eq: prffittingflux} $-$ \ref{eq: fupp}.
\begin{itemize}
\item{MIPS 24 $\mu\mathrm{m}$.\par
The MIPS PRFs are from IRSA.\footnote{\url{https://irsa.ipac.caltech.edu/data/SPITZER/docs/mips/calibrationfiles/prfs/}} To calibrate our PRF-fitting process, we compare our PRF-fitting fluxes with the cataloged fluxes for detected sources, and the PRFs are normalized so that the median $\Delta\log\mathrm{(flux)}$ is 0. For simplicity, we set the fitting region to be a square whose side length is an odd number of pixels, and the size is chosen to be the one that minimizes the normalized median absolute deviation (NMAD)\footnote{NMAD is defined as $1.4826~\times$ median absolute deviation.} of $\Delta\log\mathrm{(flux)}$. The fitting regions are determined to be $11\times11$ pixels (i.e., $13.2''\times13.2''$), and the corresponding $\mathrm{NMAD}\{\Delta\log\mathrm{(flux)}\}$ is 0.024 dex. The deviation may be caused by the variations in PRFs, the different choices between our fitting regions and the ones used in the catalog, and the fact that the real source locations may not coincide with the pixel centers. To account for these effects, we add the NMAD values to the final flux errors in quadrature.\par
We adopt $\sigma_\mathrm{conf}=0$ because it is negligible compared to $\sigma_\mathrm{inst}$ in our case. $\sigma_\mathrm{conf}$ is estimated to be $\sim0.01$ mJy in the previous literature (e.g., \citealt{Xu01, Franceschini03, Dole04}), which is $\sim10$ times smaller than $\sigma_\mathrm{inst}$. We thus only need to calibrate $C_\mathrm{corr}$. First, we mask regions around detected sources on the PRF-fitted map (i.e., the map with each pixel value being the one from Eq.~\ref{eq: prffittingflux} after calibration) and denote $\omega=f\sqrt{cov}$ on the unmasked regions, where $cov$ is the coverage; then $\omega$ is roughly normally distributed \citep{Surace05}. Similar to \citet{Smith12}, we estimate the standard deviation of $\omega$ as
\begin{align}
\sigma_\omega=\sqrt{\frac{1}{N}\sum_i(\omega_i-\mathrm{median}\{\omega_i\})^2},\label{eq: sigma_omega}
\end{align}
where the summation is restricted to $\omega_i\leq\mathrm{median}\{\omega_i\}$. The correlation correction factor is then estimated to be
\begin{align}
C_\mathrm{corr}=\mathrm{median}\left\{\frac{\sigma_\omega}{\sigma_f\sqrt{cov}}\right\},
\end{align}
and the result is 3.5.
}
\item{MIPS 70 and 160 $\mu\mathrm{m}$.\par
Following the approach for 24 $\mu\mathrm{m}$, we determine the fitting regions to be $5\times5$ and $7\times7$ pixels (i.e., $20''\times20''$ and $56''\times56''$) for 70 and 160 $\mu\mathrm{m}$, respectively; the corresponding $\mathrm{NMAD}\{\Delta\log\mathrm{(flux)}\}$ values are 0.029 and 0.026 dex. To do the error calibration, we assume $f\sim N(0, \sigma_\mathrm{tot}^2)$ for regions with $f<0$, which are not expected to be contaminated by any detectable sources. By maximizing the corresponding likelihood, we obtain $C_\mathrm{corr}=3.9$ (3.6) and $\sigma_\mathrm{conf}=1.0$ (14.3) mJy for 70 (160)~$\mu\mathrm{m}$. Our $\sigma_\mathrm{conf}$ values are consistent with the ones in the literature (e.g., \citealt{Xu01, Franceschini03, Dole04, Frayer06}) -- $\sim0.3-1.3$ mJy for 70 $\mu\mathrm{m}$ and $\sim7-19$ mJy for 160 $\mu\mathrm{m}$.
}
\item{
PACS 100 and 160 $\mu\mathrm{m}$.\par
The PACS PRFs are available on the HerMES website\footnote{\url{http://hedam.lam.fr/HerMES/}} along with the data. We follow the same approach to derive the flux upper limits as for MIPS 24 $\mu\mathrm{m}$ because PACS $\sigma_\mathrm{conf}\lesssim1$ mJy (e.g., \citealt{Berta11}), much smaller than $\sigma_\mathrm{inst}$. We adopt the fitting regions to be $9\times9$ and $7\times7$ pixels (i.e., $18''\times18''$ and $21''\times21''$) for 100 and 160 $\mu\mathrm{m}$, respectively, and the resulting $\mathrm{NMAD}\{\Delta\log\mathrm{(flux)}\}$ values are 0.018 and 0.033 dex. The correlation correction factors are calibrated to be 1.9 and 2.4. We also add additional calibration errors as 7\% of the fluxes \citep{Balog14}.
}
\item{
SPIRE 250, 350, and 500 $\mu\mathrm{m}$.\par
The SPIRE PRFs are assumed to be Gaussian with FWHMs of $18.15''$, $25.15''$, and $36.3''$ for 250, 350, and 500 $\mu\mathrm{m}$, respectively. This assumption is attested to be simple but adequate in the literature (e.g., \citealt{Roseboom10, Roseboom12, Smith12, Wang14}). Following \citet{Smith12}, we adopt $5\times5$ pixels as the PRF-fitting region. Following the same approach as for MIPS 70 and 160~$\mu\mathrm{m}$ to calibrate the errors, we obtain $C_\mathrm{corr}=1.8$, 2.1, and 2.0 and $\sigma_\mathrm{conf}=6.1$, 7.6, and 7.9 mJy for 250, 350, and 500 $\mu\mathrm{m}$, respectively. The $\sigma_\mathrm{conf}$ values are consistent with those in \citet{Smith12}. We also add a $7\%$ calibration-error term as done in \citet{Wang14}.
}
\end{itemize}

\begin{figure*}
\centering
\resizebox{\hsize}{!}{\includegraphics{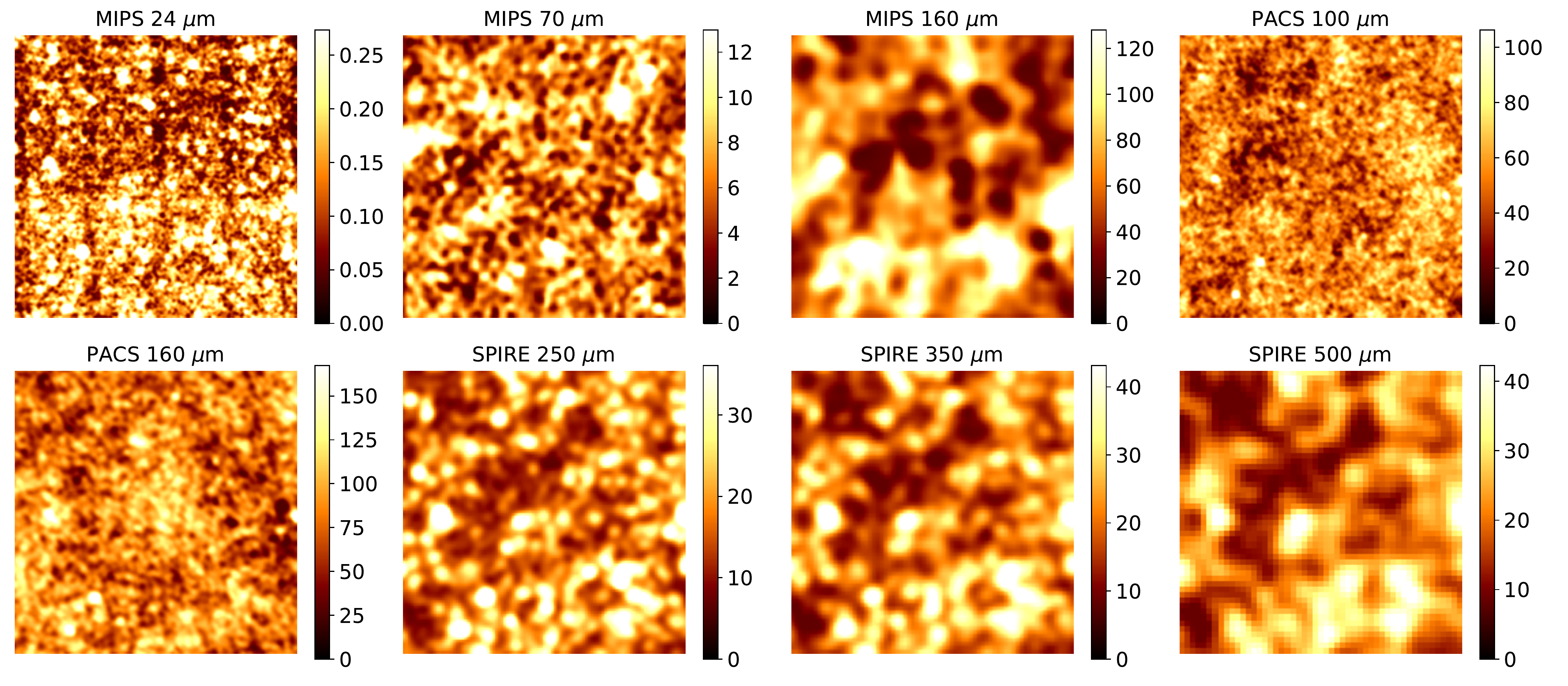}}
\caption{The $24-500~\mu\mathrm{m}$ upper-limit maps in a $10'\times10'$ region of the \mbox{W-CDF-S}, centered at J2000 $\mathrm{RA=53^\circ, Dec=-28.4^\circ}$. The map units are all mJy.}
\label{fig_fuppcutout}
\end{figure*}

As an example, we display the resulting upper-limit maps in a small region in \mbox{W-CDF-S} in Fig.~\ref{fig_fuppcutout}. The MIPS 160~$\mu\mathrm{m}$ map suffers more from source confusion than the PACS 160~$\mu\mathrm{m}$ map, but since both maps only provide flux upper-limit constraints, the source confusion does not matter, and thus we keep both the MIPS and PACS 160~$\mu\mathrm{m}$ maps.

\subsection{Galactic Extinction Correction}
We derive the Galactic extinction for a given band from FUV to $8~\mu\mathrm{m}$ as
\begin{align}
A(\mathrm{band})=2.5\log\frac{\int s_\lambda\left(\frac{\lambda}{1+z}\right)e^{-\tau_\mathrm{IGM}}T(\lambda)d\lambda}{\int s_\lambda\left(\frac{\lambda}{1+z}\right)e^{-\tau_\mathrm{IGM}}T(\lambda)10^{-0.4A(\lambda)}d\lambda},\label{eq: GalExt}
\end{align}
where $s_\lambda$ is the rest-frame intrinsic source spectrum; $\tau_\mathrm{IGM}=\tau_\mathrm{IGM}(\lambda, z)$ is the expected transmission optical depth of the intergalactic medium (IGM); $T(\lambda)$ is the filter transmission curve in energy units; and $A(\lambda)=R(\lambda)E(B-V)$ is the extinction at wavelength $\lambda$. The intrinsic source emission is absorbed by both the IGM and the Galaxy, and the above equation only corrects for the Galactic extinction for IGM-absorbed emission. The IGM absorption will be corrected during the SED fitting \citep{Boquien19}. We adopt the median spectrum in \citet{Brammer08} as a representative $s$, the IGM attenuation law in \citet{Meiksin06} as $\tau_\mathrm{IGM}$, the $E(B-V)$ values in \citet{Schlegel98}, and extinction laws in \citet{Cardelli89}, \citet{ODonnell94}, and \citet{Indebetouw05} assuming $R_V=3.1$. Generally speaking, in our case, $R(\mathrm{band})=A(\mathrm{band})/E(B-V)$ has little dependence on $E(B-V)$ and the selection of $s$. Instead, the IGM attenuation plays a more important role, especially for the NUV band because the NUV covers the 2200~\AA\ extinction bump of our Galaxy. As redshift increases from 1 to 1.9, the IGM attenuation gradually absorbs the emission around the extinction bump while keeping the emission at longer wavelengths unaffected. This significantly modifies the effective wavelength of NUV and leads $R(\mathrm{NUV})$ to be $\sim8.5$ at other redshifts but drops to as low as $\sim6.8$ at $z\sim1.9$. However, the IGM attenuation itself is highly uncertain because the number of Lyman limit systems along the line of sight is highly variable \citep{Meiksin06}, and thus Eq.~\ref{eq: GalExt} can only return typical extinctions. Fortunately, the Galactic extinctions are not severe, as listed in Table~\ref{tbl_GalExt}.

\begin{table}
\caption{Galactic extinctions}
\label{tbl_GalExt}
\centering
\begin{threeparttable}
\begin{tabular}{ccc}
\hline
\hline
Survey & Band & median\{$A(\mathrm{band})$\} (mag)\\
\hline
GALEX & FUV & 0.074\\
GALEX & NUV & 0.070\\
VOICE & $u$ & 0.044\\
VOICE & $g$ & 0.033\\
VOICE & $r$ & 0.024\\
VOICE & $i$ & 0.019\\
HSC & $g$ & 0.033\\
HSC & $r$ & 0.024\\
HSC & $i$ & 0.018\\
HSC & $z$ & 0.014\\
VIDEO & $Z$ & 0.014\\
VIDEO & $Y$ & 0.011\\
VIDEO & $J$ & 0.008\\
VIDEO & $H$ & 0.005\\
VIDEO & $K_s$ & 0.003\\
DeepDrill & $3.6~\mu\mathrm{m}$ & 0.002\\
DeepDrill & $4.5~\mu\mathrm{m}$ & 0.002\\
SWIRE & $5.8~\mu\mathrm{m}$ & 0.001\\
SWIRE & $8.0~\mu\mathrm{m}$ & 0.001\\
\hline
\hline
\end{tabular}
\end{threeparttable}
\end{table}

\subsection{Redshift}
\label{sec: redshift}
Our redshifts are from \citet{Ni21} and \citet{Zou21b}. They compiled all the available spectroscopic redshifts (spec-$z$s) for 30135 sources in \mbox{W-CDF-S}, and \citet{Zou21b} derived photo-$z$s for all the sources using \texttt{EAZY} \citep{Brammer08}. However, the photo-$z$s in \citet{Zou21b} are only valid if the optical-to-NIR emission is not dominated by an AGN, and \citet{Ni21} derived appropriate photo-$z$s for AGN-dominated sources. Therefore, we adopt the redshifts following the priority below. When available, spec-$z$s are adopted; otherwise, photo-$z$s in \citet{Ni21} are adopted; photo-$z$s in \citet{Zou21b} are used in the remaining cases. 738 photo-$z$s in \mbox{W-CDF-S} are taken from \citet{Ni21}. As discussed in \citet{Zou21b}, photo-$z$s are still appropriate for most AGNs because relatively few AGNs (sky surface density $\lesssim300~\mathrm{deg^2}$) can materially affect the observed optical-to-NIR SEDs, and most such AGN-dominated sources have been identified in \citet{Ni21} (see their Appendix~B). We thus do not need to refine further the photo-$z$s for AGN candidates (Section~\ref{sec: select_agn}).

\section{SED Fitting and Source Classification}
\label{sec: sed_step1}
In this section, we classify sources into stars, AGN candidates, BQ-galaxy candidates, or normal galaxies. The ``best'' classified categories include 42628 stars, 19612 AGN candidates, 3624 BQ-galaxy candidates, and 733743 normal galaxies, as presented in Table~\ref{tbl_fieldinfo} and Section~\ref{sec: bestsedfittingresults}. One of the main goals for performing the classification before the main SED fitting in Section~\ref{sec: bestsedfittingresults} is to reduce the computational requirements. For example, we would like to add AGN components only for AGNs, and thus we first select AGN candidates using relatively sparser parameter grids to fit all the sources with or without AGN components and then re-fit the candidates with denser grids.

\subsection{Selection of Stars}
\label{sec: select_star}
Stars are usually selected in two ways in extragalactic surveys -- by selecting point sources and by applying empirical color-color cuts (e.g., \citealt{Daddi04, Barro09, Henrion11, Malek13}). The former only works for bright sources because morphological information is limited for faint sources. In this section, we use both methods to select stars.\par
\begin{figure}
\centering
\resizebox{\hsize}{!}{\includegraphics{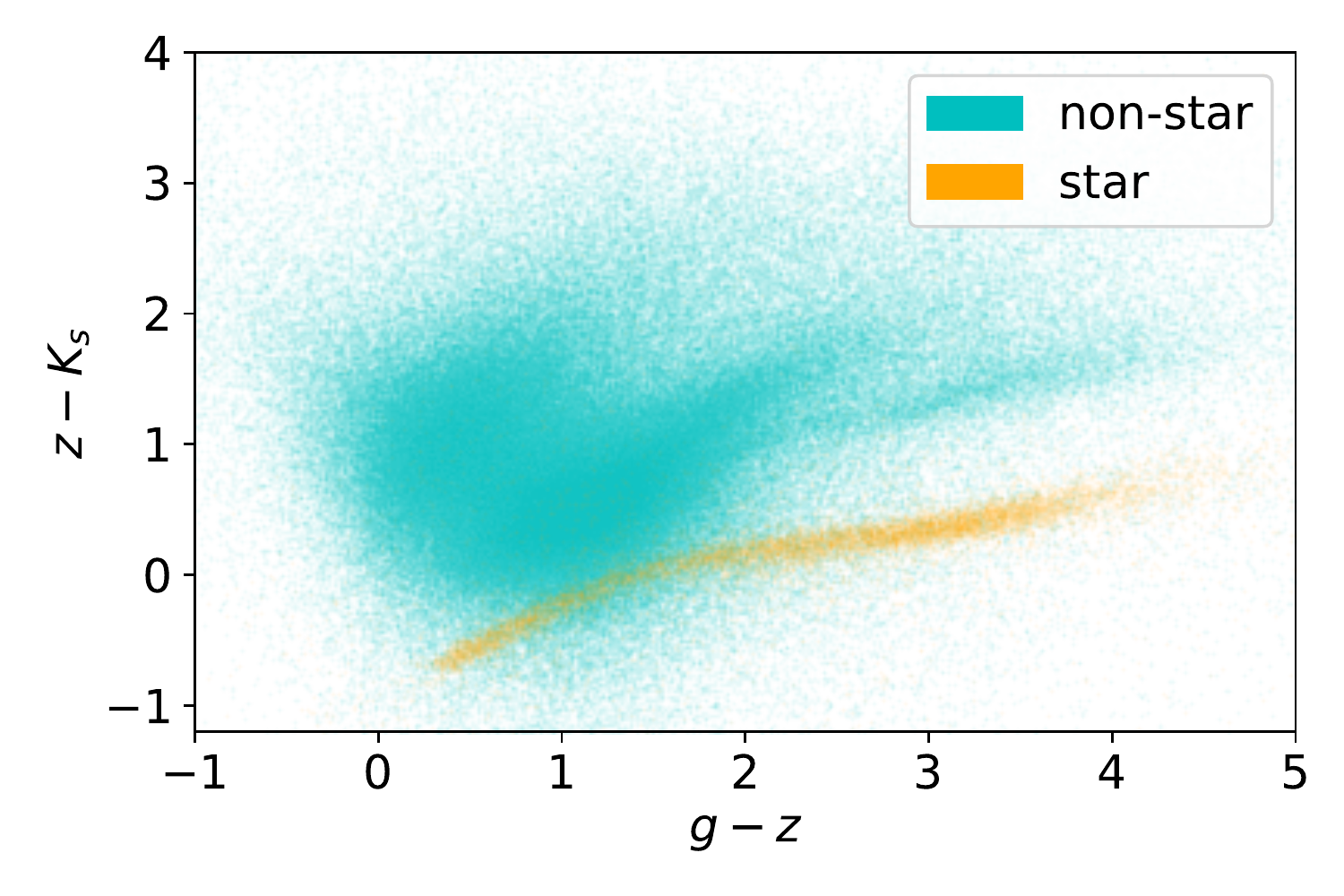}}
\caption{The $gzK$ color-color diagram for sources in \mbox{W-CDF-S}. Cyan and orange points are our selected non-stars and stars, respectively. Our selected stars form a clear stellar locus, justifying the overall reliability of the star selection. The branch lying roughly one-magnitude above the stellar locus is early-type galaxies; see, e.g., Fig.~1 in \citet{Lane07} for an example.}
\label{fig_select_star}
\end{figure}

First, we select point sources with $i$-band magnitudes brighter than 24 in HSC as stars. The reliability of the morphological selection decreases rapidly for fainter magnitudes; see \citet{Bosch18}. This selection is not applied to \mbox{X-ray} AGNs to avoid misclassifying point-like quasars, most of which are detected in \mbox{X-rays} \citep{Ni21}, as stars. A total of 21596 stars are selected in this way. Secondly, for the color-color selection, we adopt a more accurate method, SED fitting, to select stars. Similar to \citet{Laigle16} and \citet{Weaver22}, who selected stars in the COSMOS field through SED fitting with \texttt{LePhare} \citep{Arnouts99, Ilbert06}, we use the same code\footnote{\texttt{CIGALE} cannot be used to select stars because it does not have stellar templates.} to fit all of our sources with quasar, galaxy, and stellar templates and compare the resulting best-fit $\chi^2$ values for these three kinds of templates. This SED selection is not applied to extended sources in HSC. There are 39069 sources whose smallest $\chi^2$ values are from the stellar templates, and they are also selected as stars. Furthermore, 50 spectroscopic stars are also added, and \citet{Ni21} presented the details of these spectroscopic classifications. We also classify 12396 sources with statistically significant proper motions in Gaia EDR3 \citep{Gaia21} as stars. 82\% of the spectroscopic stars, 90\% of the HSC morphological stars, and 87\% of the Gaia stars are also identified by the SED selection. The positions of stars and non-stars in the $gzK$ color-color diagram are displayed in Fig.~\ref{fig_select_star}. There are 42628 stars selected in total.

\subsection{Selection of AGN Candidates}
\label{sec: select_agn}
We use different selection methods to build an AGN sample that is as complete as possible, including \mbox{X-ray}, MIR, and SED methods. We note that another important AGN selection method is based on the radio band. The analyses of radio AGNs (including their SEDs) in our fields are still ongoing and will be presented separately in Zhu et al. (in preparation), and we do not present them in this work. Besides, the AGN radio emission is not strongly correlated to other bands -- first, the radio loudness is often set to be a free parameter in SED fitting that can hardly be inferred from shorter-wavelength SEDs (e.g., \citealt{Yang22}); secondly, the shorter-wavelength (e.g., \mbox{X-ray}) AGN emission generally only shows moderate enhancements even for sources with strong radio emission (i.e., radio-loud quasars; \citealt{Zhu20}). We thus do not expect strong biases caused by ignoring radio AGNs.\par

\subsubsection{An Overview of Different Selection Methods}
\mbox{X-ray} selection is efficient at selecting pure AGN samples, and \mbox{X-ray} AGNs have already been selected in \citet{Ni21}. Especially, \mbox{X-ray} emission suffers little from starlight contamination and can penetrate through large amounts of obscuring material (see \citealt{Brandt15} for a review). However, the \mbox{X-ray} method still faces challenges when selecting highly obscured or even CT AGNs and low-luminosity AGNs at high redshifts, given the \mbox{X-ray} depth.\par
AGN candidates are also selected based on their red colors and power-law spectra in the MIR, which are approximated by Spitzer IRAC color-selection criteria \citep{Lacy04, Lacy07, Stern05, Donley12, Chang17}. The MIR method is able to select both unobscured AGNs, which may be selected by \mbox{X-ray} selection as well, and heavily obscured AGNs, which may be missed by \mbox{X-ray} selection (e.g., \citealt{Donley12}). However, it suffers from starlight contamination and thus can hardly select low-luminosity AGNs and AGNs with bright hosts. Moreover, depending upon the selection criteria, the resulting MIR AGN sample may be contaminated by star-forming galaxies, especially for the criterion in \citet{Lacy07}. The criterion in \citet{Donley12} is generally more reliable in avoiding the misclassification of star-forming galaxies as AGNs, but it may miss highly obscured AGNs (e.g., \citealt{Li20}). In this work, we select MIR AGN candidates if a source is detected in all four IRAC bands with a signal-to-noise ratio (SNR) above three and meets any criterion in \citet{Stern05}, \citet{Lacy07}, or \citet{Donley12} so that the resulting MIR AGN sample is as complete as possible. However, this will inevitably misclassify many star-forming galaxies as AGNs. We flag MIR AGNs satisfying different criteria separately in our final catalog, and users can easily select MIR AGNs based on a subset of all the three criteria depending upon their tradeoff between completeness and purity.\par
AGN candidates can also be selected through SED fitting. Depending upon the data, the SED method may also have significant drawbacks in terms of completeness and purity, and this statistical model selection often lacks apparent physical meaning. Even for sources with distinct AGN features in one band (e.g., \mbox{X-ray}), the interplay between galaxy and AGN components in other bands may still make SED fitting possibly miss such sources; the resulting SED AGN candidates may also be contaminated by large numbers of galaxies (e.g., see the bottom panel of Fig.~16 in \citealt{Yang21b}), depending upon the adopted criterion. Especially, Section~\ref{sec: sedagn} shows that in our case, if we require a high selection purity, the SED selection method can hardly select sources missed by other methods. Therefore, we mostly rely on the SED method to select AGN \textit{candidates} without trying to firmly attest that they are AGNs. We emphasize that the limitations of the SED method in our case largely originate from the data instead of the method itself. \citet{Yang21b} show that most of the drawbacks can be resolved if one has deep and continuous MIR coverage, which, however, are unavailable in our case. When putting this into a broader context of joint galaxy-AGN SED modeling, the SED selection of AGNs will be an ever-green project that requires many years of investigations of both the data and the method. Our case is mainly limited by the data, but in cases where good data or external information are available, it is equally important to develop and evaluate appropriate methods that can effectively extract information from the data. Examples include developing the \mbox{X-ray} module in \texttt{CIGALE} \citep{Yang20, Yang22} and utilizing MIR color gradients in resolved galaxies \citep{Leja18}.

\subsubsection{SED Fitting to Select AGN Candidates}
We use \texttt{CIGALE v2022.0} to do the SED fitting. We use a delayed star-formation history (SFH) because it can model both early-type and late-type galaxies \citep{Boquien19} with only two free parameters, and its general reliability in measuring SFR and $M_\star$, even for AGN host galaxies, has been well attested in previous literature (e.g., \citealt{Ciesla15, Ciesla17, Carnall19, Lower20}).\footnote{Generally, adopting different parametric SFHs can result in a systematic difference $\lesssim0.1~\mathrm{dex}$ for $M_\star$ and SFR (e.g., \citealt{Carnall19}). For example, by comparing the results for all the sources based on the delayed SFH and the truncated delayed SFH in Section~\ref{sec: select_bqgal} using the parameter settings in Table~\ref{tbl_sedpar_step1}, we obtain a systematic difference in $M_\star$ (SFR) of 0.04 (0.08) dex, and the NMADs of the differences are 0.06 and 0.11 dex for $M_\star$ and SFR, respectively.} Stellar templates are from \citet{Bruzual03}, and a Chabrier initial mass function \citep{Chabrier03} is adopted. Nebular emission is also included in a self-consistent manner using CLOUDY photoionization calculations \citep{Ferland17}, as described and implemented in \citet{Villa-Velez21}. Dust attenuation is assumed to follow \citet{Calzetti00}, and dust emission in the IR is assumed to follow templates in \citet{Dale14} for simplicity. The \mbox{X-ray} module is included, where the AGN \mbox{X-ray} emission is assumed to be moderately anisotropic following
\begin{align}
\frac{L_\mathrm{X}(\theta)}{L_\mathrm{X}(0)}=a_1\cos\theta+a_2\cos^2\theta+1-a_1-a_2,
\end{align}
where $\theta$ is the viewing angle (face-on corresponds to $0^\circ$), and the angle coefficients, $a_1$ and $a_2$, are calibrated in \citet{Yang22} to be 0.5 and 0, respectively. The UV-to-IR AGN module is based on the SKIRTOR model \citep{Stalevski12, Stalevski16} with polar-dust extinction, and the disk spectral shape is modified from \citet{Schartmann05}, as detailed in \citet{Yang22}. The polar-dust extinction law is assumed to follow that in the Small Magellanic Cloud (SMC; \citealt{Prevot84}). \citet{Mountrichas20} demonstrated that the polar-dust component can help AGN selection, and SED-fitting results are insensitive to the temperature of the polar dust. \citet{Buat21} further showed that the SMC extinction law is largely optimal for polar dust and can return reliable results even if the real polar extinction curve is different from the SMC law. The viewing angle is set to include at least one face-on (type~1 AGN) and one edge-on (type~2 AGN) system, and \citet{Padilla21} showed that the SED-fitting results are insensitive to the choice of viewing angles as long as both type~1 and type~2 representatives are included.\par
We use a two-step SED-fitting approach to select AGN candidates. In the first step, we run SED fitting for all the sources twice with coarse parameter grids -- once with the AGN module included and once without AGNs. The parameter settings are summarized in Table~\ref{tbl_sedpar_step1}. Among the parameters, the AGN fraction ($f_\mathrm{AGN}$) is defined as the fractional contribution of the AGN component to the total IR luminosity, where the IR luminosity is defined as all the dust-absorbed luminosity at shorter wavelengths. $f_\mathrm{AGN}$ is the primary parameter controlling the impact upon the SED shape from the AGN component and is hence assigned with a dense grid of possible values. The aim of this step is to narrow down all the millions of sources to a much smaller sample of \textit{raw} SED AGN candidates.\footnote{We will always include the word ``raw'' when referring to candidates selected in this step.} We compare how much the fitting is improved after adding an AGN component, as done in previous literature for selecting AGNs via SED-fitting techniques (e.g., \citealt{Chung14, Huang17, Pouliasis20b}). We adopt the Bayesian information criterion (BIC) to make the comparison, defined as $\mathrm{BIC}=2p\ln{N}-2\ln{L}$, where $p$ is the number of parameters, $N$ is the number of data points, and $L$ is the maximum likelihood of the model. Since $L=\exp(-\chi^2/2)$, $\Delta\mathrm{BIC}=2\Delta p\ln{N}+\Delta\chi^2$, where $\Delta\chi^2$ is the best-fit $\chi^2$ when not including the AGN module minus that with the AGN module. We set $\Delta p$ as $-3$, accounting for the fact that there are three free parameters in the AGN module (viewing angle, AGN fraction, and $E(B-V)$ of the polar extinction), and $N$ as the number of bands with SNR above three. We add a subscript of ``1'' to $\Delta\mathrm{BIC}$ to refer to the values derived in this step and write ``AGN'' in parentheses to mean that this is for the AGN selection. We will present the BQ-galaxy selection in Section~\ref{sec: select_bqgal}, and thus writing ``AGN'' explicitly helps distinguish the AGN selection and the BQ-galaxy selection. Raw SED AGN candidates are chosen to be those with $\Delta\mathrm{BIC_1(AGN)}>2$, which is a loose threshold so that the raw candidates are as complete as possible.\footnote{Nevertheless, the overall completeness cannot reach a near-unity level; see Sections~\ref{sec: select_results} and \ref{sec: sedagn} for more discussion.} This returns 48 thousand raw SED AGN candidates, which are only 6\% of the whole sample. This coarse-grid fitting is not designed to be perfect and tends to overestimate the actual $\Delta\mathrm{BIC}$ because the galaxy templates in Table~\ref{tbl_sedpar_step1} are limited; e.g., the number of possible values that the SFH parameters can have is small. Thus, the stellar continuum may not be well constrained, and the best-fit $\chi^2$ tends to be elevated. Some sources may be selected as raw SED AGN candidates simply because the galaxy templates are not sufficiently flexible to explain their SEDs. However, this is not necessarily a disadvantage in this step because the completeness is increased, and we will trim the sample in subsequent steps.\par

\begin{table*}
\caption{Coarse-grid \texttt{CIGALE} parameter settings used in step~1 of the AGN and BQ-galaxy selections}
\label{tbl_sedpar_step1}
\centering
\begin{threeparttable}
\begin{tabular}{cccc}
\hline
\hline
\multirow{2}{*}{Module} & \multirow{2}{*}{Parameter} & Name in the \texttt{CIGALE} & \multirow{2}{*}{Possible values}\\
&& configuration file &\\
\hline
\multirow{2}{*}{Delayed SFH} & Stellar \textit{e-}folding time & tau\_main & 0.1, 0.5, 1, 3, 5, 10 Gyr\\
& Stellar age & age\_main & 0.1, 0.5, 1, 3, 5, 10 Gyr\\
or\\
\multirow{5}{*}{Truncated delayed SFH} & Stellar \textit{e-}folding time & tau\_main & 0.1, 0.5, 1, 3, 5, 10 Gyr\\
& Stellar age & age\_main & 1, 3, 5, 7, 10 Gyr\\
& Age of the BQ episode & age\_bq & 10, 50, 100, 200, 500, 800 Myr\\
& \multirow{2}{*}{$r_\mathrm{SFR}$} & \multirow{2}{*}{r\_sfr} & 0, 0.05, 0.1, 0.3, 0.5, 0.7, 1, 2,\\
&&&5, 7, 10, 30, 50, 100\\
\hline
Simple stellar population & Initial mass function & imf & \citet{Chabrier03}\\
\citet{Bruzual03} & Metallicity & metallicity & 0.02\\
\hline
Nebular & ----- & ----- & -----\\
\hline
\multirow{3}{*}{Dust attenuation} & \multirow{2}{*}{$E(B-V)_\mathrm{line}$} & \multirow{2}{*}{E\_BV\_lines} & 0, 0.05, 0.1, 0.2, 0.3,\\
\multirow{3}{*}{\citet{Calzetti00}}&&& 0.4, 0.6, 0.8, 1, 1.5\\
& $E(B-V)_\mathrm{line}/E(B-V)_\mathrm{continuum}$ & E\_BV\_factor & 1\\
\hline
Dust emission & \multirow{2}{*}{Alpha slope} & \multirow{2}{*}{alpha} & \multirow{2}{*}{1.5, 2.0, 2.5}\\
\citet{Dale14}\\
\hline
\multirow{6}{*}{\mbox{X-ray}} & AGN photon index & gam & 1.8\\
& \multirow{2}{*}{AGN $\alpha_\mathrm{OX}$} & \multirow{2}{*}{alpha\_ox} & $-1.9$, $-1.8$, $-1.7$, $-1.6$, $-1.5$,\\
&&&$-1.4$, $-1.3$, $-1.2$, $-1.1$\\
& Maximum deviation of $\alpha_\mathrm{OX}$ & \multirow{2}{*}{max\_dev\_alpha\_ox} & \multirow{2}{*}{0.2}\\
& from the $\alpha_\mathrm{OX}-L_{\nu,2500}$ relation &\\
& AGN \mbox{X-ray} angle coefficients & angle\_coef & (0.5, 0)\\
\hline
\multirow{7}{*}{AGN (optional)} & Viewing angle & i & $30^\circ, 70^\circ$\\
\multirow{7}{*}{\citet{Stalevski12, Stalevski16}}& Disk spectrum & disk\_type & \citet{Schartmann05}\\
& Modification of the optical & \multirow{2}{*}{delta} & \multirow{2}{*}{$-0.27$}\\
& Power-law index &\\
& \multirow{2}{*}{AGN fraction} & \multirow{2}{*}{fracAGN} & 0, 0.1, 0.2, 0.3, 0.4,\\
&&& 0.5, 0.7, 0.9, 0.99\\
& $E(B-V)$ of the polar extinction & EBV & 0, 0.1, 0.3, 0.5\\
\hline
\hline
\end{tabular}
\begin{tablenotes}
\item
\textit{Notes.} Unlisted parameters are set to the default values. The AGN component and the truncated delayed SFH are only used in Section~\ref{sec: select_agn} and Section~\ref{sec: select_bqgal}, respectively. This fitting returns $\Delta\mathrm{BIC_1(AGN)}$ and $\Delta\mathrm{BIC_1(BQ)}$.
\end{tablenotes}
\end{threeparttable}
\end{table*}

In the second step, we refit our raw SED AGN candidates using denser parameter grids to refine the selection. Such fitting is not applied to the whole sample because that is too computationally intensive\footnote{To be more specific, the running time is estimated to be on a month-scale using two Intel Xeon Gold 6226R processors (16 cores and 32 threads each) or a year-scale for a typical personal computer, let alone that the requirement upon RAM is also heavy.} and also cannot provide many more insights (see Section~\ref{sec: sedagn}). Similar to the first step, we do the fitting twice using both normal-galaxy and AGN templates, and the parameter settings are summarized in Table~\ref{tbl_sedpar_step2_gal} for the normal-galaxy templates and Table~\ref{tbl_sedpar_step2_agn} for the AGN templates; we use $\Delta\mathrm{BIC_2(AGN)}$ to represent the comparison in this step. We select \textit{refined} SED AGN candidates as $\Delta\mathrm{BIC_2(AGN)}>2$. There are 16 thousand refined candidates, which is around one-third of the raw candidates. The exact $\Delta\mathrm{BIC_2(AGN)}$ threshold for the refined candidates is somewhat arbitrary and actually unimportant as long as it is reasonably good. What matters is the calibration of the SED selection, and we will present this in Section~\ref{sec: sedagn}.

\begin{table*}
\caption{Dense-grid \texttt{CIGALE} parameter settings for normal galaxies}
\label{tbl_sedpar_step2_gal}
\centering
\begin{threeparttable}
\begin{tabular}{cccc}
\hline
\hline
\multirow{2}{*}{Module} & \multirow{2}{*}{Parameter} & Name in the \texttt{CIGALE} & \multirow{2}{*}{Possible values}\\
&& configuration file &\\
\hline
\multirow{4}{*}{Delayed SFH} & \multirow{2}{*}{Stellar \textit{e-}folding time} & \multirow{2}{*}{tau\_main} & 0.1, 0.2, 0.3, 0.4, 0.5, 0.6, 0.7, 0.8, 0.9,\\
&&&1, 2, 3, 4, 5, 6, 7, 8, 9, 10 Gyr\\
& \multirow{2}{*}{Stellar age} & \multirow{2}{*}{age\_main} & 0.1, 0.2, 0.3, 0.4, 0.5, 0.6, 0.7, 0.8, 0.9,\\
&&&1, 2, 3, 4, 5, 6, 7, 8, 9, 10 Gyr\\
\hline
Simple stellar population & Initial mass function & imf & \citet{Chabrier03}\\
\citet{Bruzual03} & Metallicity & metallicity & 0.0001, 0.0004, 0.004, 0.008, 0.02, 0.05\\
\hline
Nebular & ----- & ----- & -----\\
\hline
\multirow{3}{*}{Dust attenuation} & \multirow{2}{*}{$E(B-V)_\mathrm{line}$} & \multirow{2}{*}{E\_BV\_lines} & 0, 0.05, 0.1, 0.15, 0.2, 0.25, 0.3, 0.4,\\
\multirow{3}{*}{\citet{Calzetti00}}&&& 0.5, 0.6, 0.7, 0.8, 0.9, 1, 1.2, 1.5\\
& $E(B-V)_\mathrm{line}/E(B-V)_\mathrm{continuum}$ & E\_BV\_factor & 1\\
\hline
Dust emission & \multirow{2}{*}{Alpha slope} & \multirow{2}{*}{alpha} & \multirow{2}{*}{1.0, 1.25, 1.5, 1.75, 2.0, 2.25, 2.5, 2.75, 3.0}\\
\citet{Dale14}\\
\hline
\mbox{X-ray} & ----- & ----- & -----\\
\hline
\hline
\end{tabular}
\begin{tablenotes}
\item
\textit{Notes.} Unlisted parameters are set to the default values. These are applied to all the sources.
\end{tablenotes}
\end{threeparttable}
\end{table*}

\begin{table*}
\caption{Dense-grid \texttt{CIGALE} parameter settings for AGN candidates}
\label{tbl_sedpar_step2_agn}
\centering
\begin{threeparttable}
\begin{tabular}{cccc}
\hline
\hline
\multirow{2}{*}{Module} & \multirow{2}{*}{Parameter} & Name in the \texttt{CIGALE} & \multirow{2}{*}{Possible values}\\
&& configuration file &\\
\hline
\multirow{2}{*}{Delayed SFH} & Stellar \textit{e-}folding time & tau\_main & 0.1, 0.3, 0.5, 0.8, 1, 3, 5, 8, 10 Gyr\\
& Stellar age & age\_main & 0.1, 0.3, 0.5, 0.8, 1, 3, 5, 8, 10 Gyr\\
\hline
Simple stellar population & Initial mass function & imf & \citet{Chabrier03}\\
\citet{Bruzual03} & Metallicity & metallicity & 0.02\\
\hline
Nebular & ----- & ----- & -----\\
\hline
\multirow{3}{*}{Dust attenuation} & \multirow{2}{*}{$E(B-V)_\mathrm{line}$} & \multirow{2}{*}{E\_BV\_lines} & 0, 0.1, 0.2, 0.3, 0.5,\\
\multirow{3}{*}{\citet{Calzetti00}}&&& 0.6, 0.8, 1, 1.2, 1.5\\
& $E(B-V)_\mathrm{line}/E(B-V)_\mathrm{continuum}$ & E\_BV\_factor & 1\\
\hline
Dust emission & \multirow{2}{*}{Alpha slope} & \multirow{2}{*}{alpha} & \multirow{2}{*}{1.5, 2.0, 2.5}\\
\citet{Dale14}\\
\hline
\multirow{6}{*}{\mbox{X-ray}} & AGN photon index & gam & 1.8\\
& \multirow{2}{*}{AGN $\alpha_\mathrm{OX}$} & \multirow{2}{*}{alpha\_ox} & $-1.9$, $-1.8$, $-1.7$, $-1.6$, $-1.5$,\\
&&&$-1.4$, $-1.3$, $-1.2$, $-1.1$\\
& Maximum deviation of $\alpha_\mathrm{OX}$ & \multirow{2}{*}{max\_dev\_alpha\_ox} & \multirow{2}{*}{0.2}\\
& from the $\alpha_\mathrm{OX}-L_{\nu,2500}$ relation &\\
& AGN \mbox{X-ray} angle coefficients & angle\_coef & (0.5, 0)\\
\hline
\multirow{7}{*}{AGN} & Viewing angle & i & $0^\circ, 10^\circ, 30^\circ, 50^\circ, 70^\circ, 90^\circ$\\
\multirow{7}{*}{\citet{Stalevski12, Stalevski16}}& Disk spectrum & disk\_type & \citet{Schartmann05}\\
& Modification of the optical & \multirow{2}{*}{delta} & \multirow{2}{*}{$-0.27$}\\
& power-law index &\\
& \multirow{2}{*}{AGN fraction} & \multirow{2}{*}{fracAGN} & 0, 0.05, 0.1, 0.2, 0.3, 0.4,\\
&&& 0.5, 0.6, 0.7, 0.8, 0.9, 0.99\\
& $E(B-V)$ of the polar extinction & EBV & 0, 0.05, 0.1, 0.2, 0.3, 0.4, 0.5\\
\hline
\hline
\end{tabular}
\begin{tablenotes}
\item
\textit{Notes.} Unlisted parameters are set to the default values. These are only applied to AGNs and raw SED AGN candidates with $\Delta\mathrm{BIC_1(AGN)>2}$ in order to return $\Delta\mathrm{BIC_2(AGN)}$.
\end{tablenotes}
\end{threeparttable}
\end{table*}

\subsubsection{The Selection Results}
\label{sec: select_results}
We first summarize our SED selection here. Similar to a steelmaking process, the overall SED selection undergoes multiple procedures to increase the purity step-by-step. The first-pass fitting returns raw SED AGN candidates with $\Delta\mathrm{BIC_1(AGN)}>2$, and the second-pass fitting is applied to the raw candidates and returns refined SED AGN candidates with $\Delta\mathrm{BIC_2(AGN)}>2$ (and $\Delta\mathrm{BIC_1(AGN)}>2$ by construction). As we will see in Section~\ref{sec: sedagn}, another calibration step is necessary to increase the purity further, but we only focus on the candidates in this section to obtain first insights.\par
\begin{figure*}
\centering
\resizebox{\hsize}{!}{\includegraphics{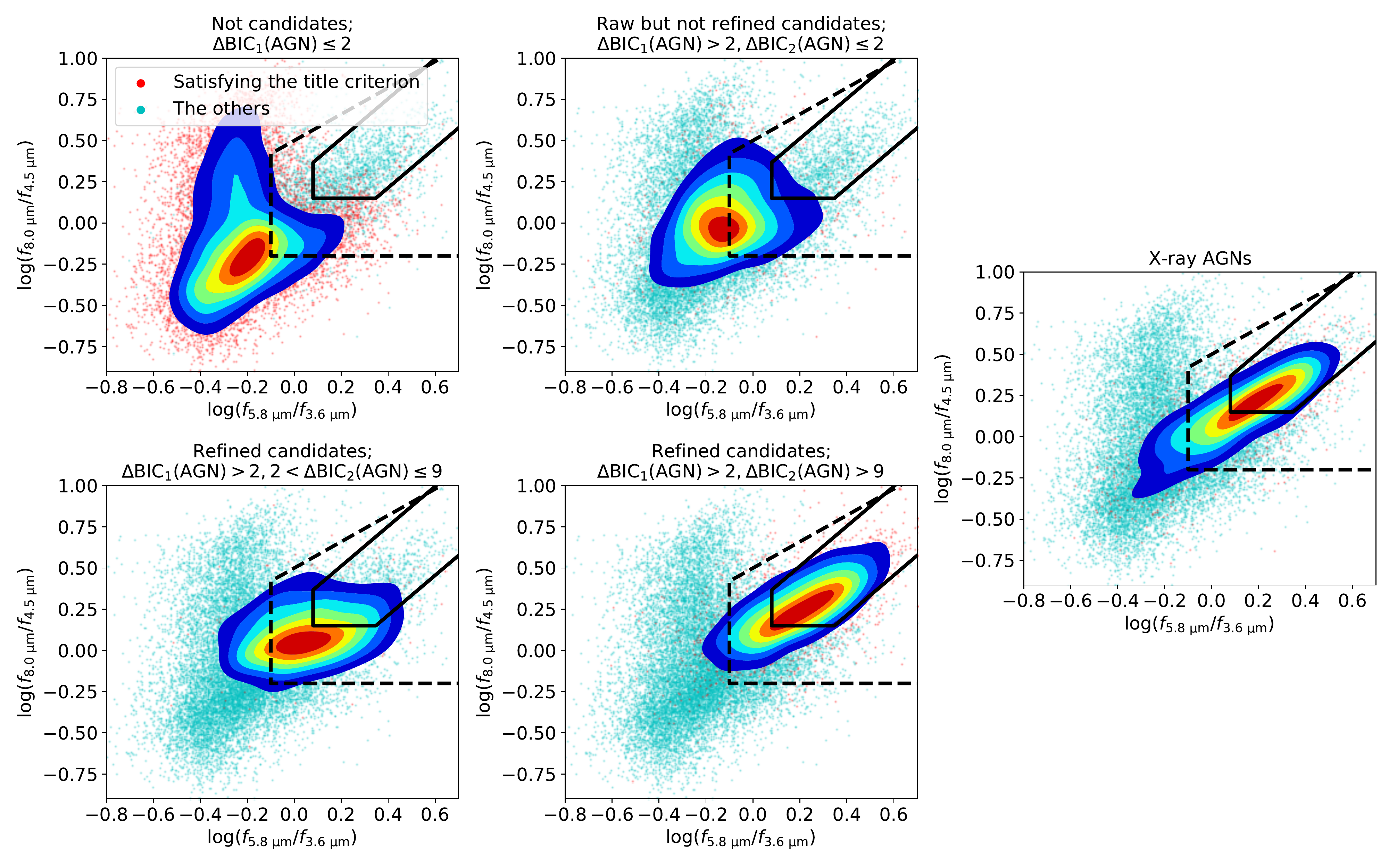}}
\caption{The distributions of sources on the IR color-color diagram. Red points are sources satisfying the criterion in the panel title, and cyan ones are the others. The kernel density estimations of the red points are plotted as the red-to-blue contour profiles. The dashed and solid lines are the AGN-selection wedges in \citet{Lacy07} and \citet{Donley12}, respectively. Among the four panels on the left, the upper left one shows galaxies that are not selected as raw SED AGN candidates; the upper right one shows sources that are selected as raw SED AGN candidates but fail to pass the refined candidate selection; the bottom two panels present the refined SED AGN candidates and divide them into two $\Delta\mathrm{BIC_2(AGN)}$ bins that roughly contain the same number of sources. As a comparison, the distribution of \mbox{X-ray} AGNs is plotted in the rightmost panel. The distribution of the red points gradually converges into the MIR AGN-selection wedges as $\Delta\mathrm{BIC}$ increases.}
\label{fig_ircolor}
\end{figure*}
\begin{figure*}
\centering
\resizebox{\hsize}{!}{\includegraphics{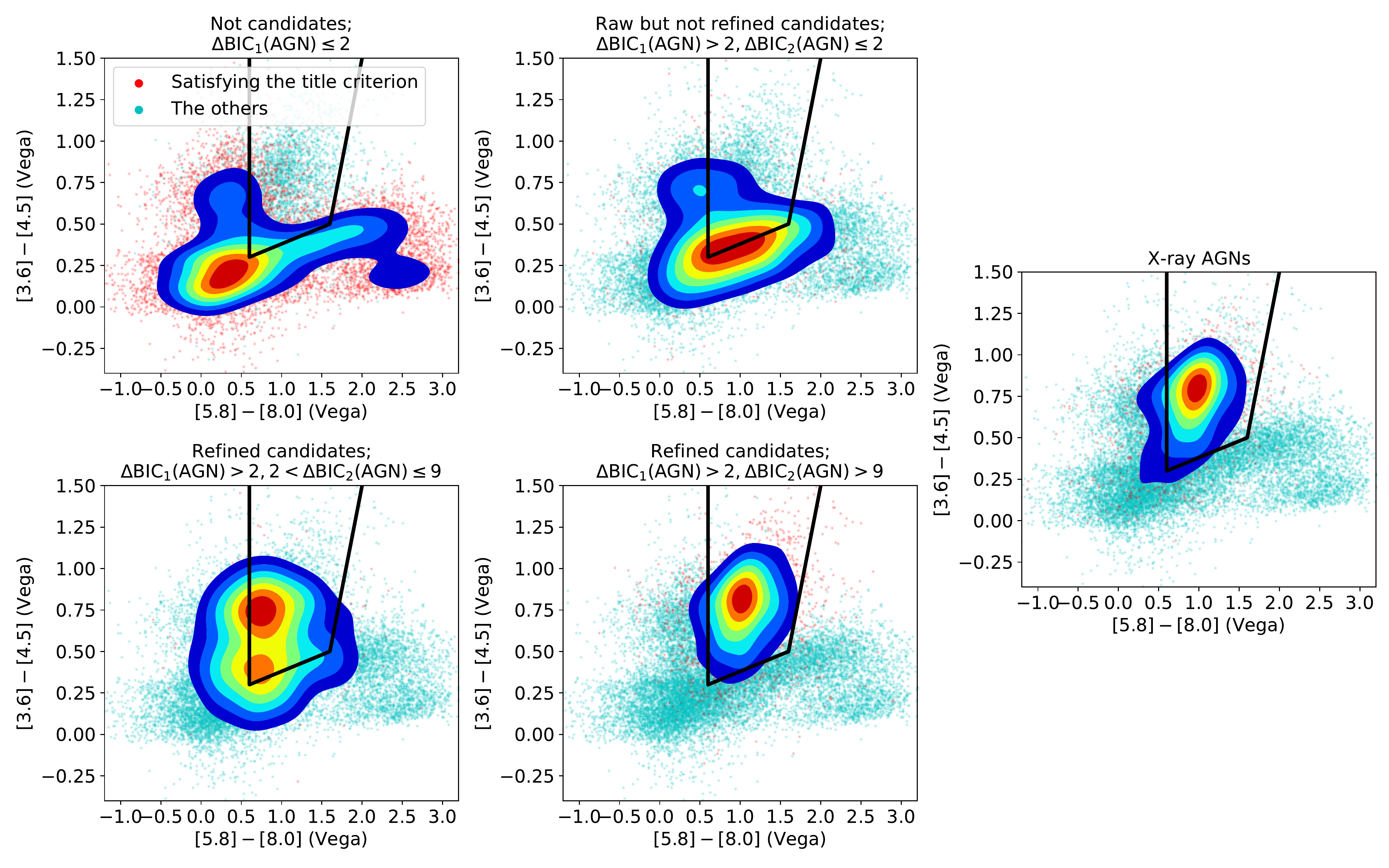}}
\caption{Same as Fig.~\ref{fig_ircolor}, but for the AGN-selection wedge in \citet{Stern05}.}
\label{fig_ircolor_S05}
\end{figure*}
\begin{figure*}
\centering
\resizebox{\hsize}{!}{\includegraphics{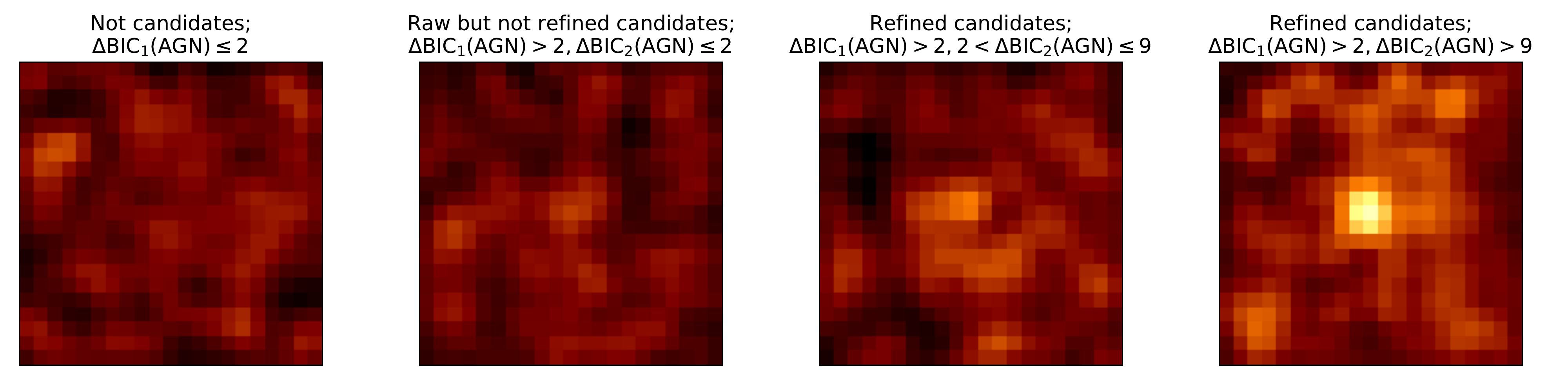}}
\caption{The stacked and smoothed \mbox{X-ray} FB net count-rate images covering $84''\times84''$. Each image is constructed from 1000 random sources satisfying the criterion in the corresponding panel title and at least $1'$ away from all the \mbox{X-ray} sources. The stacked signal visually increases with $\Delta\mathrm{BIC(AGN)}$.}
\label{fig_xraynetCR}
\end{figure*}

We compare the distributions of our sources with the MIR selection wedges \citep{Stern05, Lacy07, Donley12} in Figs.~\ref{fig_ircolor} and \ref{fig_ircolor_S05}. The distributions converge into the canonical wedges as $\Delta\mathrm{BIC(AGN)}$ increases, indicating that $\Delta\mathrm{BIC(AGN)}$ can indeed serve as an indicator for the existence of AGNs. Alternatively, excess \mbox{X-ray} emission can also indicate the existence of AGNs. The \mbox{X-ray} detection fractions of the four categories listed in the titles of the left four panels in Figs.~\ref{fig_ircolor} and \ref{fig_ircolor_S05} are 0.1\%, 1\%, 3\%, and 20\%, respectively. For the \mbox{X-ray}-undetected sources, we further perform \mbox{X-ray} stacking and present the results in Fig.~\ref{fig_xraynetCR}. For each panel of the figure, we have a list of sources satisfying the criterion at the top of the panel, and we randomly select 1000 sources that are at least $1'$ away from all the \mbox{X-ray} sources to avoid contamination. We then calculate the FB net count-rate map within a $84''\times84''$ region around each selected source and sum the signals together to obtain the stacked image, which is further smoothed and presented in Fig.~\ref{fig_xraynetCR}. The figure shows that the stacked \mbox{X-ray} signal increases toward higher $\Delta\mathrm{BIC(AGN)}$. Therefore, both the detected population and the undetected population in \mbox{X-rays} support that the AGN activity increases with $\Delta\mathrm{BIC(AGN)}$.\par

However, we caution that Figs.~\ref{fig_ircolor} $-$ \ref{fig_xraynetCR} are biased toward bright sources. For example, only 17\% of the refined SED AGN candidates have valid MIR colors, i.e., detected in all four IRAC bands with SNR above three, and thus the apparent agreements among sources with $\Delta\mathrm{BIC_2(AGN)}>9$ and the MIR wedges in Figs.~\ref{fig_ircolor} and \ref{fig_ircolor_S05} do not necessarily mean that $\Delta\mathrm{BIC_2(AGN)}>9$ is a good AGN-selection criterion for all sources. The only way to overcome this bias is to calibrate the SED selection with a complete and pure AGN sample. We present such a calibration in Section~\ref{sec: sedagn}, and it shows that faint SED AGNs are less reliable than bright SED AGNs.\par
\begin{figure*}
\centering
\resizebox{\hsize}{!}{
\includegraphics{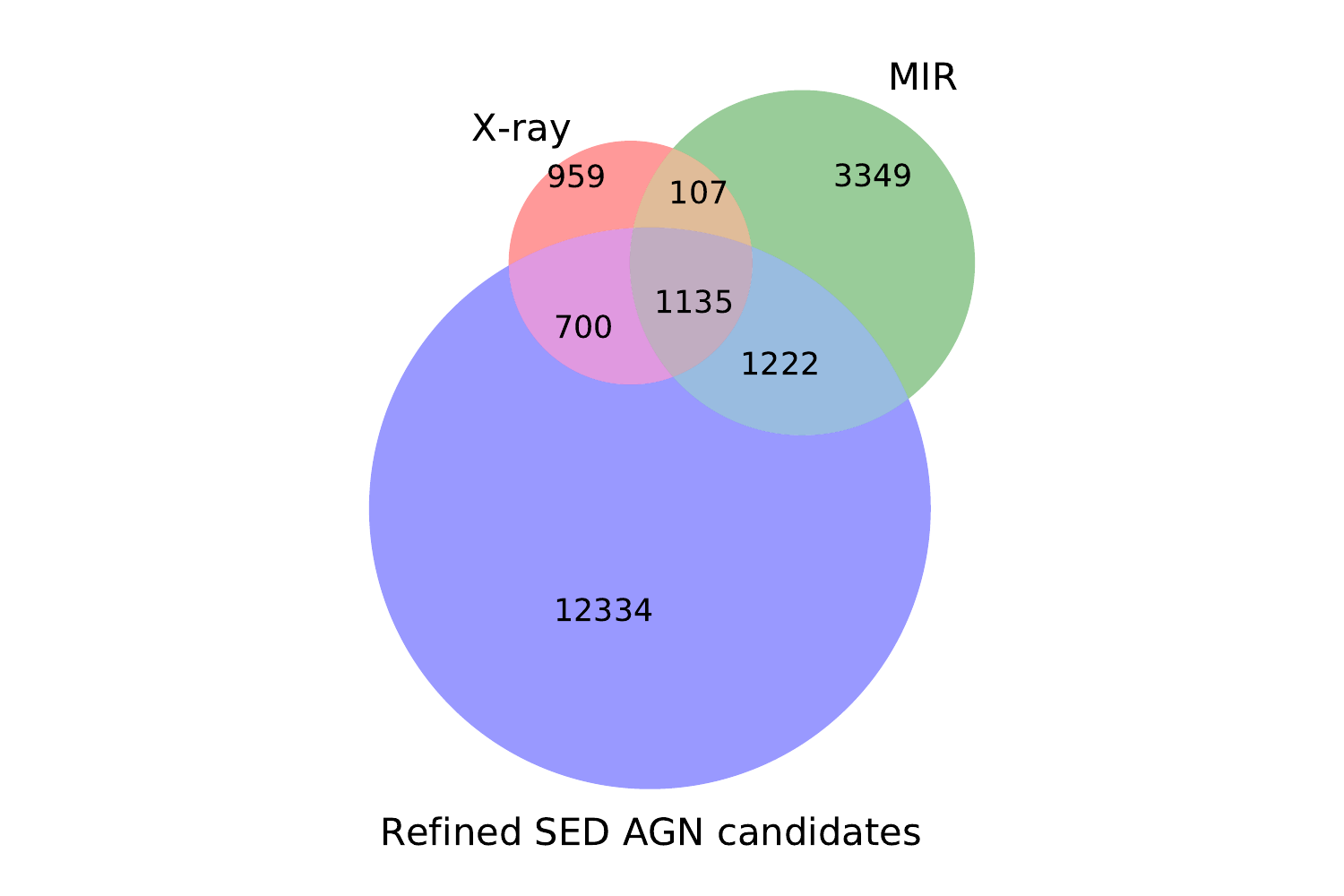}
\includegraphics{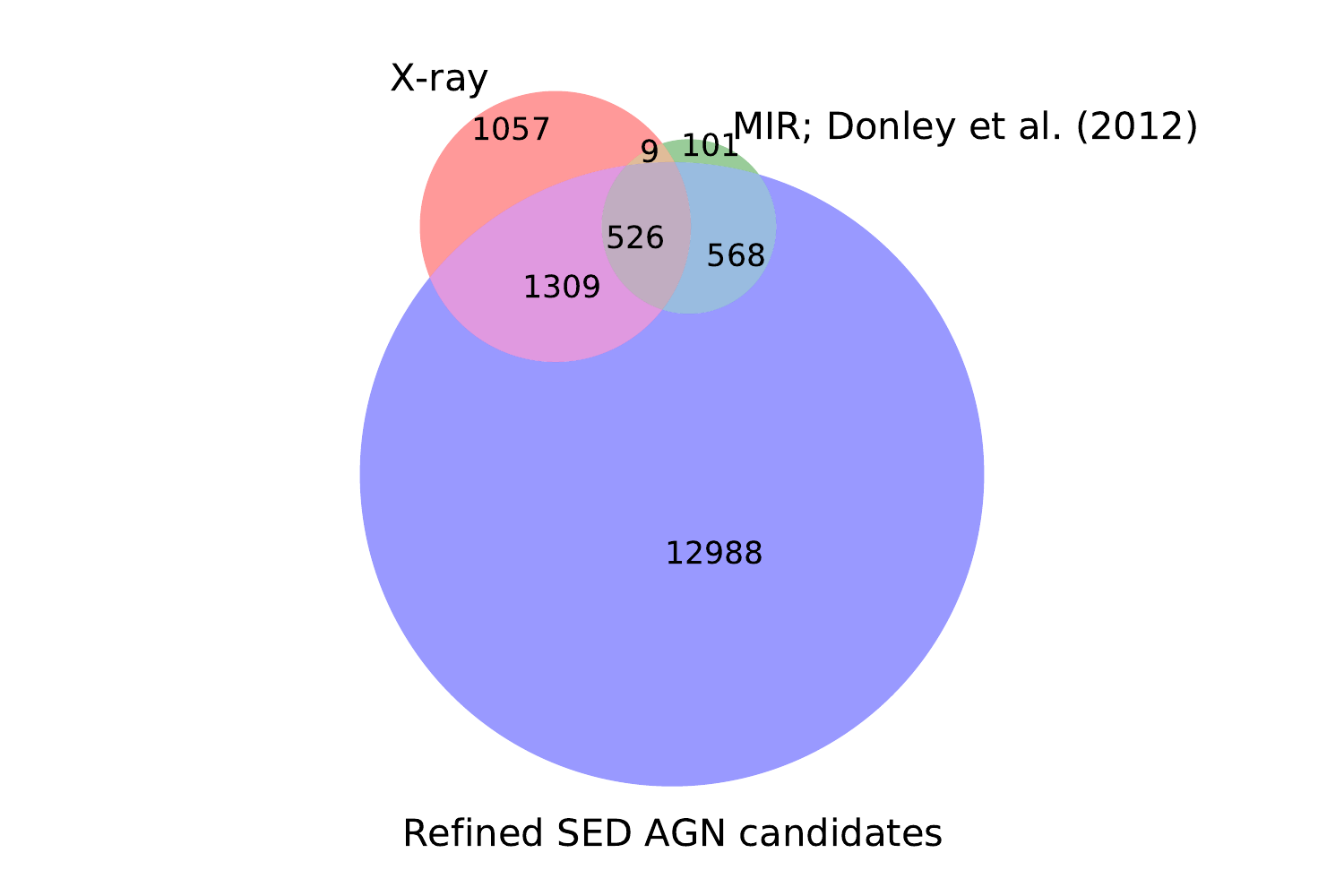}}
\caption{Venn diagrams comparing different AGN-selection results. The two panels differ for MIR AGNs, where the left panel contains MIR AGNs satisfying any criterion in \citet{Stern05}, \citet{Lacy07}, or \citet{Donley12}, while the right panel only contains MIR AGNs based on \citet{Donley12} to increase the purity. 63\% of \mbox{X-ray} AGNs, 41\% of MIR AGN, and 91\% of MIR AGNs based on \citet{Donley12} are also selected as refined SED AGN candidates, i.e., with $\Delta\mathrm{BIC_2(AGN)}>2$. There are many more SED AGN candidates than \mbox{X-ray} or MIR AGNs because of both the contamination of normal galaxies and missed AGNs by \mbox{X-ray} and MIR. The total number of AGNs and refined SED AGN candidates is slightly different from the number of sources with ``best'' results from the AGN fitting in Table~\ref{tbl_fieldinfo} because the best results of some refined SED AGN candidates are instead from the BQ-galaxy fitting (cf., Section~\ref{sec: bqagn}).}
\label{fig_venn_agn}
\end{figure*}
\begin{figure*}
\centering
\resizebox{\hsize}{!}{\includegraphics{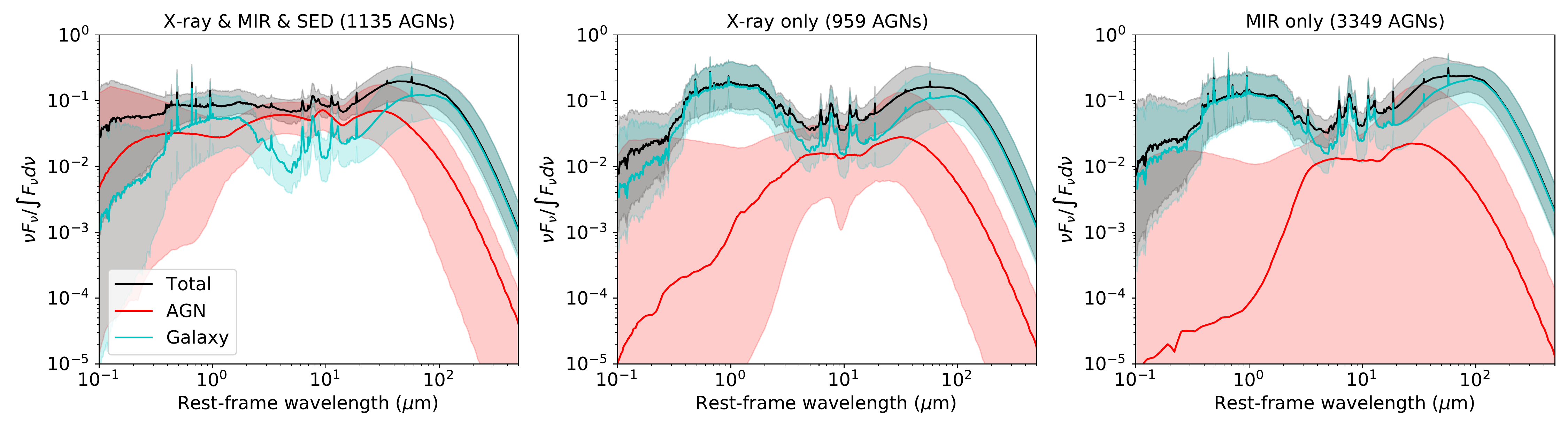}}
\caption{The typical SEDs of sources satisfying different AGN-selection conditions, as indicated by the panel titles, where ``SED'' means refined SED AGN candidates, and MIR AGNs are those satisfying any criterion in \citet{Stern05}, \citet{Lacy07}, or \citet{Donley12}. The source SEDs are normalized by their total fluxes integrated from \mbox{X-ray} to FIR, and the black, cyan, and red solid lines are the median total SEDs, galaxy, and AGN components, respectively. The shaded regions indicate $\mathrm{16^{th}-84^{th}}$ percentiles of the corresponding components. The red shaded regions become large in the optical because there are both type~1 and type~2 AGNs -- the former have bright and blue optical emission, while the latter are usually much fainter. The red shaded region for MIR-only AGNs is large because its $16^\mathrm{th}$ percentile is zero across all wavelengths, indicating that MIR-only AGNs may be largely contaminated by star-forming galaxies. The AGN components of \mbox{X-ray}- or MIR-only AGNs are generally more obscured and less dominant, which also explains why they are not selected as refined SED AGN candidates.}
\label{fig_stacksed}
\end{figure*}
\begin{figure}
\centering
\resizebox{\hsize}{!}{\includegraphics{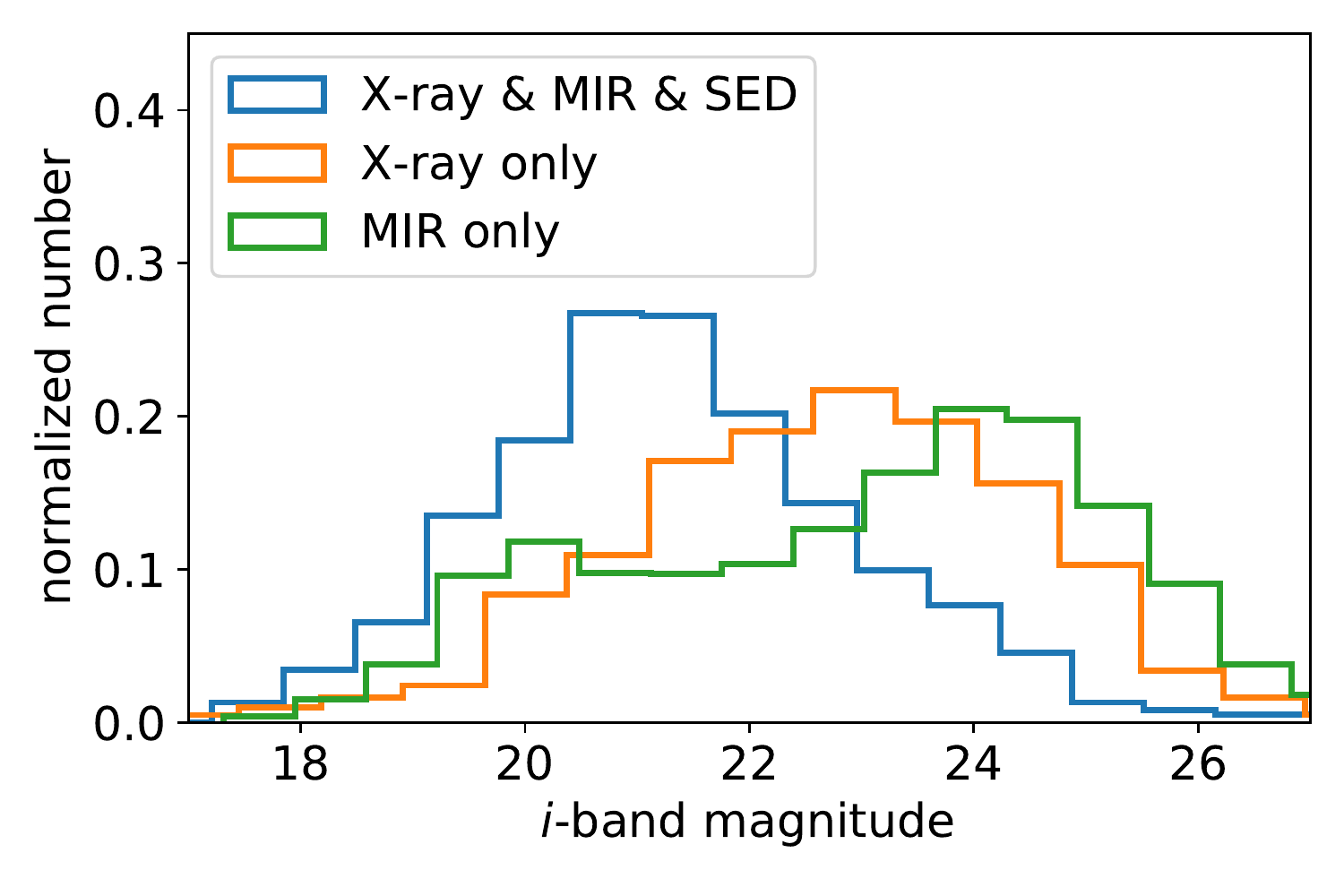}}
\caption{The $i$-band magnitude distributions of AGNs selected by different methods. The legend ``X-ray \& MIR \& SED'' means refined SED AGN candidates that are also identified by the \mbox{X-ray} and MIR selections, and ``X-ray (MIR) only'' refers to those only identified by \mbox{X-ray} (MIR) but not other methods. The \mbox{X-ray}- or MIR-only AGNs are generally fainter.}
\label{fig_imag_agns}
\end{figure}

We compare different selections in the left panel of Fig.~\ref{fig_venn_agn} using Venn diagrams. 63\% of the ground-truth \mbox{X-ray} AGNs are also identified as refined SED AGN candidates, but the total number of refined SED AGN candidates is much larger than those selected by \mbox{X-ray} or MIR, and this is because of both the contamination of galaxies to refined SED AGN candidates and missed AGNs by \mbox{X-ray} and MIR. The MIR AGNs may also be contaminated by star-forming galaxies, and this problem can be largely solved by adopting the stringent criterion in \citet{Donley12}, which is known to be able to select purer MIR AGN samples. The right panel of Fig.~\ref{fig_venn_agn} shows that most (91\%) of MIR AGNs selected by the criterion in \citet{Donley12} are also selected as refined SED AGN candidates. To probe the nature of the AGNs that are selected by only \mbox{X-ray} or MIR approaches, we show the composite SEDs of AGNs identified by different combinations of selection methods in Fig.~\ref{fig_stacksed}. The composite SEDs are defined as the median $\nu F_\nu/\int F_\nu d\nu$ curves of best-fit models as functions of rest-frame wavelength. For the composite SED of AGNs identified by all the three methods (left panel in Fig.~\ref{fig_stacksed}), the AGN component dominates in the MIR and also has considerable contributions in the optical. In contrast, the composite SEDs of AGNs selected only from \mbox{X-ray} or MIR show much weaker AGN contributions, and they tend to be more obscured, leading to the difficulty of identifying such AGNs through SED fitting. Especially, the $16^\mathrm{th}$ percentile of the composite AGN component for MIR-only AGNs is zero across all wavelengths, indicating that the MIR-only AGNs may be largely contaminated by star-forming galaxies. If we only adopt the MIR criterion in \citet{Donley12}, we will obtain similar results as in Fig.~\ref{fig_stacksed}, except that the $16^\mathrm{th}$ percentile SED of the MIR-only AGNs in the right panel of the figure will not be zero, but will look similar to that of the \mbox{X-ray}-only AGNs because there is little contamination from normal galaxies to the \citet{Donley12} AGNs. When matching $z$ and $L_\mathrm{X}$, we found that \mbox{X-ray} AGNs that are not selected as refined SED AGN candidates have larger host $M_\star$ than those that are selected as both \mbox{X-ray} and SED AGNs, also indicating that these \mbox{X-ray} AGNs are missed by the SED selection because of larger galaxy dilution. Using the $N_\mathrm{H}$ values derived in Section~\ref{sec: xrayphot}, we found that the \mbox{X-ray}-only AGNs are slightly more obscured in the \mbox{X-ray}, with a median $\log N_\mathrm{H}=22$, while the median $\log N_\mathrm{H}$ of AGNs selected by all the three methods is 21. Besides, \mbox{X-ray}-only or MIR-only AGNs are generally fainter, as illustrated by their $i$-band magnitude distributions in Fig.~\ref{fig_imag_agns}. Overall, the result that these \mbox{X-ray}-only or MIR-only AGNs are not selected by other methods may be caused by their faintness, smaller AGN contributions, and higher obscurations.\par

The incompleteness of our refined SED AGN candidates can hardly be resolved without greatly sacrificing purity. To illustrate this, we compare our raw SED AGN candidates with the \mbox{X-ray} and MIR selections, and the completeness only marginally increases compared to that of the refined SED AGN candidates -- 76\% of \mbox{X-ray} AGNs, 52\% of MIR AGNs, and 94\% of \citet{Donley12} MIR AGNs are identified as raw SED AGN candidates. Recall that the total number of raw SED AGN candidates is around three times larger than that of refined candidates, but the completeness only differs by around 10\%.

\subsubsection{Are SED AGNs reliable?}
\label{sec: sedagn}
\begin{figure}
\centering
\resizebox{\hsize}{!}{
\includegraphics{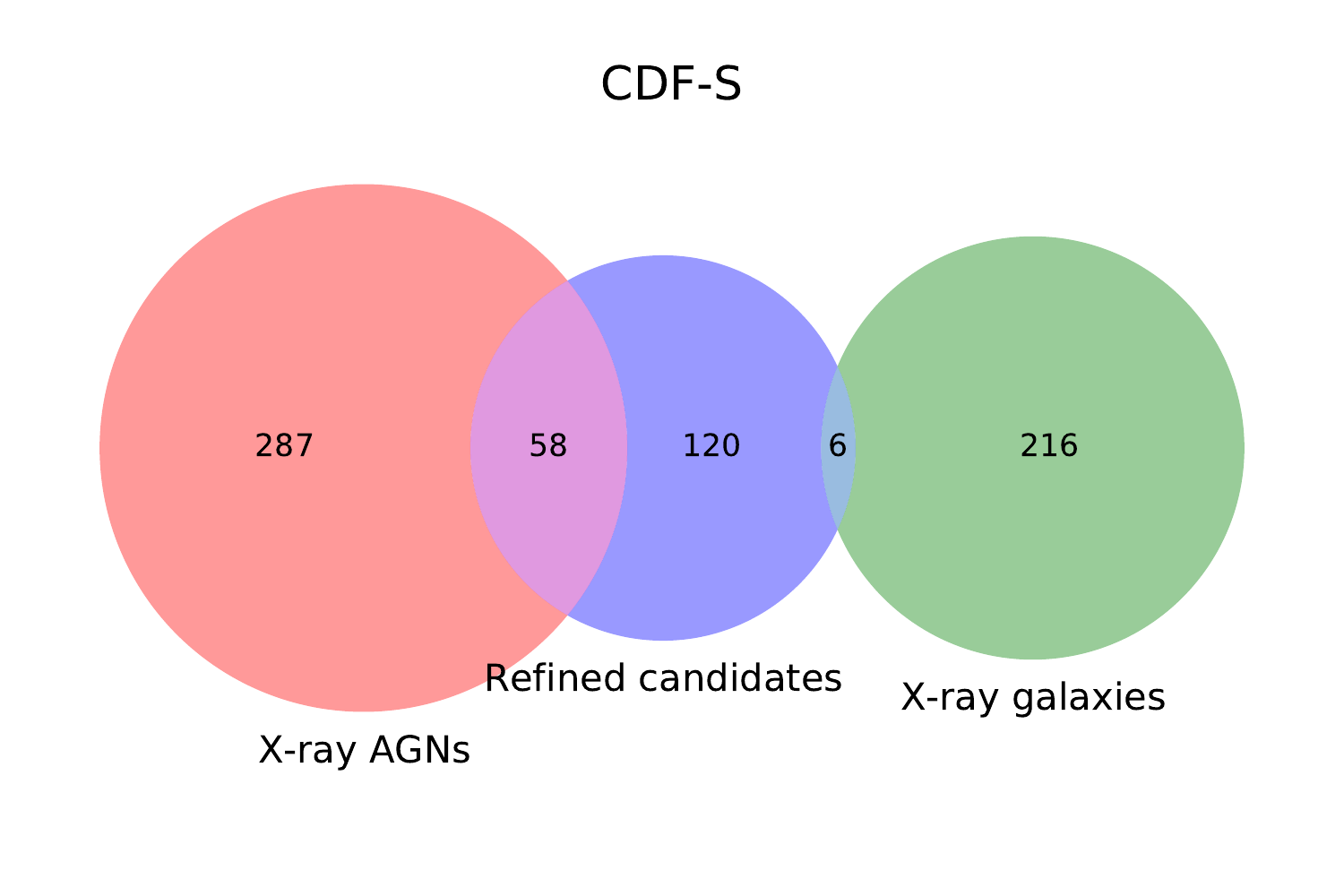}
}
\caption{A Venn diagram comparing our refined SED AGN candidates with the ground-truth 7~Ms \mbox{X-ray} sources in \mbox{CDF-S}. The overlap between our refined candidates and the \mbox{X-ray} AGNs is limited, indicating that both the purity and the completeness are not high for the refined candidates.}
\label{fig_venn_agn_candidate_cdfs_L17}
\end{figure}

We quantitatively examine the reliability of the SED AGN selection and further construct a criterion to select purer reliable SED AGNs from our refined SED AGN candidates in this section. We turn to the smaller embedded \mbox{CDF-S} field with 7~Ms Chandra observations \citep{Luo17} and ultradeep multi-wavelength observations to calibrate our SED selection. This deepest \mbox{X-ray} field ever obtained provides a largely complete pure AGN sample; that is, we do not expect our SED selection to be able to identify many AGNs missed by Chandra in \mbox{CDF-S}. Note that our multi-wavelength data have similar depths in \mbox{CDF-S} compared to the remaining parts of \mbox{W-CDF-S}, and thus the comparison should be representative for the whole \mbox{W-CDF-S} field. We focus on the central region with high Chandra exposure, i.e., within $6'$ around J2000 $\mathrm{RA=03^h32^m28.27^s, Dec=-27^\circ48'21.8''}$, and match our sources with those in \citet{Luo17}. \citet{Lambrides20} argued that the \mbox{X-ray} luminosities of faint sources in \citet{Luo17} may be underestimated due to their heavy obscuration, and thus some AGNs may be misclassified in \citet{Luo17}. Therefore, we regard a source to be an AGN if it is classified as an AGN in either \citet{Luo17} or \citet{Guo19}, where \citet{Guo19} reclassified six galaxies in \citet{Luo17} as AGNs. There are 345 AGNs and 222 \mbox{X-ray}-detected galaxies in total, and we display the Venn diagram comparing them with our refined SED AGN candidates in the left panel of Fig.~\ref{fig_venn_agn_candidate_cdfs_L17}. There are many refined SED AGN candidates undetected in \mbox{X-rays}, and they are expected to be mainly galaxies misclassified as AGNs by the SED selection. The overall purity, defined as the fraction of sources identified as \mbox{CDF-S} AGNs, of our refined SED AGN candidates is 32\%; the completeness, defined as the fraction of \mbox{CDF-S} AGNs identified as refined SED AGN candidates, is 17\%.\par

\begin{figure}
\centering
\resizebox{\hsize}{!}{
\includegraphics{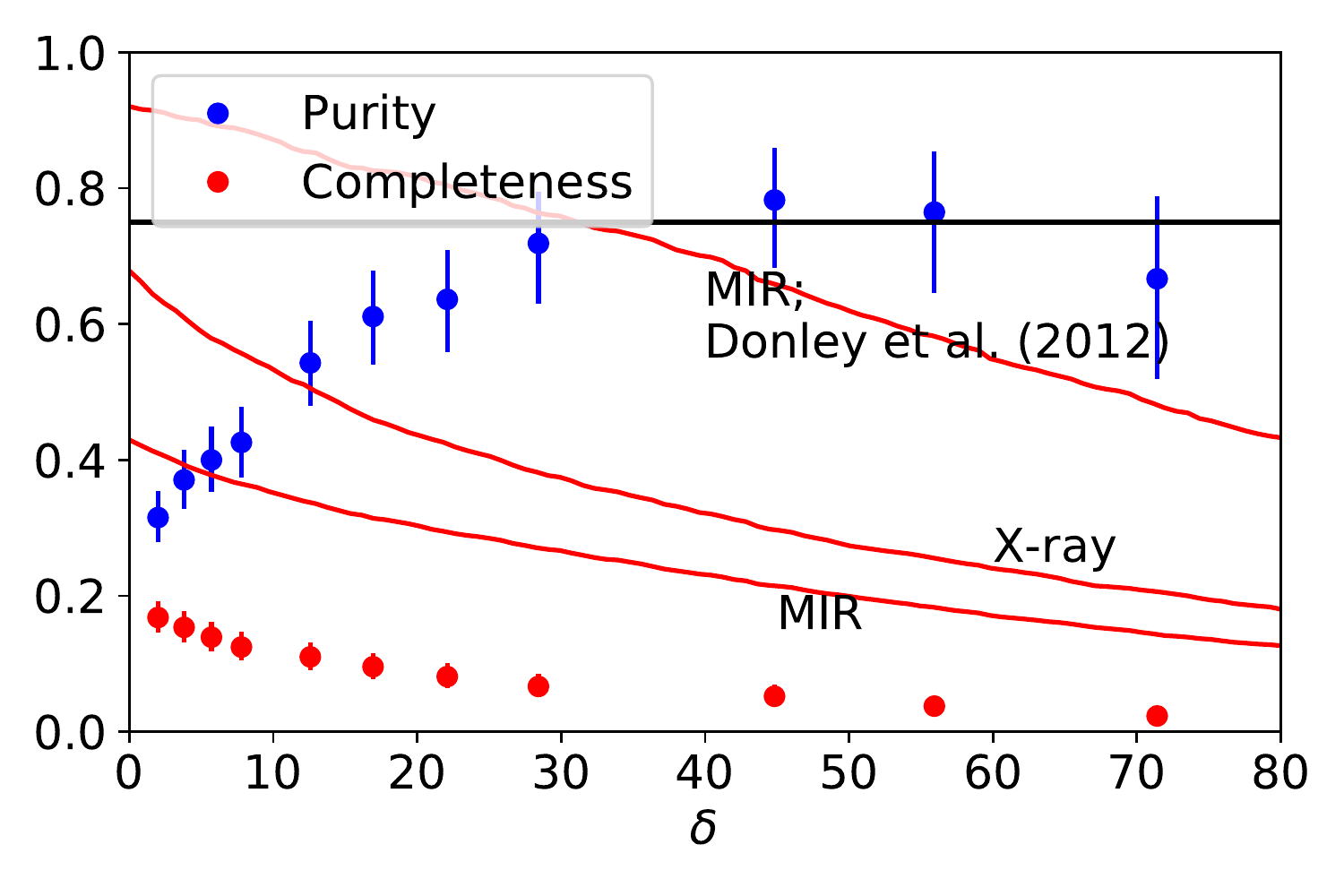}
}
\caption{The purity (blue) and completeness (red) of AGNs for a sample with $\Delta\mathrm{BIC_2(AGN)}>\delta$ as a function of the threshold $\delta$. The purity and completeness points are calculated by calibration over the \mbox{CDF-S} data, and the associated error bars represent their binomial proportion confidence intervals. The black horizontal line marks a purity of 75\%, which is roughly the plateau that the purity can reach when $\delta$ is large. The red curves are the completeness of retrieving \mbox{X-ray} or MIR AGNs in the whole \mbox{W-CDF-S} field using the criterion of $\Delta\mathrm{BIC_2(AGN)}>\delta$, where ``MIR'' means all the sources satisfying any criterion in \citet{Stern05}, \citet{Lacy07}, or \citet{Donley12}, while ``MIR; Donley et al. (2012)'' means only for MIR AGNs satisfying \citet{Donley12}.}
\label{fig_purity_complete_detbic}
\end{figure}

We further probe how the purity and completeness evolve with the threshold of $\Delta\mathrm{BIC_2(AGN)}$. For a given threshold, $\delta$, we select sources with $\Delta\mathrm{BIC_2(AGN)}>\delta$ and follow the same procedure as above to calculate the corresponding purity and completeness. The results are presented in Fig.~\ref{fig_purity_complete_detbic}. As a comparison, we also show the completeness curves of retrieving the AGNs selected by \mbox{X-ray} or MIR in the whole \mbox{W-CDF-S} field given the $\Delta\mathrm{BIC_2(AGN)}$ threshold. The curves are higher than the completeness points in \mbox{CDF-S} because the AGN samples themselves in \mbox{W-CDF-S} are incomplete. Some MIR AGNs are actually galaxies, which may lower the completeness by increasing the denominator of the completeness calculation (i.e., the total number of MIR AGNs). As discussed in the last paragraph of Section~\ref{sec: select_results}, the completeness of our candidates can hardly be improved much without greatly decreasing the purity. This is also supported by Fig.~\ref{fig_purity_complete_detbic}, which shows that the purity decreases rapidly around $\delta=2$.\par

\begin{figure*}
\centering
\resizebox{\hsize}{!}{
\includegraphics{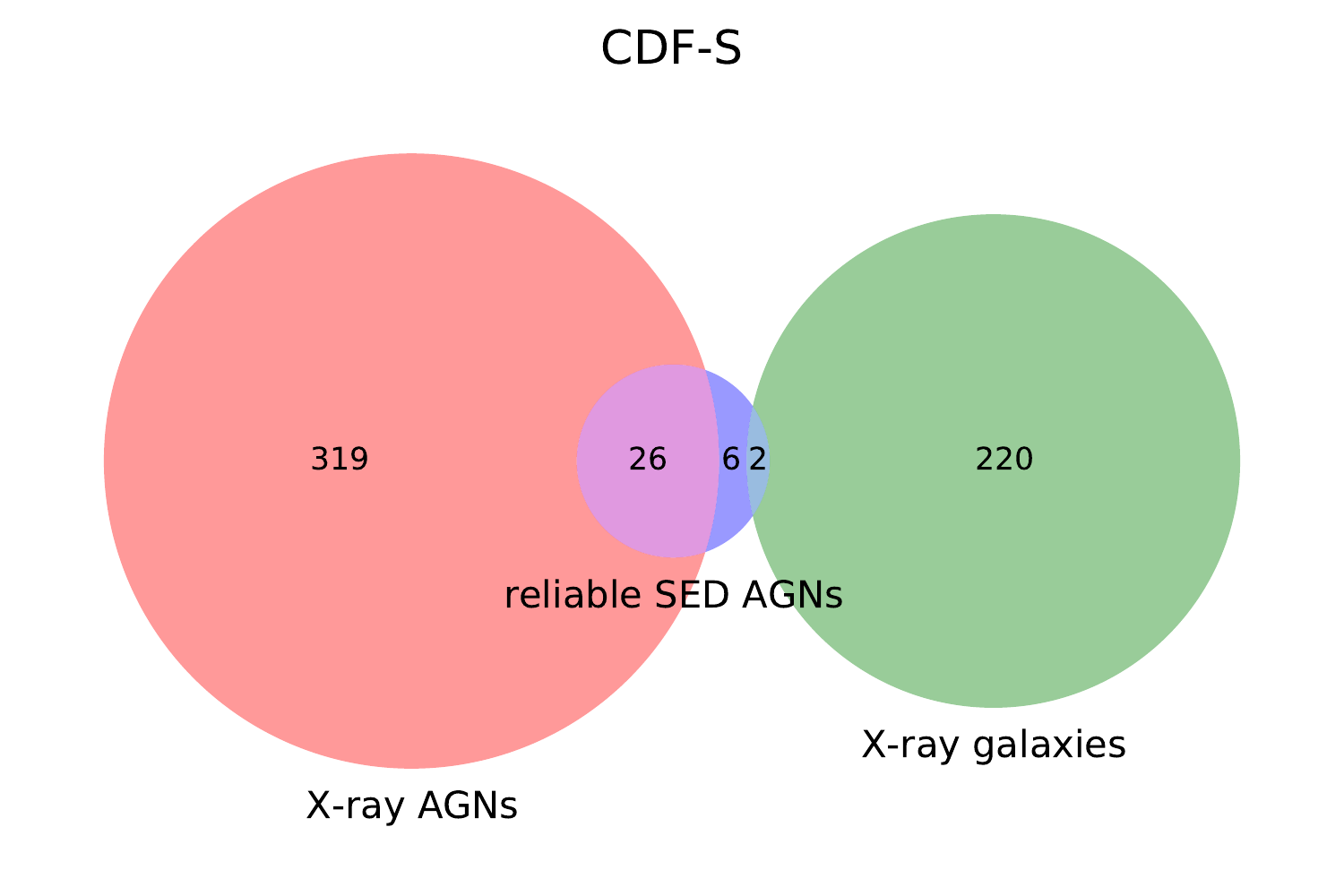}
\includegraphics{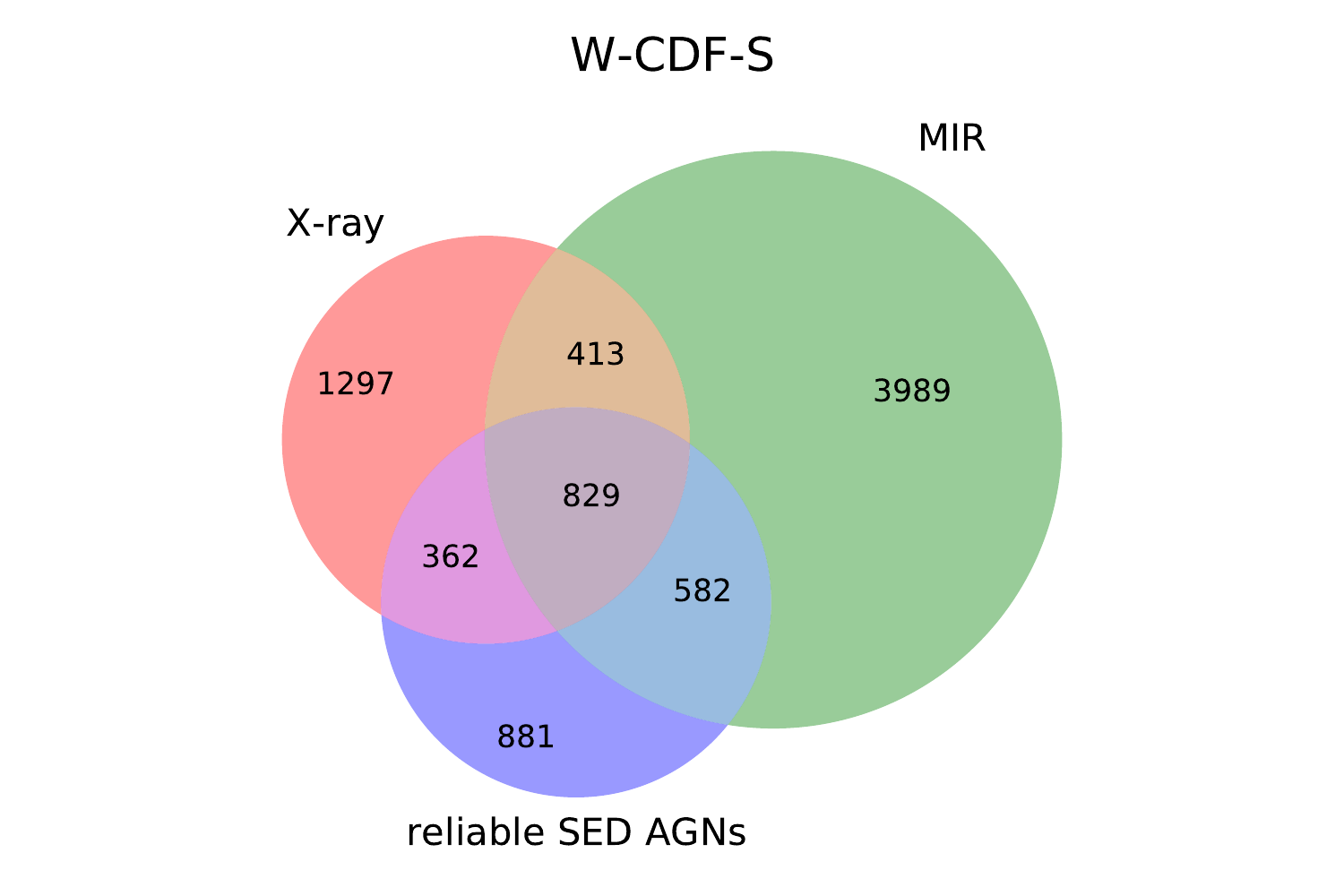}
}
\caption{Venn diagrams comparing our reliable SED AGNs (cf., Eqs.~\ref{eq: reliable_sedagn} and \ref{eq: crit_sedagn}) in \mbox{CDF-S} (\textit{left}) and \mbox{W-CDF-S} (\textit{right}). The purity of our reliable SED AGNs is $(76\pm7)\%$, according to the \mbox{CDF-S} calibration, and 69\% of these sources are also identified as \mbox{X-ray} or MIR AGNs in \mbox{W-CDF-S}.}
\label{fig_venn_reliable_agn_cdfs_L17}
\end{figure*}

Oftentimes, purity matters more than completeness, and thus we further calibrate the selection to select reliable SED AGNs with a high purity. Fig.~\ref{fig_purity_complete_detbic} shows that there is a plateau of $\approx75\%$ in purity when $\delta$ is high, and thus we adopt purity $\ge75\%$ as our requirement for reliable SED AGNs. We adopt a simple tree-like criterion and assume that a reliable SED AGN should satisfy
\begin{align}
\begin{cases}
\Delta\mathrm{BIC_2(AGN)}\ge\delta_1, \chi^2_r\leq3,~\mathrm{if}~i_\mathrm{mag}\leq i_\mathrm{break},\\
\Delta\mathrm{BIC_2(AGN)}\ge\delta_2, \chi^2_r\leq3,~\mathrm{if}~i_\mathrm{mag}>i_\mathrm{break},
\end{cases}\label{eq: reliable_sedagn}
\end{align}
where $i_\mathrm{mag}$ is the HSC $i$-band magnitude, and we found that the magnitude condition can help our selection. We require that among the sources satisfying the above criteria in each magnitude bin, at least 75\% are AGNs. By adjusting the parameters, we found that the total number of such sources is maximized when
\begin{align}
i_\mathrm{break}=23, \delta_1=4, \mathrm{and}~\delta_2=50.\label{eq: crit_sedagn}
\end{align}
This results in 34 sources, and 26 of them are labeled as AGNs in \citet{Luo17} or \citet{Guo19}, i.e., a purity of $(76\pm7)\%$. The Venn diagram under our criterion is displayed in the left panel of Fig.~\ref{fig_venn_reliable_agn_cdfs_L17}. The high purity is achieved at the expense of a high incompleteness, and the source sky density decreases from $\sim3000~\mathrm{deg^{-2}}$ for refined SED AGN candidates to $\sim600~\mathrm{deg^{-2}}$ for these reliable SED AGNs. It is worth noting that if we perform the same calibration using $\Delta\mathrm{BIC_1(AGN)}$ over the raw SED AGN candidates instead of the refined SED AGN candidates, we can obtain the following criterion.
\begin{align}
\begin{cases}
\Delta\mathrm{BIC_1(AGN)}\ge17, \chi^2_r\leq3,~\mathrm{if}~i_\mathrm{mag}\leq22.5,\\
\Delta\mathrm{BIC_1(AGN)}\ge54, \chi^2_r\leq3,~\mathrm{if}~i_\mathrm{mag}>22.5.\label{eq: reliable_sedagn_bic1}
\end{cases}
\end{align}
29 sources will be retrieved, and 22 will be AGNs. Among them, 27/29 sources selected by Eq.~\ref{eq: reliable_sedagn_bic1} are also selected by Eqs.~\ref{eq: reliable_sedagn} and \ref{eq: crit_sedagn}, and all the 22 true AGNs also satisfy the $\Delta\mathrm{BIC_2(AGN)}$ reliable SED AGN criterion in the meantime. Therefore, the reliable SED AGN sample is robust no matter whether $\Delta\mathrm{BIC_1(AGN)}$ or $\Delta\mathrm{BIC_2(AGN)}$ is adopted as long as careful calibrations are performed, except that $\Delta\mathrm{BIC_2(AGN)}$ is slightly more efficient in selecting more AGNs.\par

By far, we have been focusing only on the SED AGN candidates, and one may wonder whether we can select many more reliable SED AGNs from the non-candidates to supplement the reliable SED AGN sample selected only from the candidates. We will argue that the answer is ``no'' in this paragraph. First, as we showed in Section~\ref{sec: select_results}, sources that are raw SED AGN candidates but not refined SED AGN candidates are much more likely to host AGNs than non-candidates because the former sample has a higher \mbox{X-ray} detection fraction and MIR colors more inclined toward the AGN MIR color-color wedges. It is thus expected that the fraction of reliable SED AGNs that we can obtain among the \mbox{CDF-S} AGNs in a given sample is larger for the sources that are raw but not refined candidates compared to the non-candidates. We follow the same calibration using $\Delta\mathrm{BIC_2(AGN)}$ for the sources that are raw but not refined candidates. We have 27 true AGNs in \mbox{CDF-S} that belong to the population and can only select one reliable SED AGN out of them. There are 184 AGNs in \mbox{CDF-S} classified as non-candidates, and the expected number of retrievable reliable SED AGNs among them is thus smaller than $1/27\times184=7$. This is an expected hard limit and is much smaller than the current number of reliable SED AGNs constructed only from the SED AGN candidates. Furthermore, we can directly try using $\Delta\mathrm{BIC_1(AGN)}$ to select reliable SED AGNs among the non-candidates.\footnote{Recall that the non-candidates do not have $\Delta\mathrm{BIC_2(AGN)}$ values.} The calibration using Eq.~\ref{eq: reliable_sedagn} but with $\Delta\mathrm{BIC_1(AGN)}$ returns zero reliable SED AGNs. Our previous paragraph has justified that the reliable SED AGNs are largely insensitive to the choice of $\Delta\mathrm{BIC_1(AGN)}$ or $\Delta\mathrm{BIC_2(AGN)}$, and thus it is expected that even if we spend vast computational resources obtaining the $\Delta\mathrm{BIC_2(AGN)}$ values for all the non-candidates, they can hardly provide more reliable SED AGNs, and we hence decide not to run the dense-grid AGN-template fitting for the non-candidates.\par
We then apply the calibration results in Eqs.~\ref{eq: reliable_sedagn} and \ref{eq: crit_sedagn} to the whole \mbox{W-CDF-S} field and found that 69\% of the resulting reliable SED AGNs can be selected by \mbox{X-ray} or MIR, as shown in the right panel of Fig.~\ref{fig_venn_reliable_agn_cdfs_L17}. Recall that the expected purity is $(76\pm7)\%$ for the reliable SED AGNs, and 69\% is consistent with this expected purity. We thus conclude that the SED method can hardly \textit{reliably} identify more AGNs missed by other methods in our fields. This is not surprising because better SED selections require high-quality MIR data, and \citet{Yang21b} showed that this problem cannot be solved straightforwardly without deep and continuous MIR-band coverage from, e.g., JWST.\par
\begin{figure}
\centering
\resizebox{\hsize}{!}{\includegraphics{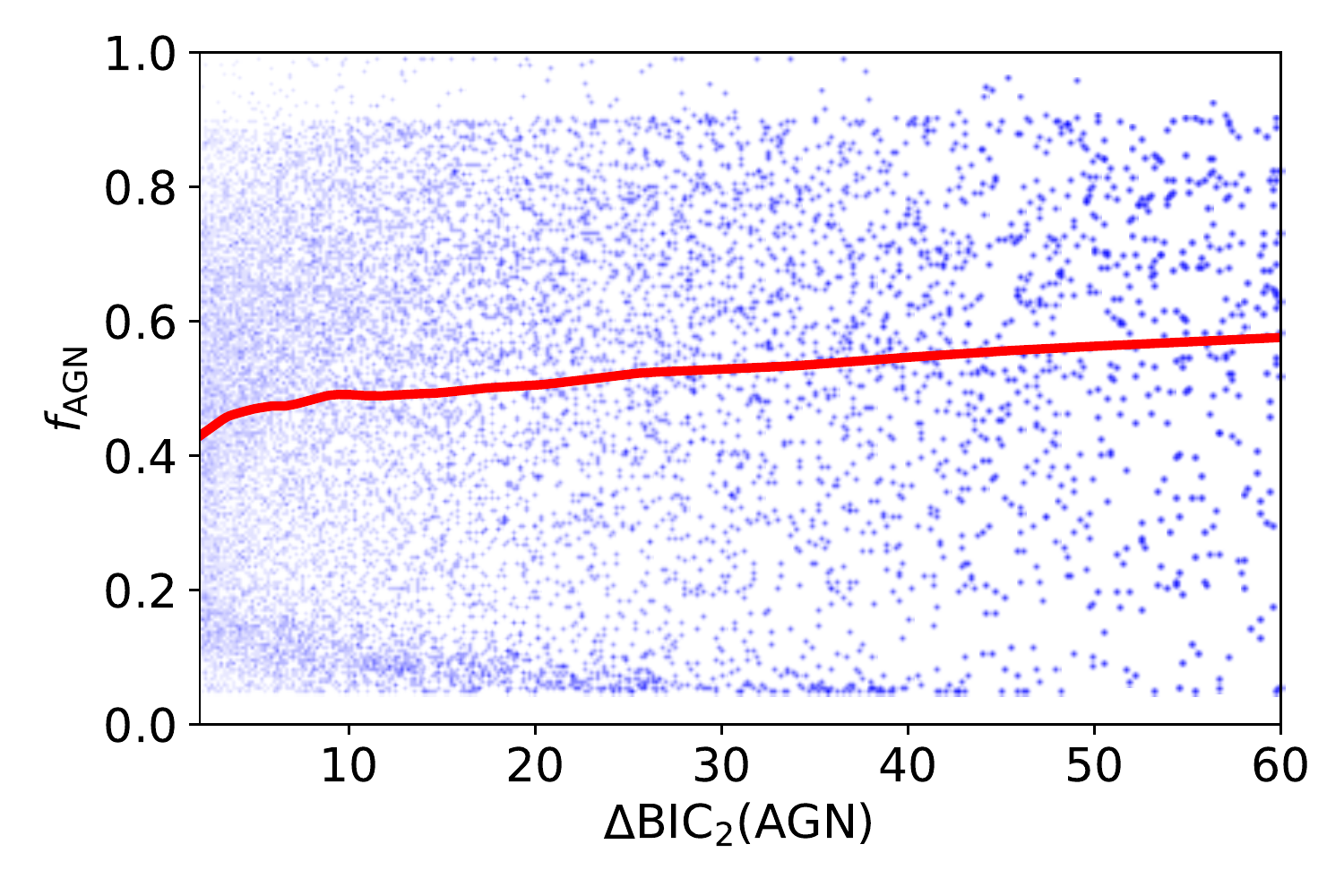}}
\caption{$f_\mathrm{AGN}$ vs. $\Delta\mathrm{BIC_2(AGN)}$ for refined SED AGN candidates. The red line is the LOESS curve of the points. The apparent horizontal point density does not reflect source number because we intentionally increase the point size and opacity at larger $\Delta\mathrm{BIC_2(AGN)}$ for better visualization. There is only a weak positive correlation between the two parameters, and $f_\mathrm{AGN}$ scatters across a wide range because it can hardly be constrained well by the available data.}
\label{fig_fracagn_detbic}
\end{figure}

Although \texttt{CIGALE} outputs $f_\mathrm{AGN}$, we do not rely on this parameter to select AGN candidates because it often has large systematic and statistical errors (e.g., \citealt{Ciesla15, Yang21b}). We present $f_\mathrm{AGN}$ versus $\Delta\mathrm{BIC_2(AGN)}$ in Fig.~\ref{fig_fracagn_detbic}. To depict the general trend of our sources, we also plot the locally estimated scatterplot smoothing (LOESS; e.g., Chapter~6 of \citealt{Feigelson12} and references therein) curve. The LOESS technique is effectively similar to the running mean or median in nonparametrically drawing a rough trend for scattered points, but LOESS provides smoother curves and avoids arbitrarily choosing abscissa bins. We will consistently use LOESS in Section~\ref{sec: sedfitting}. Fig.~\ref{fig_fracagn_detbic} shows that the two parameters only have a weak positive correlation, and $f_\mathrm{AGN}$ is largely a random number spanning a wide range regardless of $\Delta\mathrm{BIC_2(AGN)}$. This is because $f_\mathrm{AGN}$ generally cannot be constrained well by the current data \citep{Yang21b}, and thus we should not directly use $f_\mathrm{AGN}$ to select AGNs. \citet{Thorne22} is a recent example supporting our argument. They selected SED AGN candidates in COSMOS by requiring an AGN fraction between $5-20~\mu\mathrm{m}$ above 0.1. 42\% of their sources were regarded as AGN candidates, and they successfully classified 69\% of the \citet{Donley12} MIR AGNs as SED AGN candidates. In contrast, our refined SED AGN candidates only constitute 2\% of all the sources but include up to 91\% of MIR AGNs from \citet{Donley12}, and thus using AGN fraction to select AGNs may misclassify many normal galaxies as AGNs and/or miss real AGNs.

\subsection{Selection of BQ-galaxy Candidates}
\label{sec: select_bqgal}
We select BQ galaxies in this section. Our main goal is only to select BQ-galaxy candidates in an economical manner to improve their SED-fitting results within the \texttt{CIGALE} framework, and detailed characterizations of these sources are left for future works.\par
Broad-band SEDs can be used to select BQ galaxies by checking if the modeled SFH has undergone a rapid change within several hundred million years. In \texttt{CIGALE}, such galaxies can be modeled by a truncated delayed SFH \citep{Ciesla16}, formulated as the following:
\begin{align}
\mathrm{SFR}(t)\propto
\begin{cases}
t\exp(-t/\tau),~t\leq t_\mathrm{trunc}\\
r_\mathrm{SFR}\mathrm{SFR(t_\mathrm{trunc})},~t>t_\mathrm{trunc}
\end{cases},\label{eq: delayedsfhbq}
\end{align}
where the formula at $t\leq t_\mathrm{trunc}$ is the normal delayed SFH with an \textit{e}-folding time of $\tau$, and the SFR is assumed to instantaneously change by a factor of $r_\mathrm{SFR}$ at $t_\mathrm{trunc}$ and then remain constant until the current age. A normal delayed SFH is thus modeled by $r_\mathrm{SFR}=1$ to a first-order approximation.\par
Similar to Section~\ref{sec: select_agn} and \citet{Ciesla18}, we use $\Delta\mathrm{BIC}$ between the fitting with normal and truncated delayed SFHs, and the candidates are selected in two steps. In the first step, we use the coarse-grid setting in Table~\ref{tbl_sedpar_step1} to calculate $\Delta\mathrm{BIC_1(BQ)}$ for all the sources and obtain 51 thousand sources with $\Delta\mathrm{BIC_1(BQ)}>2$; in the second step, we only fit sources with $\Delta\mathrm{BIC_1(BQ)}>2$ using the dense-grid settings in Tables~\ref{tbl_sedpar_step2_gal} and \ref{tbl_sedpar_step2_bqgal} to measure $\Delta\mathrm{BIC_2(BQ)}$. We do not add AGN components for simplicity. Six thousand BQ-galaxy candidates are selected with the criterion of $\Delta\mathrm{BIC_2(BQ)}>2$. The selection is done for all the sources no matter whether they are selected as SED AGN candidates or not (also see Section~\ref{sec: bqagn} for further discussion). When fitting BQ galaxies, the age of the BQ episode is set to be between 10 and 800~Myr. BQ episodes happening within $\sim100$~Myr are generally hard to detect with broad-band SEDs, but there are galaxies with strong bursts within a few tens of millions of years producing strong H$\alpha$ emission (e.g., \citealt{Broussard19}) that happen to reside in and dominate one of the observed bands (see Fig.~\ref{fig_example_sed} for an example). H$\alpha$ emission traces the star formation on a time scale of $\sim10$~Myr, and thus it is still helpful to include a few possible values between 10 to 100~Myr for the BQ episode age to better represent these bursting galaxies. Nevertheless, it is generally difficult to measure the BQ episode age reliably, as discussed in previous works (e.g., \citealt{Ciesla16, Ciesla21}), and thus this parameter should not be over-interpreted.\par
It is worth noting that unlike the AGN selection, which judges whether an additive component from the AGN emission is necessary, the BQ-galaxy selection judges whether the SED shape of the normal-galaxy templates is satisfactory. As we discussed in Section~\ref{sec: sedagn}, using $\Delta\mathrm{BIC_1(AGN)}$ alone can also return fairly reliable SED AGN results as long as calibrations are performed, and the $\Delta\mathrm{BIC_1(AGN)}$ results are only slightly less efficient than the $\Delta\mathrm{BIC_2(AGN)}$ results. This indicates that the AGN selection is largely insensitive to whether the galaxy templates are sufficiently inclusive because the difference between AGN and galaxy SEDs is large. However, the BQ-galaxy selection is subject to more subtle differences, and it is hence more important to have good normal-galaxy templates. We indeed found that most $\Delta\mathrm{BIC_1(BQ)}$ values are dominated by the imperfect galaxy templates, and only 11\% of sources with $\Delta\mathrm{BIC_1(BQ)}>2$ pass the criterion of $\Delta\mathrm{BIC_2(BQ)}>2$. Again, the limitation of $\Delta\mathrm{BIC_1(BQ)}$ is not necessarily a disadvantage as it returns a more complete sample. We will only focus on $\Delta\mathrm{BIC_2(BQ)}$ hereafter. Unlike introducing the terms of ``raw SED AGN candidates'' and ``refined SED AGN candidates'', we use a single term of ``BQ-galaxy candidates'' to describe sources with $\Delta\mathrm{BIC_2(BQ)}>2$ for simplicity.\par

\begin{table*}
\caption{Dense-grid \texttt{CIGALE} parameter settings for BQ-galaxy candidates}
\label{tbl_sedpar_step2_bqgal}
\centering
\begin{threeparttable}
\begin{tabular}{cccc}
\hline
\hline
\multirow{2}{*}{Module} & \multirow{2}{*}{Parameter} & Name in the \texttt{CIGALE} & \multirow{2}{*}{Possible values}\\
&& configuration file &\\
\hline
\multirow{7}{*}{Truncated delayed SFH} & \multirow{2}{*}{Stellar \textit{e-}folding time} & \multirow{2}{*}{tau\_main} & 0.1, 0.2, 0.3, 0.5, 0.7, 0.8,\\
&&&1, 2, 3, 5, 7, 8, 10 Gyr\\
& Stellar age & age\_main & 1, 2, 3, 5, 7, 8, 10 Gyr\\
& \multirow{2}{*}{Age of the BQ episode} & \multirow{2}{*}{age\_bq} & 10, 50, 100, 200, 300, 400\\
&&&500, 600, 700, 800 Myr\\
& \multirow{2}{*}{$r_\mathrm{SFR}$} & \multirow{2}{*}{r\_sfr} & 0, 0.05, 0.1, 0.15, 0.2, 0.3, 0.4, 0.5, 0.7, 0.8,\\
&&&1, 1.5, 2, 3, 5, 7, 8, 10, 20, 50, 70, 100\\
\hline
Simple stellar population & Initial mass function & imf & \citet{Chabrier03}\\
\citet{Bruzual03} & Metallicity & metallicity & 0.0001, 0.0004, 0.004, 0.008, 0.02, 0.05\\
\hline
Nebular & ----- & ----- & -----\\
\hline
\multirow{3}{*}{Dust attenuation} & \multirow{2}{*}{$E(B-V)_\mathrm{line}$} & \multirow{2}{*}{E\_BV\_lines} & 0, 0.05, 0.1, 0.15, 0.2, 0.25, 0.3, 0.4,\\
\multirow{3}{*}{\citet{Calzetti00}}&&& 0.5, 0.6, 0.7, 0.8, 0.9, 1, 1.2, 1.5\\
& $E(B-V)_\mathrm{line}/E(B-V)_\mathrm{continuum}$ & E\_BV\_factor & 1\\
\hline
Dust emission & \multirow{2}{*}{Alpha slope} & \multirow{2}{*}{alpha} & \multirow{2}{*}{1.0, 1.25, 1.5, 1.75, 2.0, 2.25, 2.5, 2.75, 3.0}\\
\citet{Dale14}\\
\hline
\mbox{X-ray} & ----- & ----- & -----\\
\hline
\hline
\end{tabular}
\begin{tablenotes}
\item
\textit{Notes.} Unlisted parameters are set to the default values. These are only applied to sources with $\Delta\mathrm{BIC_1(BQ)>2}$ and return $\Delta\mathrm{BIC_2(BQ)}$.
\end{tablenotes}
\end{threeparttable}
\end{table*}

\begin{figure}
\centering
\resizebox{\hsize}{!}{\includegraphics{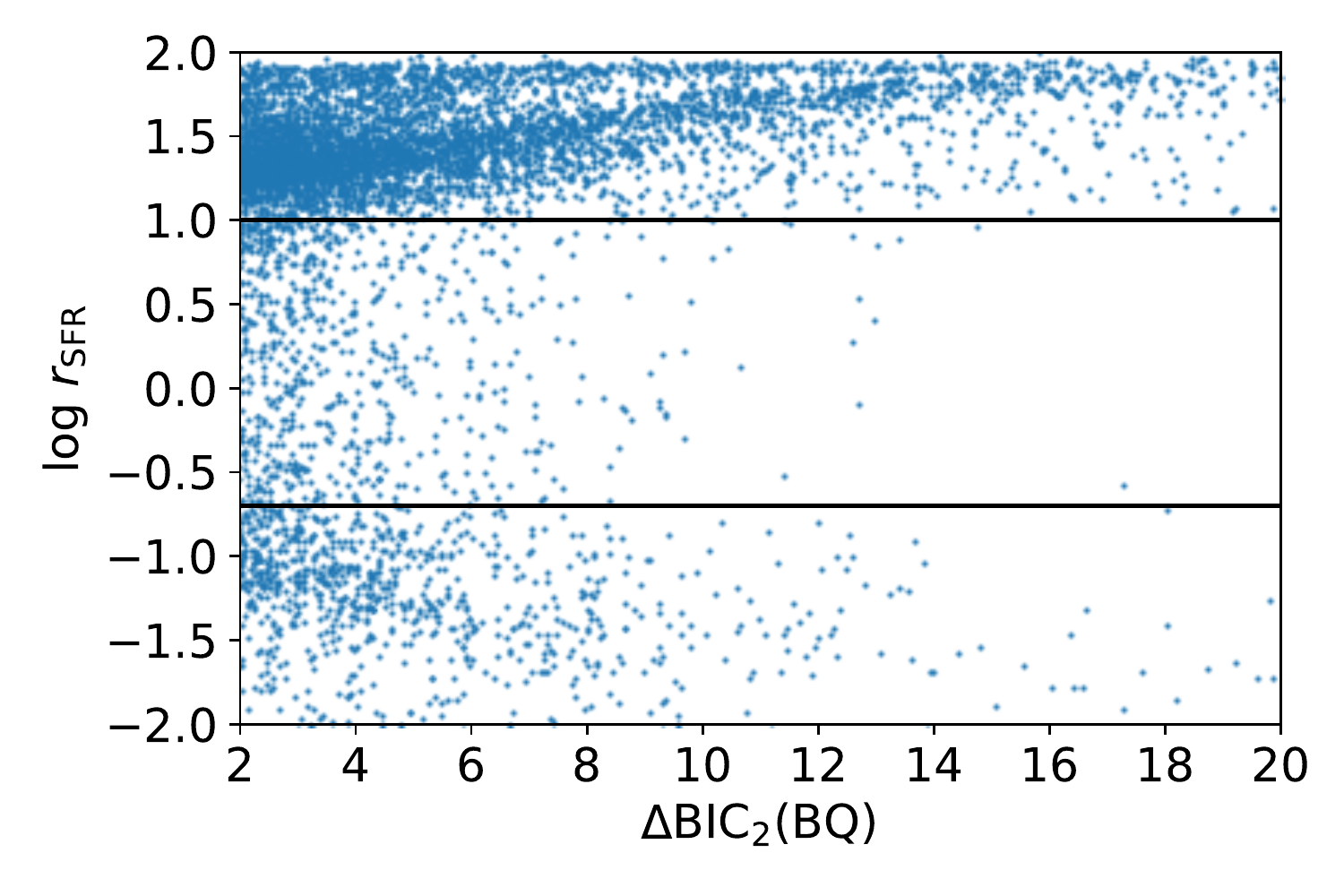}}
\caption{$r_\mathrm{SFR}$ vs. $\Delta\mathrm{BIC_2(BQ)}$ for our BQ-galaxy candidates. The $r_\mathrm{SFR}$ distribution is bimodal. The black horizontal lines represent $r_\mathrm{SFR}=0.2$ and 10, from which we empirically select quenching or bursting subpopulations.}
\label{fig_rsfr_detbic}
\end{figure}

We show $r_\mathrm{SFR}$ versus $\Delta\mathrm{BIC_2(BQ)}$ in Fig.~\ref{fig_rsfr_detbic}, and $r_\mathrm{SFR}$ is clearly bimodal. The bimodality increases with $\Delta\mathrm{BIC_2(BQ)}$. This indicates that the BQ-galaxy candidates include both quenching ($r_\mathrm{SFR}\ll1$) and bursting ($r_\mathrm{SFR}\gg1$) galaxies. Based on the figure, we empirically set $r_\mathrm{SFR}=0.2$ and 10 as the thresholds for the quenching and bursting subpopulations, respectively, where the quenching threshold is from \citet{Ciesla18}. While there is some subjectiveness in defining numerical cutoffs, these thresholds are chosen to ensure that these galaxies experience large changes in SFR which we expect to leave a clear observational signal. However, there is not a clear precedent for these choices in the galaxy-formation literature. Simulations have shown that normal galaxies can commonly have SFR variations up to 0.5~dex or even more within hundreds of millions of years, but the exact variability amplitude depends on both $M_\star$ and the simulation setup (e.g., \citealt{Iyer20}); a reasonable $r_\mathrm{SFR}$ threshold should thus be larger than 0.5~dex to distinguish from normal SFR fluctuations. Meanwhile, (post-)starburst galaxies are often defined by their observational features instead of their SFHs, and an exact mapping between the observational classification of such galaxies and their SFH parameters has not been fully constructed. \citet{Ciesla21} further showed that the recovery of $r_\mathrm{SFR}$ for normal galaxies with $r_\mathrm{SFR}\approx1$ can span a range of $\sim1$~dex. Due to these reasons, exact $r_\mathrm{SFR}$ thresholds are difficult to obtain, but we have confirmed that our qualitative results do not depend on the adopted values as long as they are reasonable. We will briefly analyze the quenching subpopulation and bursting subpopulation in the following paragraphs.\par

\begin{figure}
\centering
\resizebox{\hsize}{!}{\includegraphics{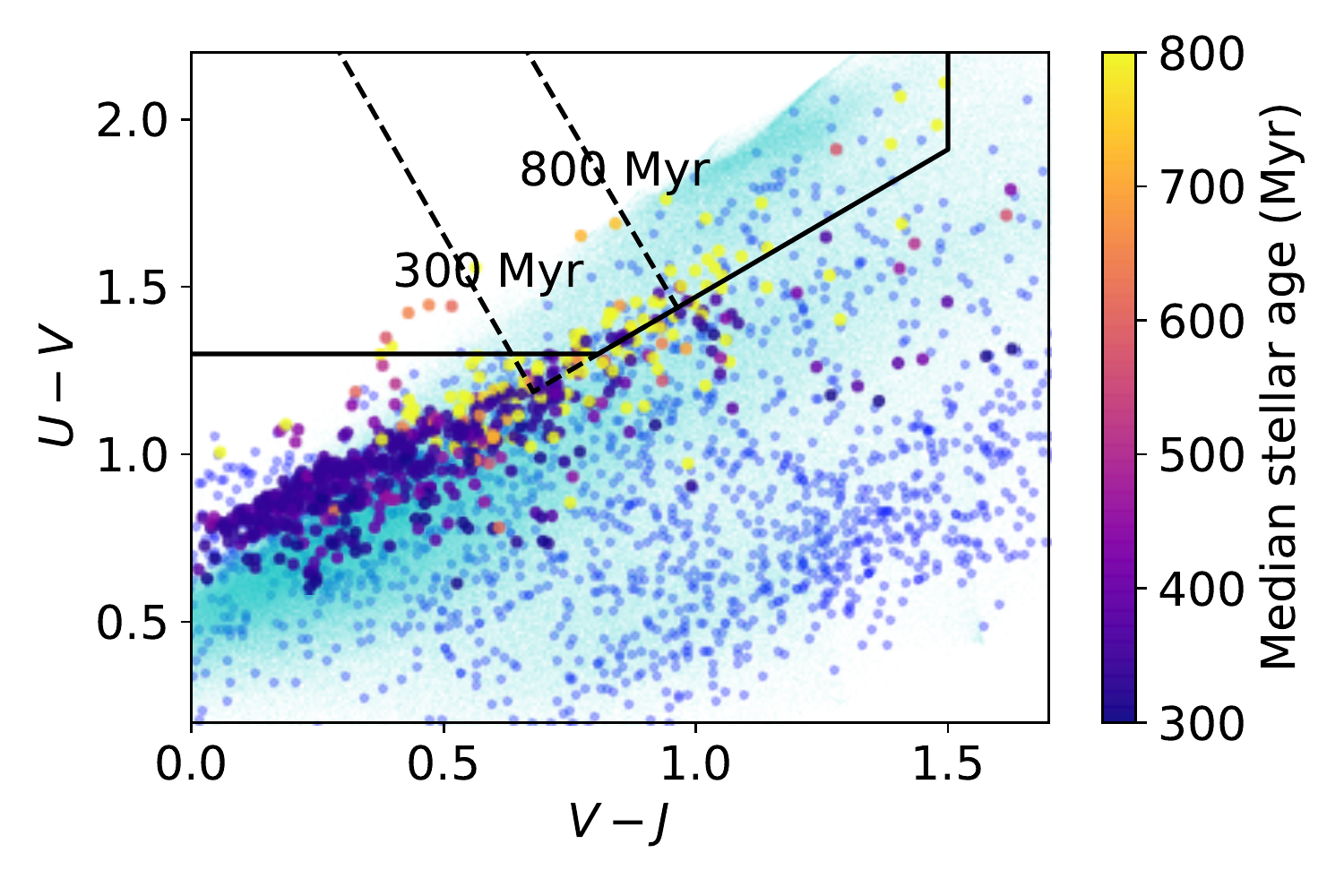}}
\caption{The distributions of sources in the $UVJ$ color-color plane. The large points are likely quenching galaxies color-coded by their median stellar age, the blue points are likely bursting galaxies, and the cyan points, plotted as a comparison, are our whole sample in \mbox{W-CDF-S}. The solid line is the boundary expected to enclose quiescent galaxies in \citet{Muzzin13} at $z>1$. The dashed line is the post-starburst region in \citet{Belli19}, which is defined as the region for which the expected median stellar age is between 300 and 800~Myr, as labeled in the figure. The likely quenching galaxies are generally in the star-forming region but on top of the main star-forming locus and have an age gradient toward the quiescent region. The likely bursting galaxies are scattered across the star-forming region.}
\label{fig_UVJ}
\end{figure}

First, we define ``likely quenching galaxies'' as those with $r_\mathrm{SFR}<0.2$ and $\chi_r^2\leq3$. This results in 639 sources. We show their distribution in the rest-frame $UVJ$ color-color diagram in Fig.~\ref{fig_UVJ}, where the $UX$- and $V$-band definitions in \citet{Bessell90} and $J$-band definition in \citet{Tokunaga02} are adopted. The $UVJ$ diagram is widely used to identify quiescent and star-forming galaxies (e.g., \citealt{Williams09, Whitaker12, Muzzin13, Leja19c}), and Fig.~\ref{fig_UVJ} reveals that our likely quenching galaxies generally locate in the star-forming region but on top of the main star-forming locus and form a line pointing toward the quiescent region. The line formed by these sources is generally parallel with the age-color relation in \citet{Belli19}, who defined the post-starburst region as median stellar age between 300 and 800~Myr (but see \citealt{Wu20} for a different conclusion). To compare with the relation in \citet{Belli19}, we color-code our likely quenching galaxies by their median stellar age in Fig.~\ref{fig_UVJ}, and they indeed show an age gradient such that the age generally increases toward the upper-right direction. Our median stellar ages are slightly larger than the relation in \citet{Belli19}, possibly because of the different choice of SFH and the fact that our redshifts are generally smaller than those in \citet{Belli19}. Nevertheless, the locations of our sources in Fig.~\ref{fig_UVJ} indicate that they should have undergone quenching very recently (within a few hundreds of millions of years) so that they have not entered the quiescent region, as also found in \citet{Ciesla18} (see their Fig.~8). Therefore, we are capturing quenching star-forming galaxies (i.e., those that are transitioning from the star-forming phase to the quiescent phase) instead of quenched quiescent galaxies because the latter generally do not strongly require a truncated delayed SFH to model their SEDs, even though they may have undergone (slow or rapid) quenching gigayears ago.\par

Similar to likely quenching galaxies, we define ``likely bursting galaxies'' as those satisfying $r_\mathrm{SFR}>10$ and $\chi_r^2\leq3$. We further require that they are not selected as AGNs or refined SED AGN candidates to avoid degeneracies (see Section~\ref{sec: bqagn} for more discussion). We also empirically exclude sources with $z=0.01$. 0.01 is the minimum photo-$z$ value allowed in \citet{Zou21b}, and photo-$z$s reaching this boundary are often caused by failures in photo-$z$ measurements and are hence unreliable. Such sources usually ``pile up'' at a single $z=0.01$ value. These cases are rare (3\%) for all the sources, but we found that they are enhanced and can account for 11\% of the bursting subpopulation. These $z=0.01$ bursting galaxies may still be real bursting galaxies at low redshifts ($z\lesssim0.2$), and their bursting nature may be the actual cause for why their photo-$z$s are inaccurate as the templates used in deriving photo-$z$s may not be able to fit their bursting SEDs.\footnote{Besides the limitations of the templates, there are also other reasons that can cause the $z=0.01$ solution for the general galaxy population, such as unreliable photometry (e.g., due to large angular sizes of low-redshift galaxies) and peculiar motions that are comparable to the Hubble flow.} The main problem caused by their small photo-$z$s is that their $M_\star$ and SFRs are hence highly underestimated. For example, the distance at $z=0.01$ is ten times smaller than that at $z=0.1$, even if the redshift difference is small and is thus not expected to cause material difference in the observed SED shape given a rest-frame SED. After removing these sources, we obtain a total of 1899 likely bursting galaxies.\par
We plot the likely bursting galaxies in the $UVJ$ plane in Fig.~\ref{fig_UVJ}, and they generally scatter across the star-forming region. These likely bursting galaxies are possible candidates for starburst and/or rejuvenating galaxies. J. Zhang et al. (in preparation) found that rejuvenating galaxies generally cover a similar region in color-color diagrams as normal star-forming galaxies, explaining the large scatter of our likely bursting galaxies (J. Zhang 2022, private comm). Rejuvenating galaxies are still largely poorly understood and worthy of probing more carefully (e.g., \citealt{Chauke19, Mancini19}), and we leave such analyses to the future.\par
Similar to the AGN selection, the BQ-galaxy selection and Eq.~\ref{eq: delayedsfhbq} may also face challenges. For example, $r_\mathrm{SFR}$ is difficult to constrain \citep{Ciesla16, Ciesla18}, and fluctuations of star formation, especially for low-mass galaxies, may mimic quenching (e.g., \citealt{El-Badry16}); how these factors may affect the selection results and Fig.~\ref{fig_UVJ} are still unknown. Efforts to improve the BQ-galaxy selection can be made in the future. For example, \citet{Aufort20} presented a machine-learning-based approximate Bayesian computation algorithm to select BQ galaxies and successfully applied it to the COSMOS field. Their method may help improve the BQ-galaxy selections in our fields. Another worthwhile project is to use \texttt{Prospector-$\alpha$} \citep{Leja17} to do the SED fitting. One of the main advantages of \texttt{Prospector-$\alpha$} compared to \texttt{CIGALE} is that the former allows non-parametric SFHs, and thus should be able to provide better measurements for SFHs. However, \texttt{Prospector-$\alpha$} is much more computationally demanding and cannot fit millions of SEDs with common computational resources. Our BQ-galaxy candidates, whose total number is much smaller than the number of all the sources, can thus significantly reduce the computational requirements by serving as a parent sample for the future \texttt{Prospector-$\alpha$} fitting, which may select BQ galaxies more accurately.

\subsection{Normal galaxies}
For the majority of sources that are not selected as stars, \mbox{X-ray} or IR AGNs, refined AGN candidates, or BQ-galaxy candidates, we call them ``normal galaxies''. They generally do not need specialized SED-fitting methods and thus will be treated together in the same manner. We use the parameter settings in Table~\ref{tbl_sedpar_step2_gal} to derive their properties.

\subsection{SED-fitting Results}
\label{sec: bestsedfittingresults}
\begin{figure*}
\centering
\resizebox{\hsize}{!}{
\includegraphics{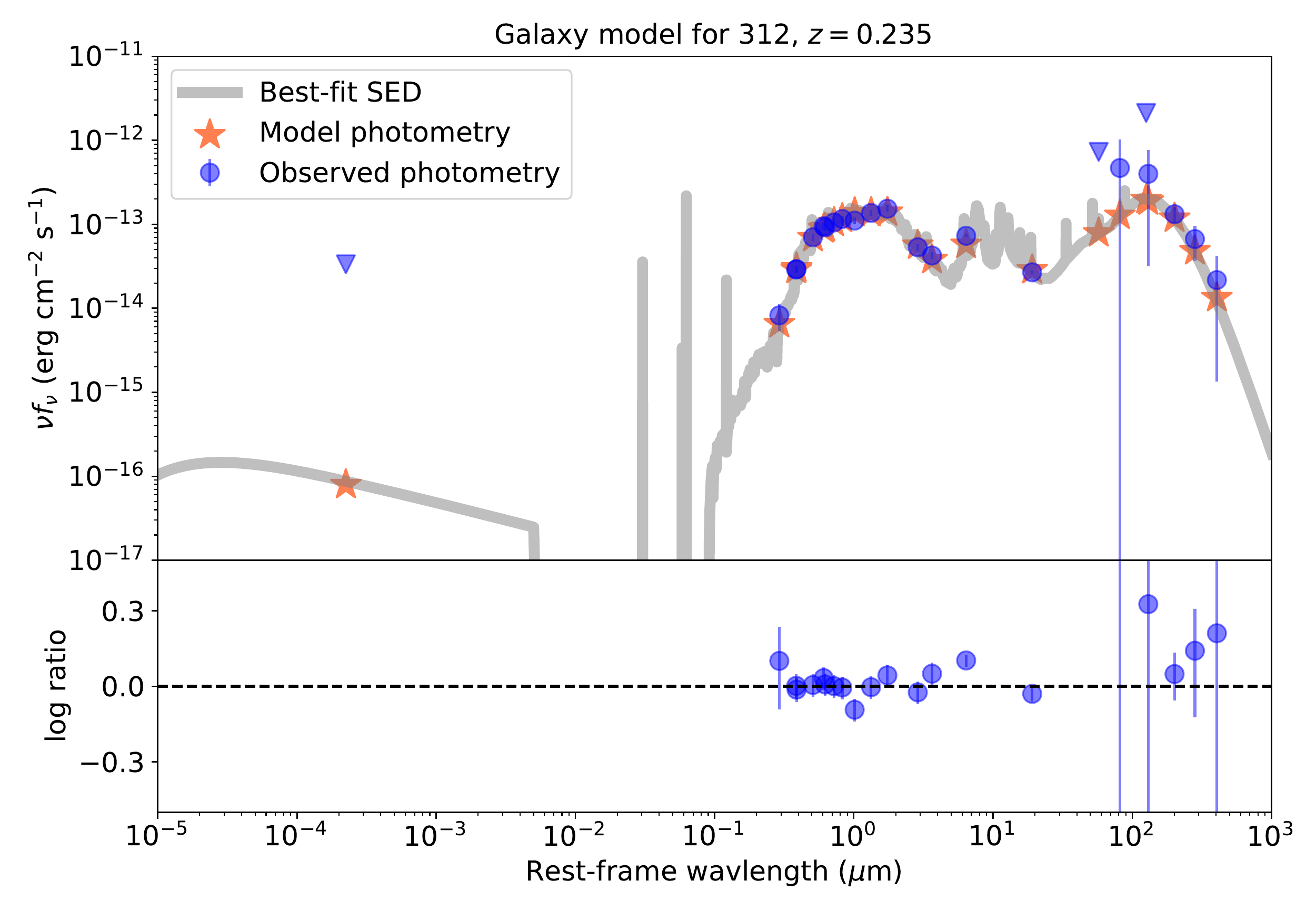}
\includegraphics{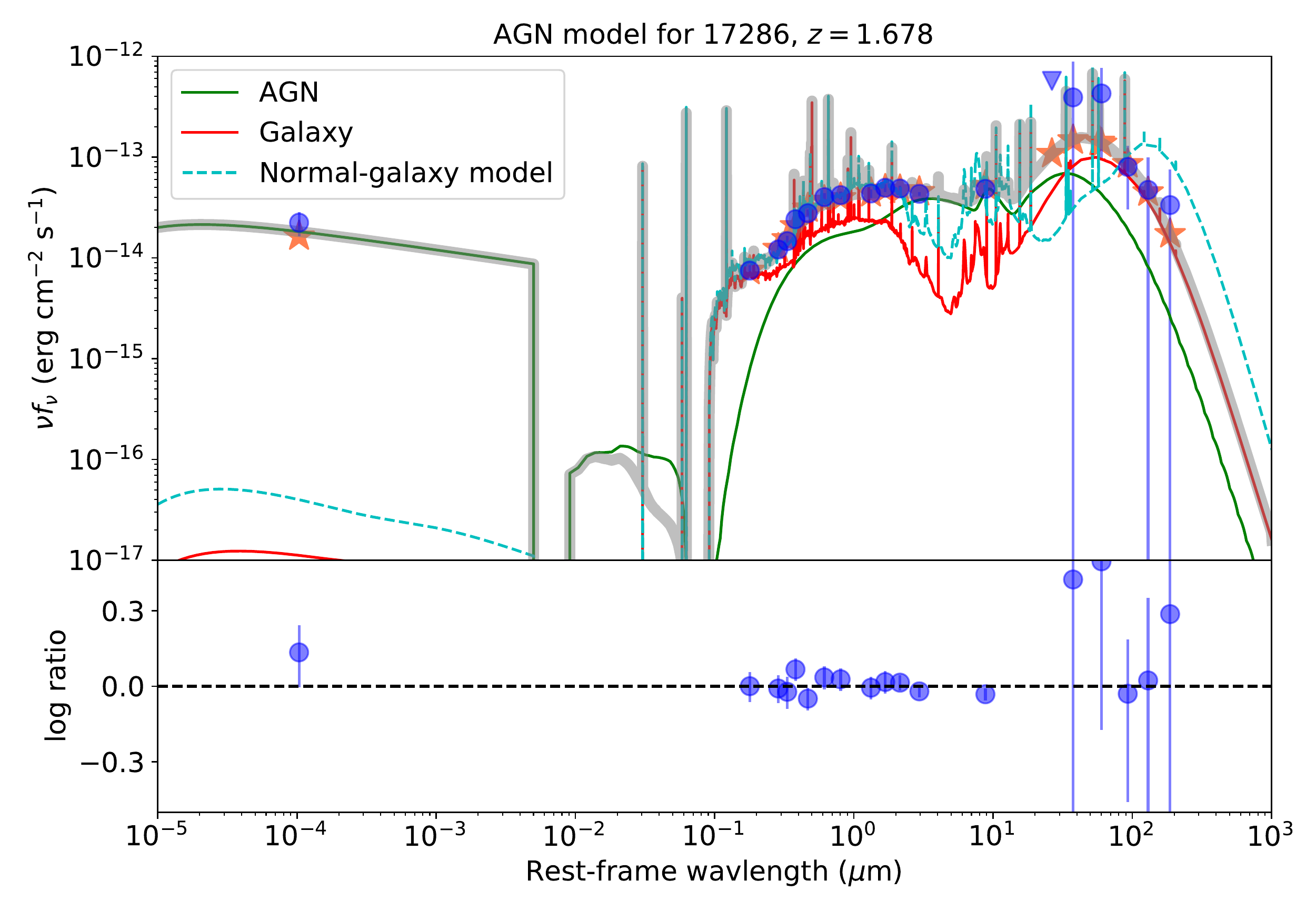}
}
\resizebox{\hsize}{!}{
\includegraphics{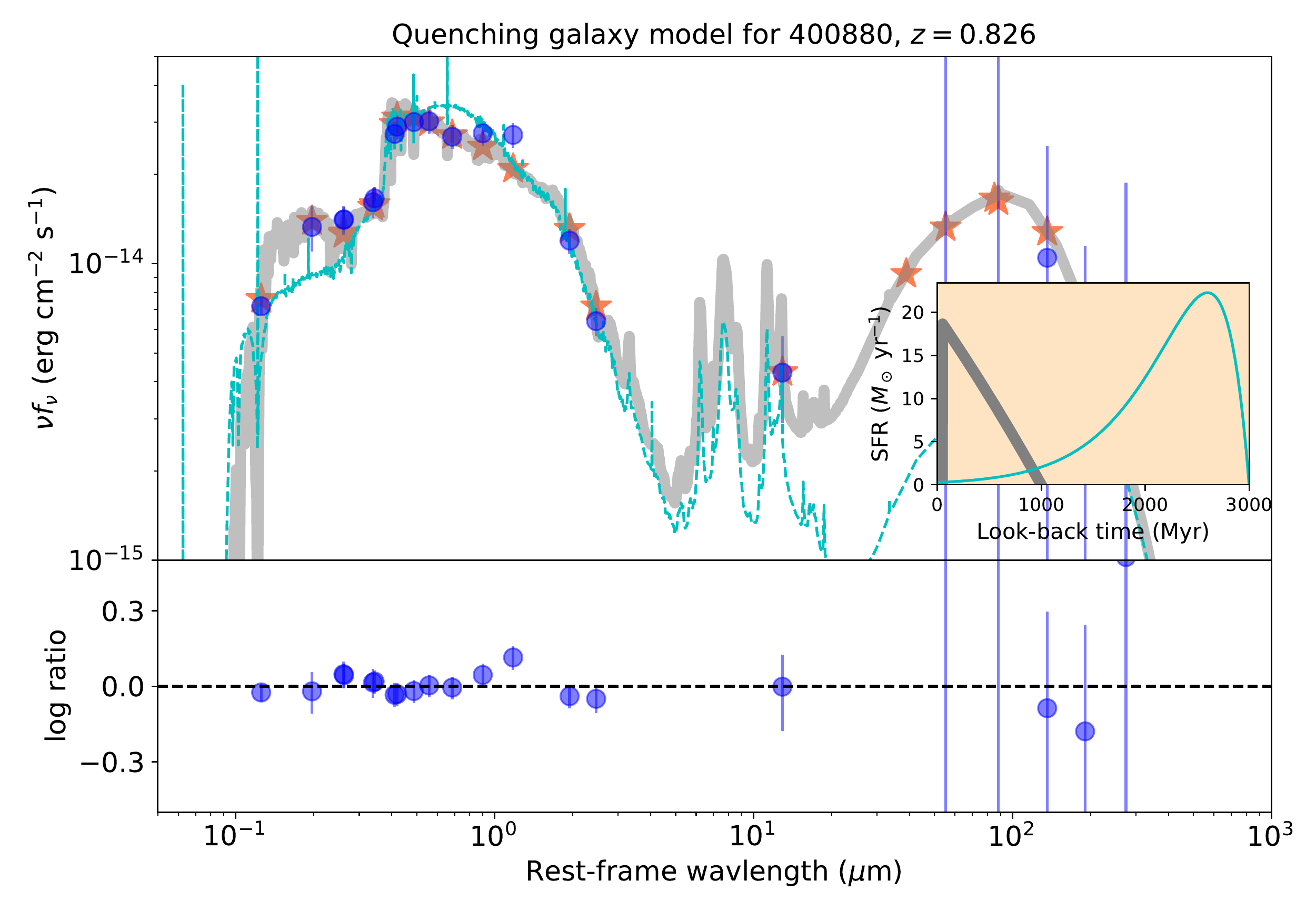}
\includegraphics{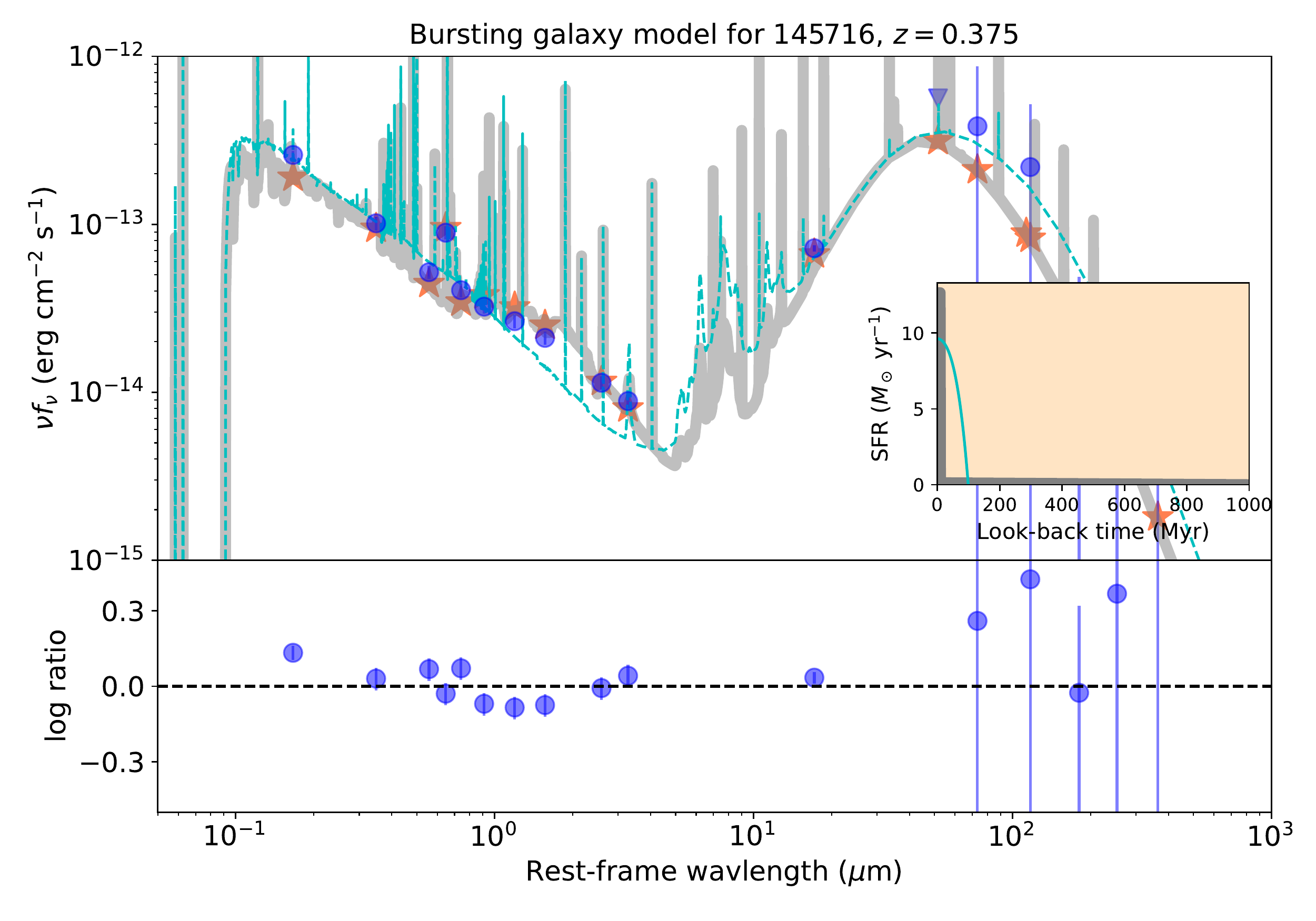}
}
\caption{Example rest-frame SEDs fitted by normal-galaxy (\textit{top left}), AGN (\textit{top right}), quenching-galaxy (\textit{bottom left}), and bursting-galaxy (\textit{bottom right}) templates. The blue points and downward triangles are the observed photometry and upper limits, respectively. The orange stars are the best-fit modeled photometry in the given bands, and the thick grey lines are the best-fit models. The bottom sub-panel of each panel shows the logarithm of the ratios of the observed fluxes over the model fluxes. The best-fit model for the AGN example in the top right panel is decomposed into an AGN component (green) and a galaxy component (red). In the panels other than the top left one, we also show the best-fit models with normal-galaxy templates as cyan dashed lines, and they cannot provide acceptable fits to the data. The abscissa axes of the bottom panels are truncated in the UV to help focus on the difference between the BQ and normal-galaxy models, for which the \mbox{X-ray} data cannot provide useful constraints. The inset plots in the bottom panels compare the best-fit SFHs of the BQ-galaxy model (grey) and normal-galaxy model (cyan). For the quenching galaxy in the bottom-left panel, the normal-galaxy model tends to assign most star formation to an early stage, and thus it cannot explain its blue optical color and red UV color simultaneously. Its quenching SFH indicates that it is generally star-forming before the quenching but the SFR has dropped significantly recently. For the bursting galaxy in the bottom-right panel, the normal-galaxy model predicts that all the stars were formed recently and hence cannot explain the excess NIR emission; in contrast, the bursting model retains low-level star formation before the burst (i.e., the part of the grey SFH that visually overlaps with the abscissa axis), which contributes to the NIR emission. Another feature of this bursting galaxy is that it has a strong H$\alpha$ line dominating its fourth photometric data point (counted from left to right), and we found that the normal-galaxy model cannot fully explain the excess. H$\alpha$ represents the star-formation activity within $\sim10~\mathrm{Myr}$, and this excessive H$\alpha$ feature does support a strong recent starburst (e.g., \citealt{Broussard19}).}
\label{fig_example_sed}
\end{figure*}

We summarize our SED fitting in this subsection. The parameter settings are listed in Tables~\ref{tbl_sedpar_step2_gal}, \ref{tbl_sedpar_step2_agn}, and \ref{tbl_sedpar_step2_bqgal} for normal-galaxy, AGN, and BQ-galaxy templates, respectively. Fig.~\ref{fig_example_sed} presents example SEDs. The AGN-template fitting is only run for MIR AGNs, raw SED AGN candidates, and \mbox{X-ray}-detected sources. The BQ-galaxy fitting is only run for sources with $\Delta\mathrm{BIC_1(BQ)}>2$. Normal-galaxy fitting is run for all sources regardless of their classifications. We set ``best'' results as NaN for stars. For a given non-stellar source, we adopt its ``best'' result following the criteria below:
\begin{itemize}
\item{If it is not a refined SED AGN candidate, \mbox{X-ray} AGN, IR AGN, or a BQ-galaxy candidate, we adopt the result from the normal-galaxy templates.}
\item{If it is a refined SED AGN candidate but not a BQ-galaxy candidate, or it is an \mbox{X-ray}, IR, or reliable SED AGN, we adopt the AGN fitting result.}
\item{If it is a BQ-galaxy candidate but not a refined SED AGN candidate or \mbox{X-ray} or IR AGN, we adopt the BQ-galaxy fitting result.}
\item{In the remaining case, i.e., it is both a BQ-galaxy candidate and a refined SED AGN candidate (Section~\ref{sec: bqagn}), but not an \mbox{X-ray}, IR, or reliable SED AGN, we take the best result as the one with a smaller $\chi_r^2$ between the AGN and BQ-galaxy fitting results.}
\end{itemize}
There are 733743, 19612, and 3624 sources whose ``best'' results are from normal-galaxy, AGN, and BQ-galaxy fits, respectively. We reiterate that the candidates may be significantly contaminated by normal galaxies. Section~\ref{sec: sedagn} shows that most refined SED AGN candidates do not satisfy the calibrated, reliable SED AGN selection. The purity of our candidates is not guaranteed, and appropriate caution should be taken when analyzing them. Especially, when $\Delta\mathrm{BIC}$ is small, different models can hardly be distinguished, and the best category may be unreliable. We thus include the normal-galaxy fitting results for all the sources in our catalog so that users can choose what they need.\par

We note that aside from $M_\star$ and SFR, other physical galaxy parameters generally cannot be reliably measured through our broad-band SED fitting. For example, the inferred detailed SFH and galaxy age often have large biases for our parametric SFH settings (e.g., \citealt{Carnall19}), dust attenuation suffers from internal biases and degeneracies (e.g., \citealt{Qin22}), and exact AGN contributions often cannot be constrained well (Section~\ref{sec: sedagn}). We thus mainly focus on $M_\star$ and SFR, which are often the most important parameters in extragalactic studies, as our primary results.

\section{Analyses of the SED-fitting Results}
\label{sec: sedfitting}
We further investigate the SED-fitting results in Section~\ref{sec: bestsedfittingresults} from various perspectives in this section.

\subsection{Galaxy Colors}
\label{sec: galcolor}
\begin{figure*}
\centering
\resizebox{\hsize}{!}{
\includegraphics{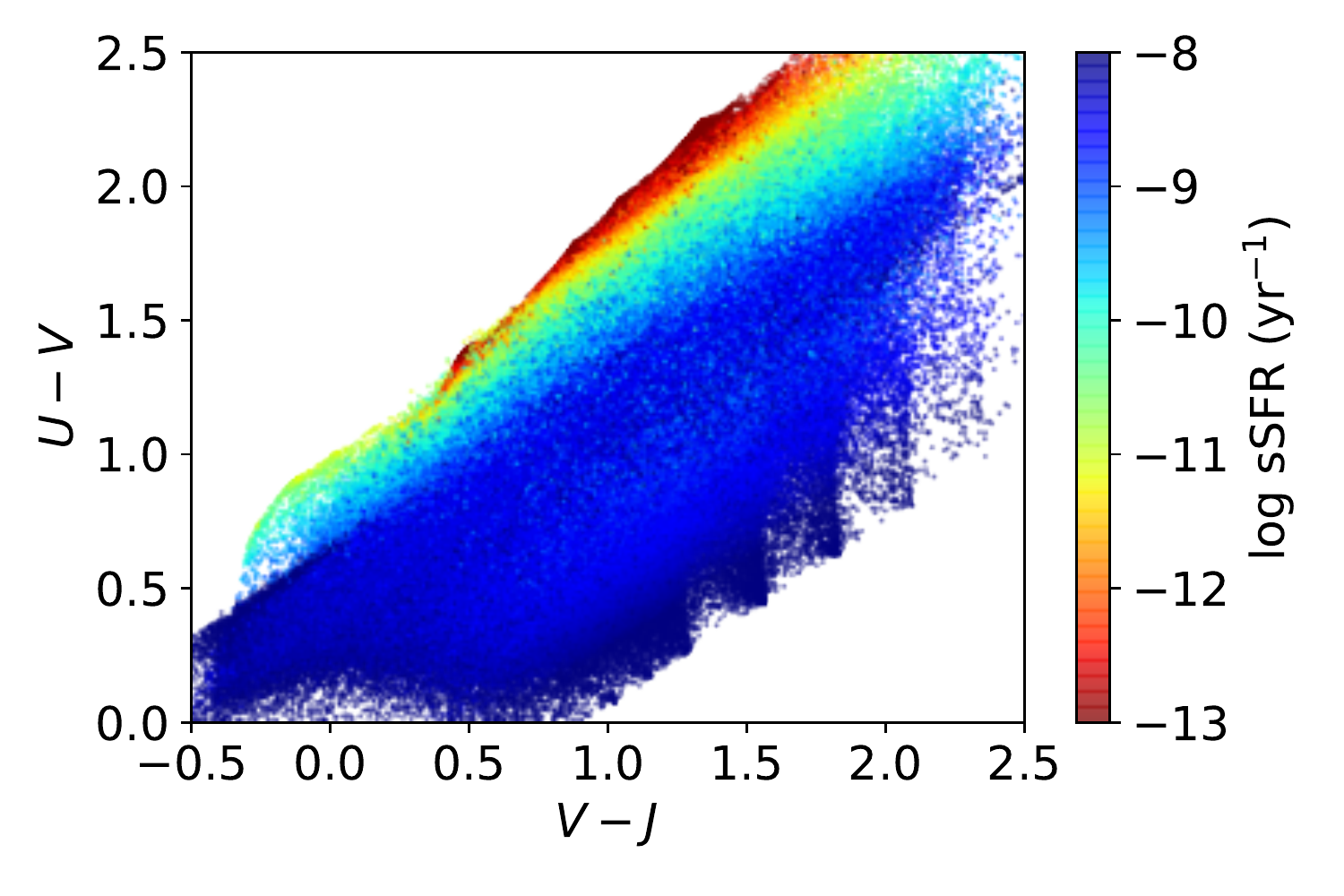}
\includegraphics{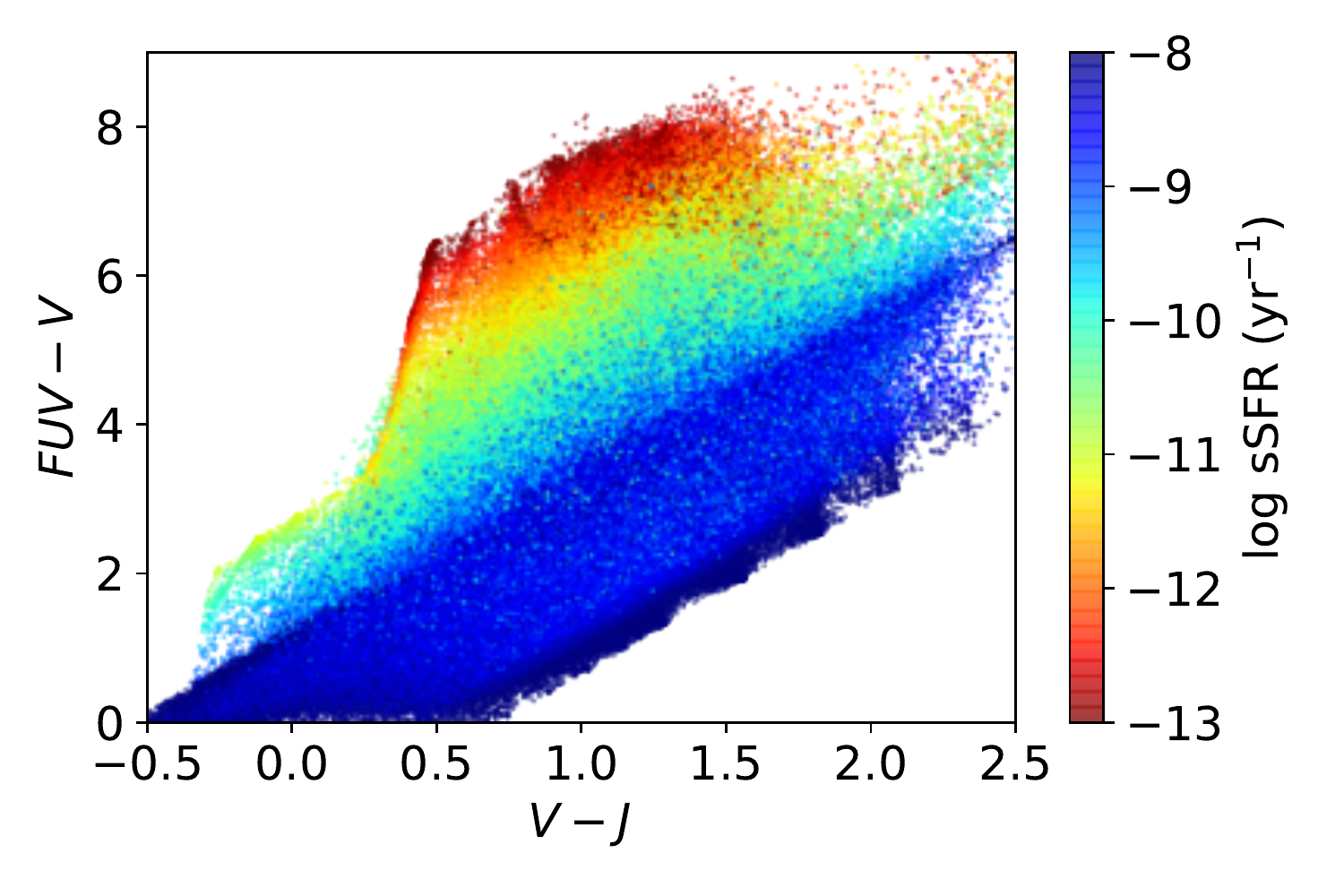}}
\caption{The rest-frame $UVJ$ (\textit{left}) and $FUVVJ$ (\textit{right}) color-color diagrams of our sources, color-coded by their $\log\mathrm{sSFR}$ in $\mathrm{yr^{-1}}$. The $FUVVJ$ diagram has a larger dynamical range to separate quiescent galaxies with different levels of sSFR.}
\label{fig_galcolor}
\end{figure*}

We show the rest-frame $UVJ$ and $FUVVJ$ color-color diagrams in Fig.~\ref{fig_galcolor}, which are color-coded by the specific SFR ($\mathrm{sSFR=SFR}/M_\star$). The traditional $UVJ$ diagram can be used to distinguish quiescent galaxies from star-forming galaxies (e.g., \citealt{Williams09, Muzzin13}) but cannot reliably separate quiescent galaxies with different levels of sSFR. Especially, the $UVJ$ diagram begins to saturate at $\sim10^{-10.5}~\mathrm{yr^{-1}}$ \citep{Leja19c}, below which all the galaxies tend to reside in the same region in the $UVJ$ diagram. In contrast, the $FUVVJ$ diagram, as proposed in \citet{Leja19c}, provides a larger dynamical range to separate efficiently different levels of sSFR and can thus help understand how the quiescent phase evolves during cosmic time. Furthermore, the inclusion of FUV can also help probing more complicated SFHs (e.g., \citealt{Akhshik21}), and we thus provide both $UVJ$ and $FUVVJ$ color information in our released catalog.

\subsection{The $M_\star$-SFR Plane}
\begin{figure}
\centering
\resizebox{\hsize}{!}{\includegraphics{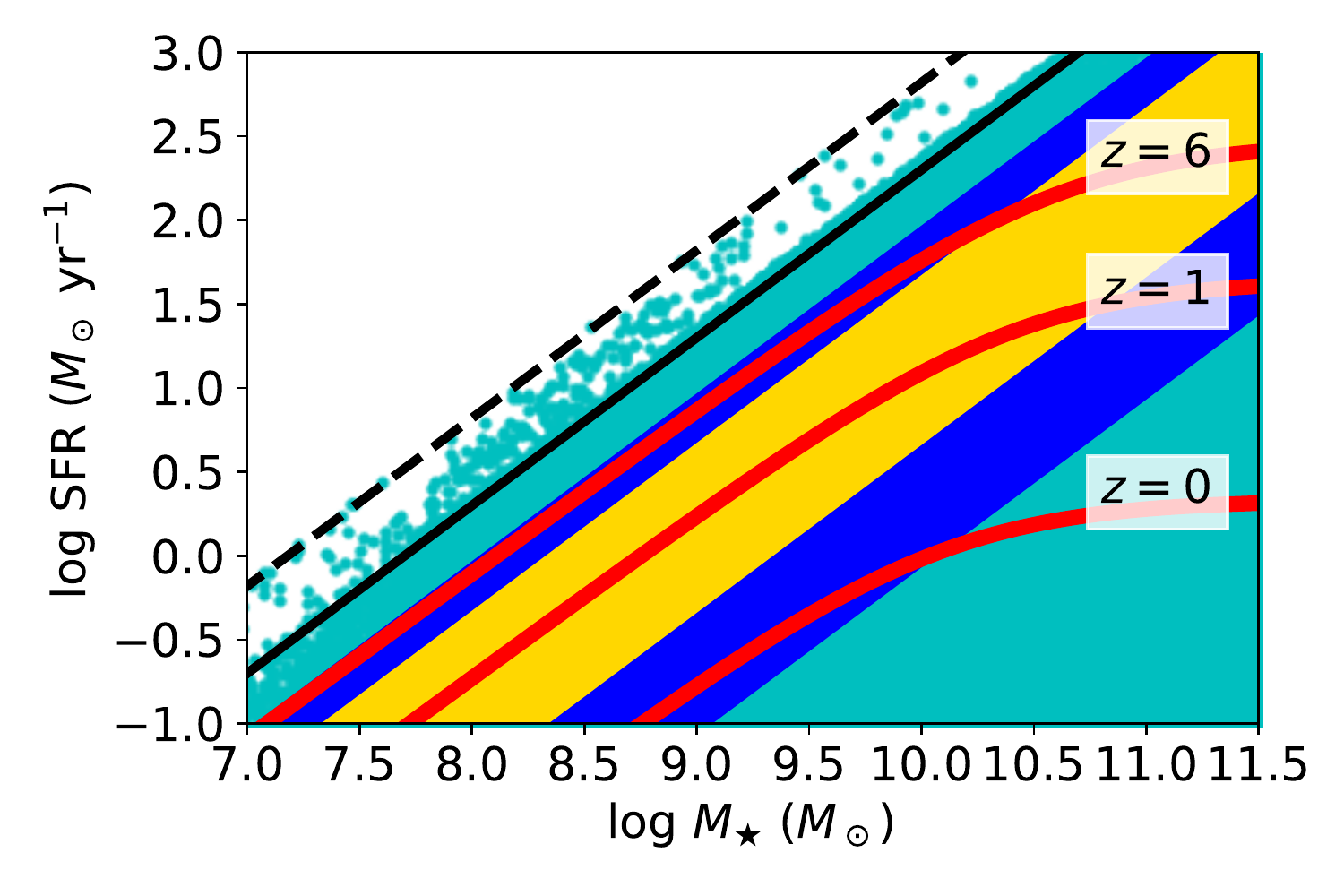}}
\caption{The cyan points are all of our sources in the $M_\star$-SFR plane, which ``pile up'' together and visually form linear cuts as the black lines. The black solid and dashed lines correspond to the maximum sSFR allowed ($10^{-7.70}$ and $10^{-7.18}~\mathrm{yr^{-1}}$) for our normal-galaxy and BQ-galaxy SFHs, respectively. When only plotting the results from the normal-galaxy fits, all the points will be constrained to lie below the black solid line. The yellow and blue bands represent the $16^\mathrm{th}-84^\mathrm{th}$ and $2.5^\mathrm{th}-97.5^\mathrm{th}$ percentile ranges of sSFR for our star-forming galaxies, respectively. The red curves are the MS at $z=0$, 1, and 6. All the bands and MS curves are far below the sSFR limits, and thus the apparent sSFR cut does not cause problems.}
\label{fig_mstar_sfr_pileup}
\end{figure}

When plotting all of our sources together in the $M_\star$-SFR plane, one finds that there is a linear ``cut'', above which there are no points, as shown in Fig.~\ref{fig_mstar_sfr_pileup}. However, this is not indicative of any material problem. The reason for this phenomenon is that there is (inevitably) a maximum sSFR allowed given our SFH settings. There are certainly some sources reaching the sSFR limit, and such points will ``pile up'' and visually form a linear cut when plotting millions of sources without a small point transparency. It can be shown that the maximum sSFR is $10^{-7.70}$ and $10^{-7.18}~\mathrm{yr^{-1}}$ for the normal-galaxy (Table~\ref{tbl_sedpar_step2_gal}) and BQ-galaxy (Table~\ref{tbl_sedpar_step2_bqgal}) settings, respectively. These values are sufficiently high and thus not problematic. To illustrate this, we first select star-forming galaxies based on the criteria in \citet{Lee18} and derive the $16^\mathrm{th}-84^\mathrm{th}$ and $2.5^\mathrm{th}-97.5^\mathrm{th}$ percentile ranges of sSFR. The sSFR ranges are plotted as the yellow and blue bands in Fig.~\ref{fig_mstar_sfr_pileup}, which are both far below the sSFR limits. We also display the MS from \citet{Popesso22} at $z=0$, 1, and 6 in Fig.~\ref{fig_mstar_sfr_pileup}, all of which are safely below the sSFR limits. The MS normalization is known to monotonically increase with $z$, and thus most galaxies are not expected to be above the $z=6$ curve; especially, $z=1$ is roughly the median redshift of our sources, and the corresponding MS is over one~dex below the sSFR limits. The sSFR limit of the BQ-galaxy setting is at least $1-2$~dex higher than the MS and hence is also sufficiently high for starburst galaxies.\par

Nevertheless, \citet{Ciesla17} argued that an exponentially rising SFH might be better than a delayed SFH for star-forming galaxies with $z>2$ because the former allows a much higher sSFR limit (theoretically, able to reach infinity), and high-redshift galaxies tend to have higher sSFR values. We tried that for a smaller random sample of sources spanning $z=0-6$, and the systematic differences of $M_\star$ and SFR are both confined within $\sim0.1$~dex, which further indicates that the sSFR limit from our SFH settings does not cause material biases.\par
\begin{figure}
\centering
\resizebox{\hsize}{!}{\includegraphics{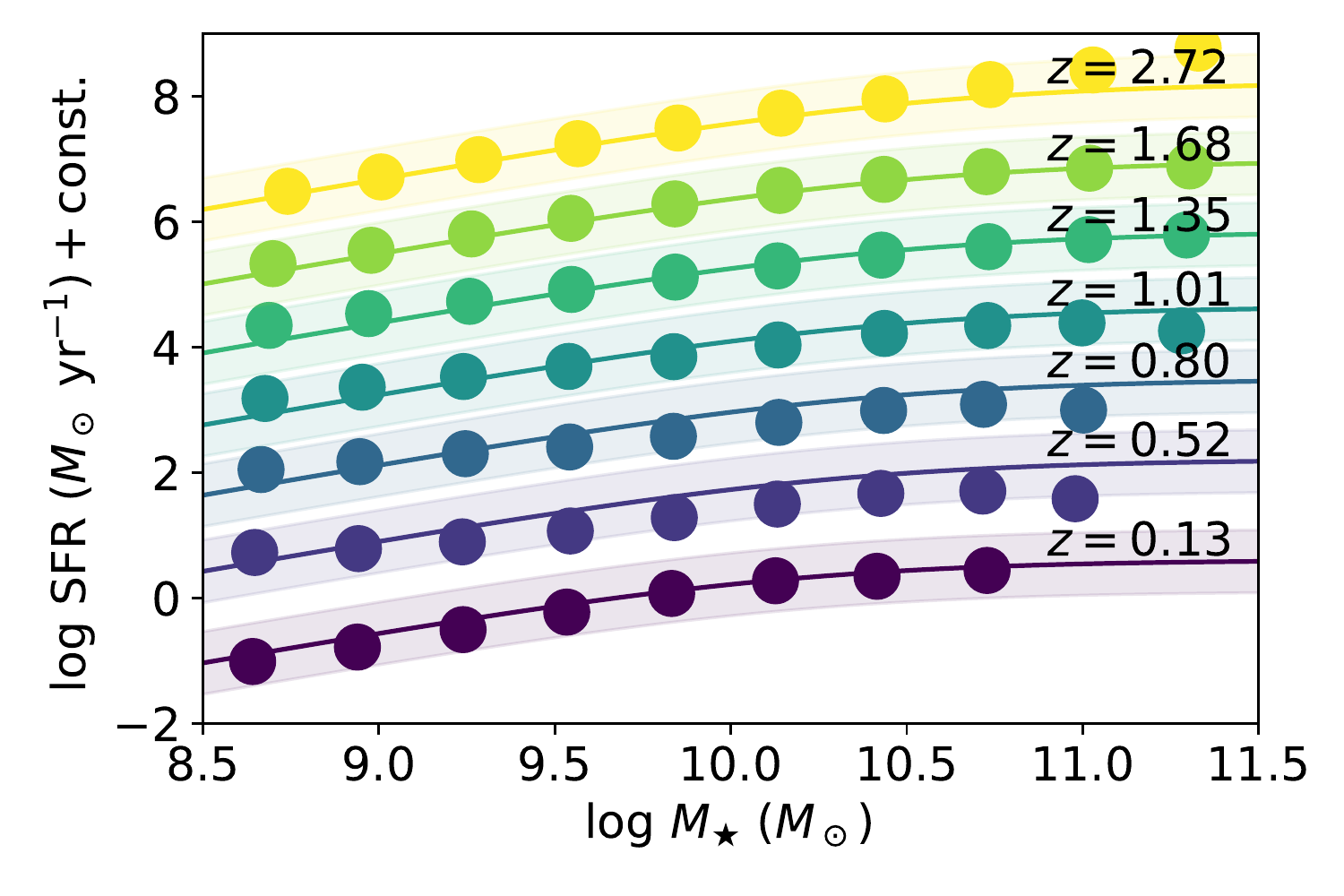}}
\caption{Comparison between our $M_\star$ and SFR results for star-forming galaxies and the MS in \citet{Popesso22} in seven redshift bins. The SFRs at the lowest redshift are not shifted (i.e., const. = 0), and the subsequent SFRs at higher redshifts are progressively shifted upward by one dex for a better visualization. The solid lines are the MS at the median redshift of each redshift bin, as marked explicitly in the figure, and the transparent bands represent $\leq0.5$~dex offset from the MS. The points are the median $M_\star$ and shifted SFR values in several $M_\star$ bins of our star-forming galaxies and are generally consistent with the MS curves.}
\label{fig_mstar_sfr_ms}
\end{figure}

We compare our star-forming galaxies with the MS in Fig.~\ref{fig_mstar_sfr_ms}, where we equally divide the sources into seven redshift bins and plot the MS from \citet{Popesso22} for comparison. They are consistent, even out to the high-redshift bins, further supporting the general reliability of our results.

\subsection{MIR-X-Ray Relations for AGNs}
\label{sec: LxL6um}
\begin{figure}
\centering
\resizebox{\hsize}{!}{\includegraphics{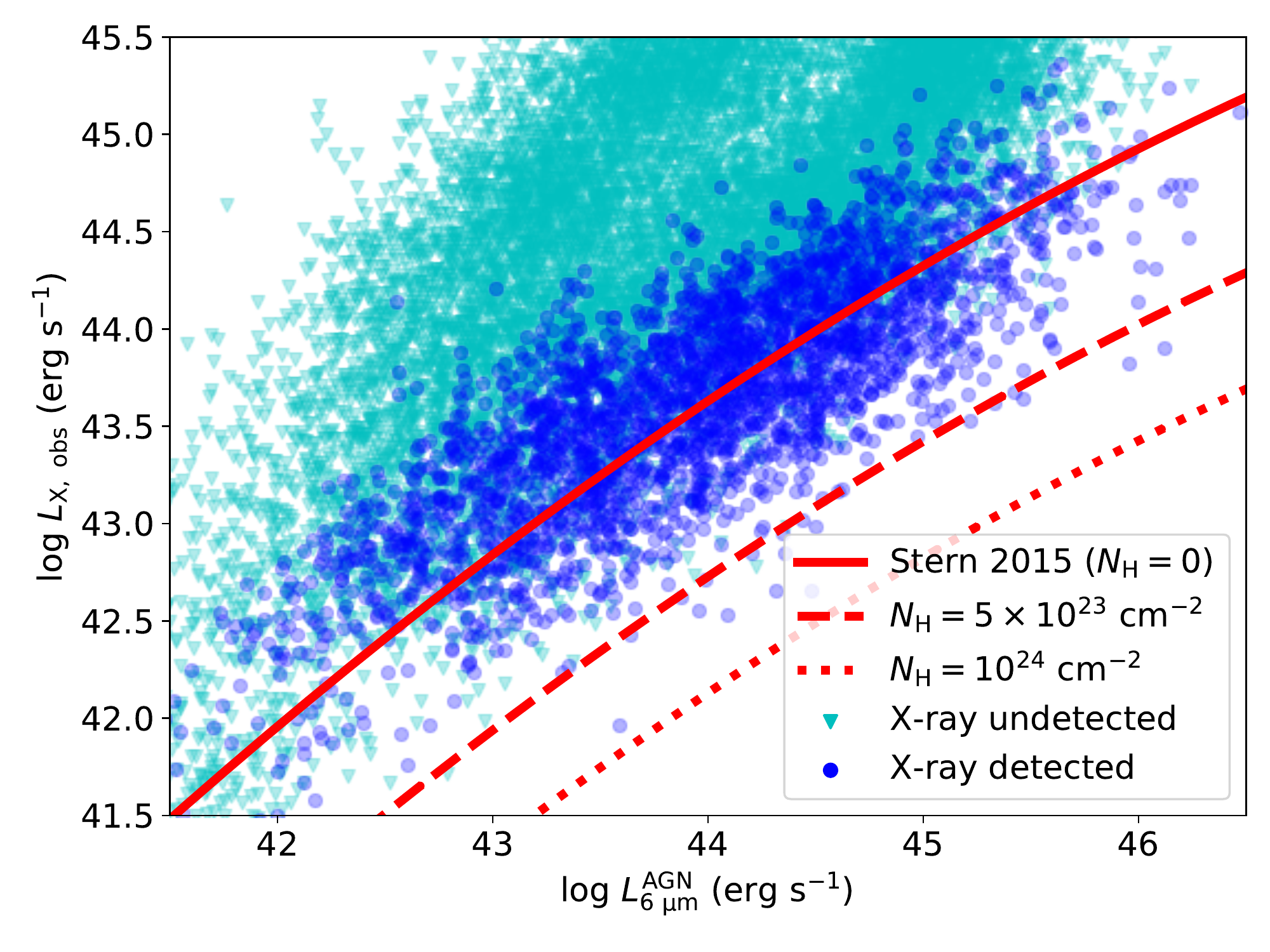}}
\caption{The rest-frame $L_{6~\mu\mathrm{m}}^\mathrm{AGN}-L_\mathrm{X,~obs}$ relation of AGNs and refined SED AGN candidates, where the $L_\mathrm{X,~obs}$ upper limits are adopted for sources undetected in \mbox{X-ray}. Only sources whose best-fit SEDs have non-zero AGN contributions are shown. The standard relation in \citet{Stern15} is also displayed as a comparison.}
\label{fig_MIR-X-ray_relation}
\end{figure}

The AGN rest-frame $6~\mu\mathrm{m}$ luminosity ($L_{6~\mu\mathrm{m}}^\mathrm{AGN}$) is known to be tightly correlated with $L_\mathrm{X}$ (intrinsic $2-10~\mathrm{keV}$ luminosity; e.g., \citealt{Stern15, Chen17}), and we examine this relation for our sources in this section. Here, we use the observed \mbox{X-ray} luminosities ($L_\mathrm{X,~obs}$), instead of $L_\mathrm{X}$, mainly for three reasons -- first, this can help roughly reveal the number of heavily obscured AGNs detected (see next paragraph); second, this can reduce the impact of the internal connections between \mbox{X-ray} and $6~\mu\mathrm{m}$ luminosities arising from the SED fitting because $L_\mathrm{X}$ is directly adopted in the SED fitting to decompose the AGN component; third, for \mbox{X-ray}-detected sources, the typical difference between $L_\mathrm{X,~obs}$ and $L_\mathrm{X}$ ($\sim0.1~\mathrm{dex}$; Fig.~\ref{fig_xraycorr}) is smaller than the intrinsic scatter of the $L_{6~\mu\mathrm{m}}^\mathrm{AGN}-L_\mathrm{X}$ relation ($\sim0.4~\mathrm{dex}$) as well as the systematic differences of the relation among different papers ($\sim0.1-0.2~\mathrm{dex}$). We measure $L_{6~\mu\mathrm{m}}^\mathrm{AGN}$ from the decomposed best-fit SEDs and compare it with $L_\mathrm{X,~obs}$ in Fig.~\ref{fig_MIR-X-ray_relation} for all the AGNs and refined SED AGN candidates (see Section~\ref{sec: select_agn}). $L_\mathrm{X,~obs}$ is derived from the observed fluxes in \citet{Ni21}, assuming a power-law model with a photon index of $\Gamma_\mathrm{eff}$. Sources detected in both the SB and the HB have $\Gamma_\mathrm{eff}$ estimations in \citet{Ni21}. $\Gamma_\mathrm{eff}$ is generally chosen to be 1.9 for sources detected in the SB but undetected in the HB, 0.6 for those detected in the HB but undetected in the SB, and 1.4 for those only detected in the FB, but there are exceptions. We refer readers to Section~3.5 of \citet{Ni21} for more details about the choice of $\Gamma_\mathrm{eff}$. We derive $L_\mathrm{X,~obs}$ upper limits for \mbox{X-ray}-undetected sources using the HB flux upper-limit map in Fig.~\ref{fig_fupp_HB} and Eq.~\ref{eq_def_fx}, where $\eta$ and $\Gamma$ in Eq.~\ref{eq_def_fx} are set to 1 and 1.4,\footnote{$\Gamma_\mathrm{eff}=1.4$ is the typical power-law index of the cosmic \mbox{X-ray} background (e.g., \citealt{Marshall80})} respectively. Our sources agree well with the relation in the literature, indicating that the SED decompositions are generally reliable.\par
When the source obscuration is high, the $L_\mathrm{X,~obs}$ value will be suppressed, and thus a large downward deviation from the $L_{6~\mu\mathrm{m}}^\mathrm{AGN}-L_\mathrm{X}$ relation may indicate a high obscuration level. We derive the suppression factor of $L_\mathrm{X,~obs}$ using the zeroth-order edge-on spectrum with a photon index of 1.8 in \texttt{MYTorus} \citep{Murphy09}, and the suppressed relations of \citet{Stern15} corresponding to $N_\mathrm{H}=5\times10^{23}~\mathrm{cm^{-2}}$ and $10^{24}~\mathrm{cm^{-2}}$ are also shown in Fig.~\ref{fig_MIR-X-ray_relation}. Note that the $L_{6~\mu\mathrm{m}}^\mathrm{AGN}-L_\mathrm{X}$ relation itself has a scatter of $\sigma\approx0.4~\mathrm{dex}$, and the suppressed curve at $N_\mathrm{H}=5\times10^{23}~\mathrm{cm^{-2}}$ roughly corresponds to the downward $2\sigma$ boundary of the $L_{6~\mu\mathrm{m}}^\mathrm{AGN}-L_\mathrm{X}$ relation. Therefore, it would be unreliable to identify obscured sources with $N_\mathrm{H}\lesssim5\times10^{23}~\mathrm{cm^{-2}}$ using the $L_{6~\mu\mathrm{m}}^\mathrm{AGN}-L_\mathrm{X}$ relation. There are few sources ($\lesssim20$) whose $L_\mathrm{X,~obs}$ values or upper limits are below the $N_\mathrm{H}=5\times10^{23}~\mathrm{cm^{-2}}$ curve, and no obvious sources are below the $N_\mathrm{H}=10^{24}~\mathrm{cm^{-2}}$ curve. Detailed \mbox{X-ray} spectral analyses are needed to reliably measure their $N_\mathrm{H}$ values and other \mbox{X-ray} spectral features (e.g., Fe K$\alpha$ lines) that are prevalent among heavily obscured AGNs. More detailed selection and analyses of heavily obscured and CT AGNs will be presented in Yan et al. (in preparation).\par

There are two caveats worth noting. First, Fig.~\ref{fig_MIR-X-ray_relation} may be biased against CT AGNs. CT AGNs have very hard \mbox{X-ray} spectra, but the typical power-law index adopted to calculate $L_\mathrm{X,~obs}$ for these sources is 0.6 \citep{Ni21}. This value may be too soft for CT AGNs, and thus may lead to overestimations of $L_\mathrm{X,~obs}$. Solving this issue requires \mbox{X-ray} spectral fitting, and Yan et al. (in preparation) will correct this bias. Secondly, there are inevitable connections between AGN \mbox{X-ray} and $6~\mu\mathrm{m}$ luminosities resulting from the SED fitting. Nevertheless, the effect upon Fig.~\ref{fig_MIR-X-ray_relation}, compared to other luminosity-luminosity relations, has been minimized because the MIR-\mbox{X-ray} relation is only secondary \citep{Yang20, Brandt21} and $L_\mathrm{X,~obs}$, instead of $L_X$ (which is directly utilized in the SED fitting), is used.

\subsection{Host galaxies of quasars}
\label{sec: qsohost}
\begin{figure}
\centering
\resizebox{\hsize}{!}{\includegraphics{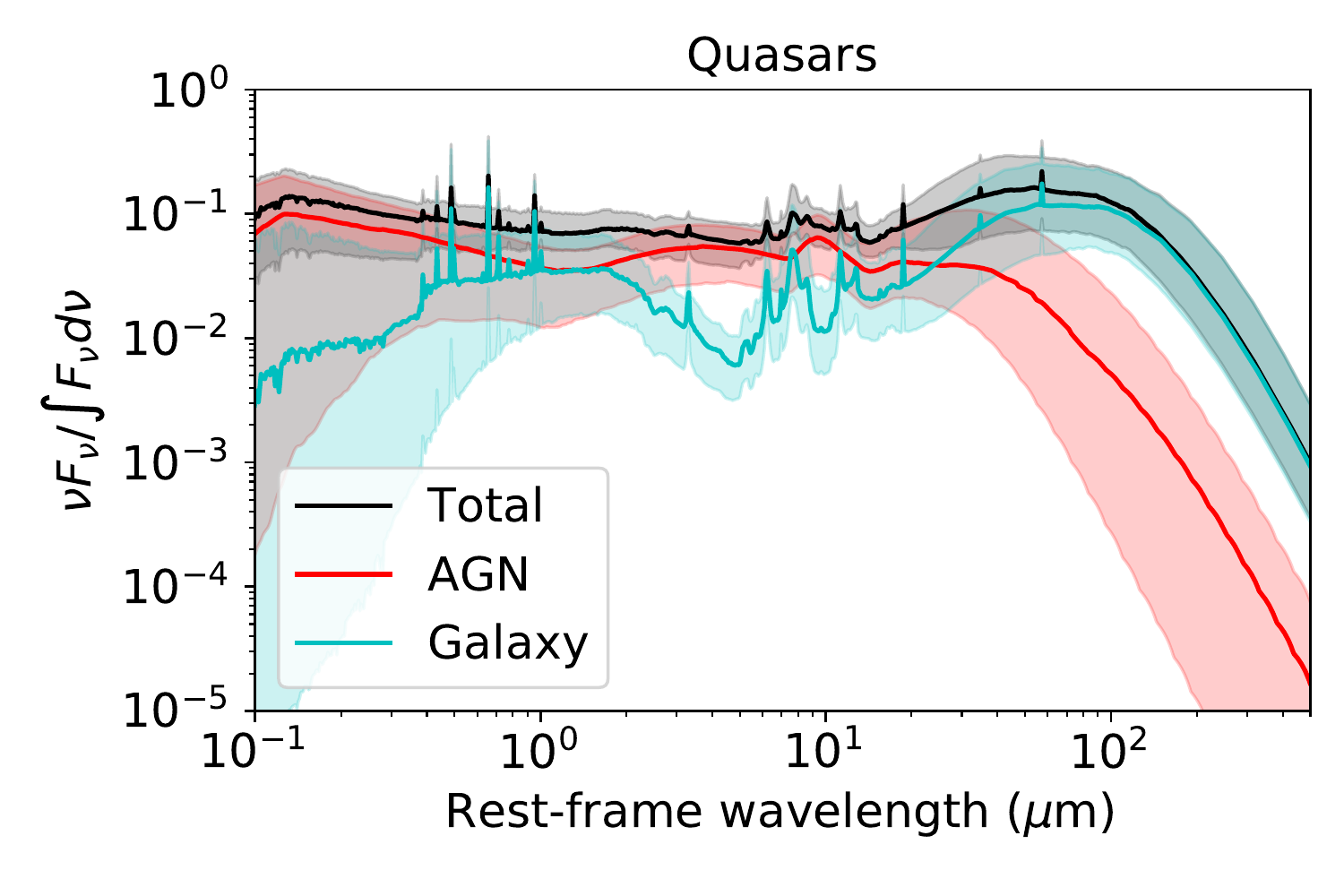}}
\caption{The typical decomposed SED of quasars, constructed from 424 BL quasars in \mbox{W-CDF-S}. This SED is plotted in the same way as for Fig.~\ref{fig_stacksed}, where the cyan and red solid lines represent the galaxy and AGN components, respectively.}
\label{fig_sed_qso_wcdfs}
\end{figure}

Our \mbox{W-CDF-S} and XMM-LSS fields overlap with those of the SDSS-V Black Hole Mapper project \citep{Kollmeier17}, where optically luminous quasars will be studied in detail via reverberation mapping. However, our $M_\star$ measurements cannot be safely utilized for such quasars. To illustrate this, we show the typical SED of quasar-like reliable broad-line (BL) AGNs (i.e., \texttt{SED\_BLAGN\_FLAG = 1} in \citealt{Ni21}) in \mbox{W-CDF-S} in Fig.~\ref{fig_sed_qso_wcdfs}. Its AGN component generally dominates the emission from UV to MIR. Particularly, the NIR emission, which is important for measuring $M_\star$ and usually dominated by starlight for the general AGN population, is also significantly contaminated by the AGN emission for these quasars. The galaxy emission still generally dominates in the FIR, and thus SFRs can be reliably estimated for quasars detected in the FIR (see Section~\ref{sec: compare_results} for measurements of FIR-based SFRs), and 15\% of them are detected in the FIR.\par

Decomposing the optical-to-IR host-galaxy components for quasars and measuring their $M_\star$ often requires specialized methods, such as imaging decomposition (e.g., \citealt{Yue18, Li21}), which are beyond the scope of this work.

\subsection{Sources Selected as Both Refined SED AGN Candidates and BQ-galaxy Candidates}
\label{sec: bqagn}
2159 sources are selected as both refined SED AGN candidates and BQ-galaxy candidates, which constitutes over one-third of BQ-galaxy candidates. This mainly originates from their enhanced rest-frame UV emission. When matching the optical-to-NIR SEDs, the truncated delayed SFH may lead to larger UV emission than the normal delayed SFH, and type~1 AGNs can also increase the UV emission relative to the optical due to their blue UV-to-optical colors. Therefore, enhanced UV emission may be either explained by a truncated delayed SFH or a type~1 AGN, but it is hard to distinguish which one is correct without further information. Indeed, the $\chi_r^2$ distributions from BQ-galaxy-model fitting and AGN-model fitting are similar for sources selected as both refined SED AGN candidates and BQ-galaxy candidates. This difficulty is also known among (hot) dust-obscured galaxies, whose UV emission sometimes shows an excess compared to the optical (e.g., \citealt{Assef16, Assef19}). This excess can be explained by both ``leaked'' AGN emission (e.g., broad UV emission lines; \citealt{Zou20}) and unusual star-formation, and SEDs alone cannot reliably determine its origin.\par
To break the degeneracy, we usually need other indicators that can independently and firmly classify such sources into one category, including \mbox{X-ray} and MIR information that we are using to classify AGNs in this work. Optical-to-NIR spectra should also be valuable as they can provide direct diagnostics for both AGNs and BQ galaxies, including locations in Baldwin-Phillips-Terlevich diagrams \citep{Baldwin81} and age-sensitive Balmer absorption lines. However, simultaneous co-analyses of both photometry and spectra are not directly supported in \texttt{CIGALE}, although some efforts have been devoted to including spectral information in SED fitting in \texttt{CIGALE} (e.g., \citealt{Boselli16, Villa-Velez21}).\par
There may be some AGNs whose host galaxies are undergoing rapid (within several hundreds of millions of years) changes in SFR. For example, \citet{Alatalo17} showed that the enhanced MIR emission of post-starburst galaxies may indicate the prevelance of (mainly low-luminosity) AGNs, and \citet{Greene20} showed that their intermediate-redshift massive post-starburst galaxies are much more likely to host AGNs than quiescent galaxies. However, it is generally challenging and may have a danger of over-interpretation to select such sources solely based on SEDs, and thus we do not try to identify them in this work. Nevertheless, if type~2 AGNs can be selected by other methods (e.g., through optical spectra), it is still possible to safely characterize their recent SFHs \citep{Smethurst16}, which can then be used to probe the connection between type~2 AGN activity and rapid quenching. This will be left to future work.\par
Given the aforementioned challenges, we adopt our best results for these sources as the AGN-fitting ones if they are \mbox{X-ray}, IR, or reliable SED AGNs, or their AGN-fitting $\chi_r^2$ values are smaller than their BQ-galaxy-fitting values; otherwise, their best results are set to be the BQ-galaxy-fitting ones. Sources selected as both refined SED AGN and BQ-galaxy candidates can be identified by requiring $\texttt{detBIC2\_agn}>2~\mathrm{AND}~\texttt{detBIC2\_bqgal}>2~\mathrm{AND}~\texttt{flag\_star}==0$ in our catalog (see Section~\ref{sec: release}).

\subsection{The X-Ray Data Points}
\label{sec: xraydatapoint}
\mbox{X-ray} data are important for AGNs because they can directly constrain the AGN emission. However, the statistical contributions of the \mbox{X-ray} data to SED fitting may be ``diluted'' by dozens of longer-wavelength bands, and a direct consequence is that not all (though most) \mbox{X-ray} AGNs are selected as SED AGN candidates. One direct way to overcome this issue is via weighting the \mbox{X-ray} data points. This is analogous to simultaneous SED fitting for both photometry and spectra, where the contributions of the photometric data and spectroscopic data should be separated to prevent the significant statistical dilution from a large number of spectroscopic data points to the photometry, whose total number is usually much more limited (e.g., \citealt{Chilingarian12, LopezFernandez16, Thomas17}).\par
\begin{figure*}
\centering
\resizebox{\hsize}{!}{\includegraphics{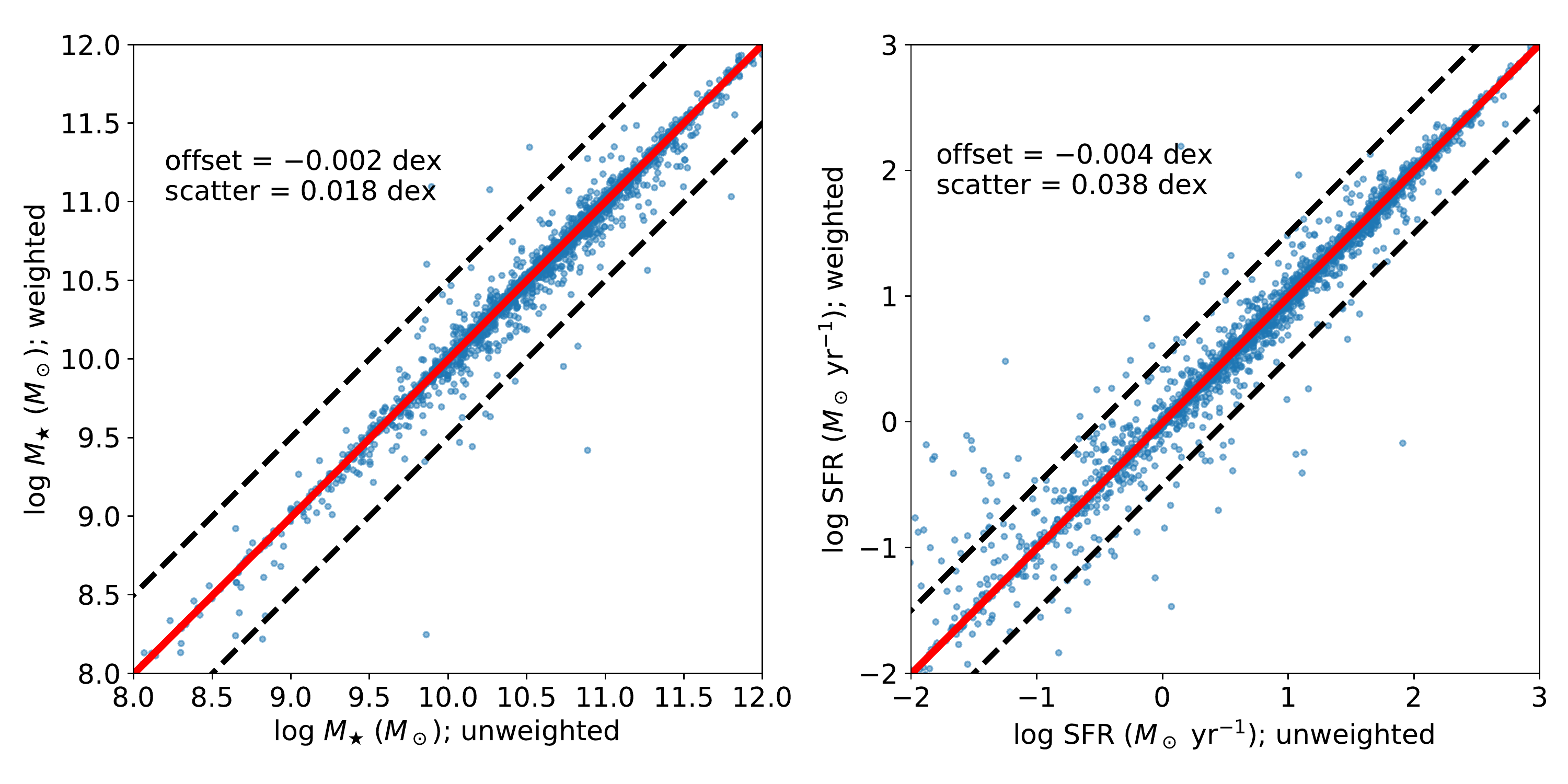}}
\caption{Comparison of $M_\star$ (\textit{left}) and SFR (\textit{right}) between the SED fitting with weighted and unweighted \mbox{X-ray} data for \mbox{X-ray} AGNs. The black dashed lines represent 0.5~dex offsets from the one-to-one relationships, and the red lines, which visually overlap with the one-to-one lines, are LOESS curves of the points. The inferences about host-galaxy properties are generally not affected by the weighting.}
\label{fig_comp_mstar_sfr_xeq}
\end{figure*}

Generally, there are no guidelines to choose the weight, and here we try a weighting that makes the contribution of the \mbox{X-ray} photometry roughly equal to that from all the other bands. We set the \mbox{X-ray} flux errors to be
\begin{align}
\sigma_\mathrm{X}=\frac{f_\mathrm{X}}{\sqrt{\sum_{i\in\mathrm{\{UV\ to\ FIR\}}}\left(\frac{f_i}{\sigma_i}\right)^2}},
\end{align}
where $f$ and $\sigma$ are the flux and uncertainty, respectively. This analysis is only applied to \mbox{X-ray} AGNs because the equation above cannot be applied to undetected ones. We then re-do the SED fitting and compare the results with the unweighted ones. The weighting causes $\Delta\mathrm{BIC}$ to lose its statistical meaning, and nearly all the sources satisfy $\Delta\mathrm{BIC}>2$ or even more stringent criteria, as expected. We thus only focus on the resulting $M_\star$ and SFRs. The comparisons of $M_\star$ and SFRs for \mbox{X-ray} AGNs are displayed in Fig.~\ref{fig_comp_mstar_sfr_xeq}. There are almost no systematic differences, and the scatters are also small, and thus we conclude that the weighting generally does not affect the $M_\star$ and SFR measurements. Note that, throughout this paper, we still adopt the results based on the original, unweighted \mbox{X-ray} data.\par

The analyses above indicate that the \mbox{X-ray} data may not significantly influence the $M_\star$ and SFR measurements, even for \mbox{X-ray} AGNs. We conduct an additional test of removing the \mbox{X-ray} data from our SED fitting. In this way, we can also examine the results for \mbox{X-ray}-undetected refined SED AGN candidates. The comparisons are similar to those in Fig.~\ref{fig_comp_mstar_sfr_xeq}, i.e., the difference is small -- for \mbox{X-ray} AGNs, the offset of $M_\star$ (SFR) is 0.002 (0.001) dex, and the scatter is 0.008 (0.017) dex; for \mbox{X-ray}-undetected refined SED AGN candidates, the offset of $M_\star$ (SFR) is 0.002 ($-0.001$) dex, and the scatter is 0.006 (0.010) dex.\par
\begin{figure}
\centering
\resizebox{\hsize}{!}{\includegraphics{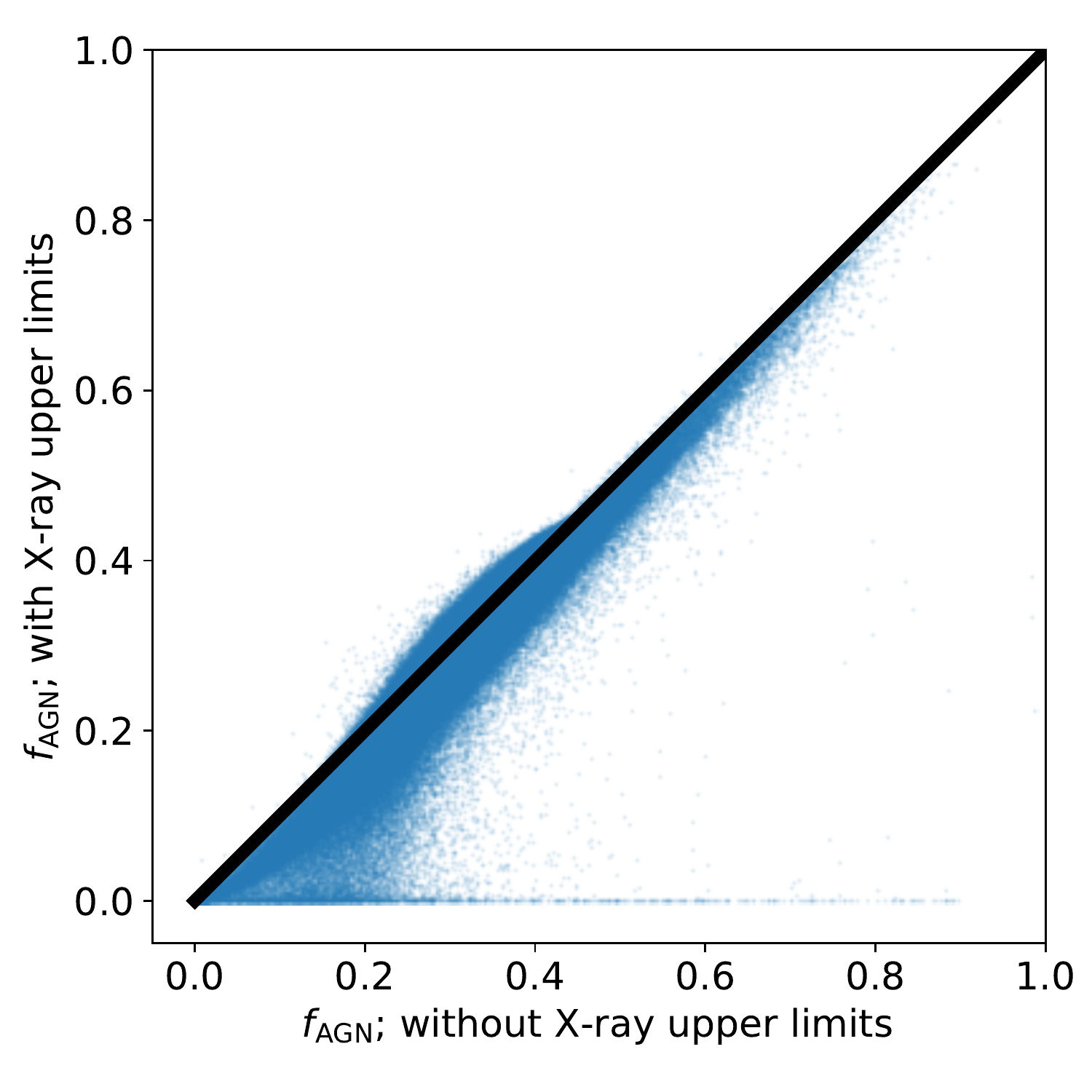}}
\caption{Comparison of $f_\mathrm{AGN}$ between the SED fitting with and without \mbox{X-ray} upper limits for all the \mbox{X-ray}-undetected sources. The black line is the one-to-one relationship. \mbox{X-ray} upper limits can help constrain the decomposed AGN power to a lower level.}
\label{fig_compare_fagn_xray}
\end{figure}

However, this does not mean that the \mbox{X-ray} data are useless in this respect. \mbox{X-rays}, including \mbox{X-ray} upper limits, are mainly used to constrain the AGN component in SEDs. This has been thoroughly discussed in previous works (e.g., \citealt{Yang20, Mountrichas20}). We only briefly present one test here. We remove the \mbox{X-ray} upper limits for all the \mbox{X-ray}-undetected sources and re-fit their SEDs using the AGN parameter settings in Table~\ref{tbl_sedpar_step1} (but without the \mbox{X-ray} module). The simpler settings in Table~\ref{tbl_sedpar_step1}, instead of Table~\ref{tbl_sedpar_step2_agn}, are adopted to reduce the computational requirements. We compare $f_\mathrm{AGN}$ with and without \mbox{X-ray} upper limits for all the \mbox{X-ray}-undetected sources in Fig.~\ref{fig_compare_fagn_xray}, which demonstrates that \mbox{X-ray} upper limits can reduce $f_\mathrm{AGN}$ because models with strong AGN emission may violate the \mbox{X-ray} upper-limit constraint. Note that most sources in Fig.~\ref{fig_compare_fagn_xray} are galaxies and should have $f_\mathrm{AGN}=0$. The SED fitting hence systematically overestimates $f_\mathrm{AGN}$. Such an overestimation is larger without \mbox{X-ray} upper limits and is expected to be much smaller with deeper \mbox{X-ray} coverage (e.g., from Athena or AXIS in the future; \citealt{Yang20}).

\subsection{Comparison with Other $M_\star$ and SFR Measurements}
\label{sec: compare_results}
To assess the reliability of our results, we compare them with other measurements in this section. All the numerical values are summarized in Table~\ref{tbl_compare_results}, and more details are illustrated in the following text.\par

\begin{table*}
\caption{Numerical comparison results between our $M_\star$ and SFR measurements with others}
\label{tbl_compare_results}
\centering
\begin{threeparttable}
\begin{tabular}{cccccccc}
\hline
\hline
& & \multirow{2}{*}{HELP} & \multirow{2}{*}{\texttt{Prospector-$\alpha$}} & \multirow{2}{*}{\citet{Guo19}} & \multirow{2}{*}{$\mathrm{SFR_{FIR}}$} & \texttt{Prospector-$\alpha$} & $\mathrm{SFR_{FIR}}$\\
& & & & & & (corrected) & (corrected)\\
\hline
\multirow{5}{*}{$\log M_\star$} & galaxy offset & 0.02 & $-0.12$ & 0.11 & ----- & $-0.02$ & -----\\
& galaxy NMAD & 0.22 & 0.22 & 0.15 & ----- & 0.15 & -----\\
& AGN offset & $-0.10$ & $-0.14$ & 0.10 & ----- & $-0.13$ & -----\\
& AGN NMAD & 0.24 & 0.20 & 0.19 & ----- & 0.16 & -----\\
& relevant figure & \ref{fig_comp_wcdfs_master} & \ref{fig_comp_wcdfs_master} & \ref{fig_comp_wcdfs_master} & ----- & \ref{fig_comp_correction_Leja} & -----\\
\hline
\multirow{5}{*}{log SFR} & galaxy offset & $-0.15$ & 0.20 & $-0.01$ & $-0.33$ & $-0.02$ & 0.00\\
& galaxy NMAD & 0.24 & 0.37 & 0.49 & 0.41 & 0.28 & 0.20\\
& AGN offset & $-0.17$ & 0.01 & 0.20 & $-0.29$ & $-0.26$ & 0.02\\
& AGN NMAD & 0.25 & 0.37 & 0.67 & 0.38 & 0.30 & 0.22\\
& relevant figure & \ref{fig_comp_wcdfs_master} & \ref{fig_comp_wcdfs_master} & \ref{fig_comp_wcdfs_master} & \ref{fig_comp_sfr_fir} & \ref{fig_comp_correction_Leja} & \ref{fig_comp_sfr_fir}\\
\hline
\hline
\end{tabular}
\begin{tablenotes}
\item
\textit{Notes.} The median and NMAD values of the differences (defined as our values minus the reference ones) between our $M_\star$ and (SED-based) SFR measurements and others, as clarified in the column heads. The values are in dex. ``\texttt{Prospector-$\alpha$} (corrected)'' means the comparison between our corrected values based on Eqs.~\ref{eq: corr_mstar_Joel} and \ref{eq: corr_sfr_Joel} and the \texttt{Prospector-$\alpha$} ones. ``$\mathrm{SFR_{FIR}}$ (corrected)'' represents the comparison between our corrected FIR-based SFRs based on Eq.~\ref{eq: correct_fir_sfr} and SED-based SFRs. Note that these values are global, and more subtle trends of the difference are plotted in the relevant figures listed in the table.
\end{tablenotes}
\end{threeparttable}
\end{table*}

We first compare our results with those from the HELP project across the whole \mbox{W-CDF-S} field, \texttt{Prospector-$\alpha$} results in the small 3D-HST \mbox{CDF-S} field \citep{Leja19b, Leja20, Leja21}, and \citet{Guo19} for Chandra sources in the smaller \mbox{CDF-S}. SED measurements in HELP are mainly limited to bright sources (see \citealt{Shirley21} for more details), and we provide SED-fitting results for $10-100$ times more sources than HELP in our fields. The \texttt{Prospector-$\alpha$} results and those in \citet{Guo19} are expected to be better than our results because their multi-wavelength data are deeper in the small \mbox{CDF-S} region.\footnote{Recall that \mbox{CDF-S} constitutes only 3\% of the whole \mbox{W-CDF-S} field.} More importantly, \texttt{Prospector-$\alpha$} enables highly flexible SED fitting with nonparametric SFHs, and millions of CPU hours were devoted to the SED fitting in 3D-HST to overcome the bottleneck that a systematic factor-of-two uncertainty generally exists among different SED-fitting results \citep{Leja19b}. We thus regard the \texttt{Prospector-$\alpha$} results as the ``ground truth'', at least for non-AGN galaxies. The \texttt{Prospector-$\alpha$} results for AGNs are not necessarily more reliable than our \texttt{CIGALE} results because \texttt{CIGALE} has more advanced AGN templates that have been extensively explored (e.g., \citealt{Yang20, Yang22, Mountrichas21, Mountrichas20, Buat21, Padilla21}) and can directly utilize the \mbox{X-ray} data. Therefore, the comparisons with \texttt{Prospector-$\alpha$} results and \citet{Guo19} are mainly for galaxies and AGNs, respectively.\par
\begin{figure*}
\centering
\resizebox{\hsize}{!}{
\includegraphics{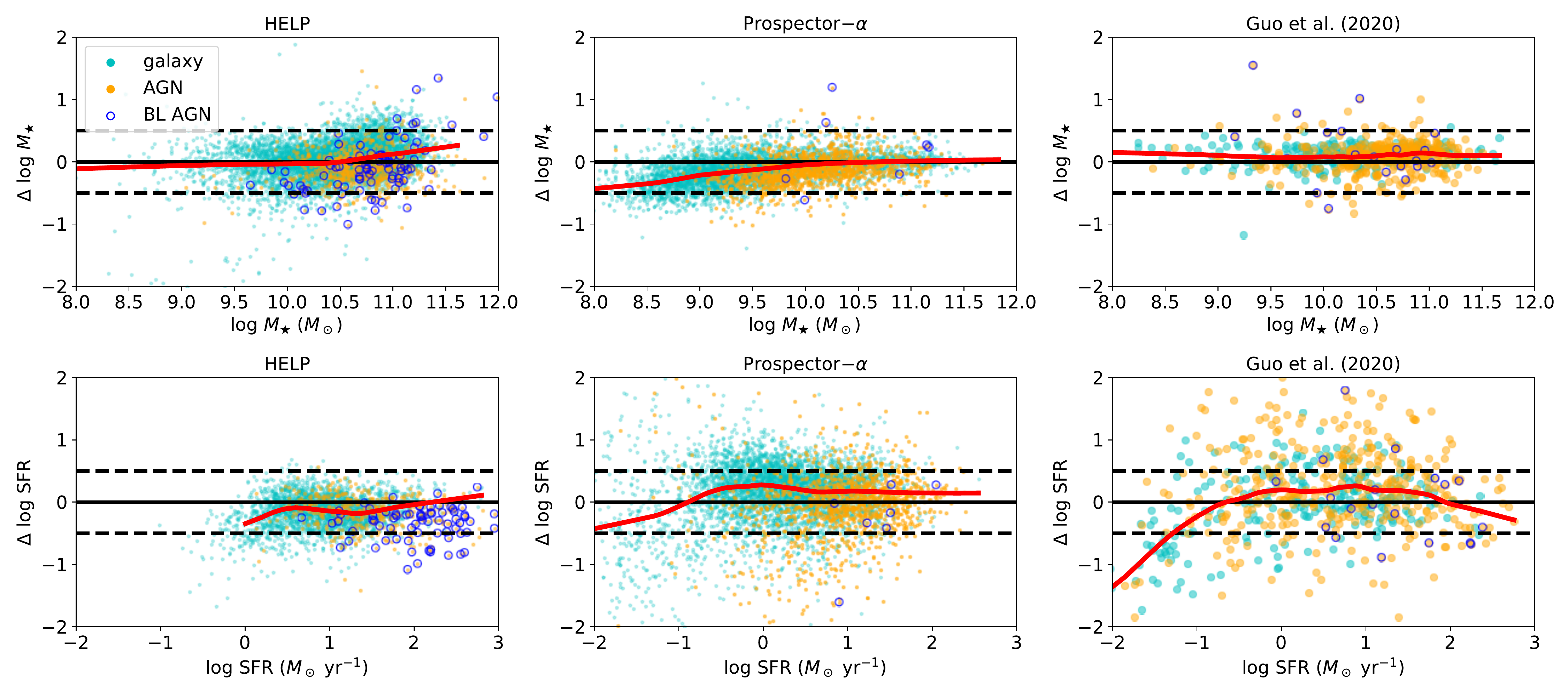}
}
\caption{Comparisons of our $M_\star$ (\textit{top}) and SFR (\textit{bottom}) measurements with the HELP values (\textit{left}), the \texttt{Prospector-$\alpha$} values (\textit{middle}), and those in \citet{Guo19} (\textit{right}). The black solid lines are zero-difference relationships, and the black dashed lines represent 0.5~dex offsets. The red lines are LOESS curves for galaxies in the left and middle panels and all the sources in the right panel. The numerical comparison results are displayed in Table~\ref{tbl_compare_results}. Our measurements are generally consistent with others, though a small mass-dependent offset exists between ours and the \texttt{Prospector-$\alpha$} values.}
\label{fig_comp_wcdfs_master}
\end{figure*}

We show the comparisons of $M_\star$ in the top panels of Fig.~\ref{fig_comp_wcdfs_master}, in which we explicitly mark AGNs and BL AGNs, where AGNs are defined as \mbox{X-ray}, IR, or reliable SED AGNs, and BL AGNs are compiled in \citet{Ni21}. Only sources with consistent redshifts between the compared catalog and ours are shown, i.e., $|\Delta z|/(1+z)<0.15$. Denoting $X_1$ and $X_2$ as our measurements and the comparison ones, respectively, the abscissa axes in Fig.~\ref{fig_comp_wcdfs_master} are defined as $(X_1+X_2)/2$, and the ordinates are their difference, $X_1-X_2$. The choice of adopting $(X_1+X_2)/2$, instead of $X_1$ or $X_2$, as the abscissa axes can be easily explained as follows. Since $\{X_1\}$ and $\{X_2\}$ roughly span the same range, $\mathrm{Var}(\{X_1\})\approx\mathrm{Var}(\{X_2\})$.\footnote{Curly braces are used outside $X_1$ and $X_2$ to indicate that we are considering the collection of all the data points. For example, $\mathrm{Var}(\{X_1\})$ means the variance of all the $X_1$ values, but $\mathrm{Var}(X_1)$ may represent the square of the measurement uncertainty of a single data point.} Therefore, $\mathrm{Cov}(\{X_1\}, \{X_2-X_1\})=\mathrm{Cov}(\{X_1\}, \{X_2\})-\mathrm{Var}(\{X_1\})\leq\sqrt{\mathrm{Var}(\{X_1\})\mathrm{Var}(\{X_2\})}-\mathrm{Var}(\{X_1\})\approx0$, and $\mathrm{Cov}(\{(X_1+X_2)/2\}, \{X_2-X_1\})=[\mathrm{Var}(\{X_2\})-\mathrm{Var}(\{X_1\})]/2\approx0$. This indicates choosing $X_1$ as the abscissa axes will artificially introduce negative global trends (or positive trends for $X_2$), but adopting $(X_1+X_2)/2$ will not.\par
The $M_\star$ measurements of galaxies are generally accurate, and systematic differences among different works are $\lesssim0.3~\mathrm{dex}$ (e.g., \citealt{Ni20}). Most $M_\star$ measurements agree with each other within $\sim0.5~\mathrm{dex}$. However, the $M_\star$ measurements of AGNs, especially BL AGNs, are more scattered. The comparison with \citet{Guo19} (upper-right panel of Fig.~\ref{fig_comp_wcdfs_master}) indicates that most of our $M_\star$ measurements of AGNs still agree with theirs within $\sim0.5~\mathrm{dex}$. However, unobscured AGN contributions are not considered in HELP SED fitting,\footnote{They included intermediate-type and type~2 AGN contributions, but not type~1 AGN contributions.} and thus their $M_\star$ measurements for (BL) AGNs are systematically larger than ours in the upper-left panel of Fig.~\ref{fig_comp_wcdfs_master}.

We further compare SFRs in the bottom panels of Fig.~\ref{fig_comp_wcdfs_master}. We empirically exclude sources with $\mathrm{sSFR}\leq10^{-9.8}~\mathrm{yr^{-1}}$ in HELP because their SFR measurements ``saturate'' for low-sSFR galaxies. Their SED-fitting parameter settings are mainly for star-forming galaxies, and the smallest sSFR allowed is $10^{-10.2}~\mathrm{yr^{-1}}$, causing over-estimations of SFR for sources with small sSFR. Indeed, we found that the HELP SFRs are much larger than \texttt{Prospector-$\alpha$} values when $\mathrm{sSFR}\lesssim10^{-10}~\mathrm{yr^{-1}}$. The SFR measurements of galaxies are generally more scattered than for $M_\star$, especially when $\mathrm{SFR}\lesssim10^{-1}~M_\odot~\mathrm{yr^{-1}}$. The typical systematic difference of SFR is $\lesssim0.5~\mathrm{dex}$, and most SFR measurements agree within $\sim1~\mathrm{dex}$. Compared with \citet{Guo19}, our SFRs of (BL) AGNs are still generally consistent with theirs, although the scatter is larger than for galaxies.\par
\begin{figure*}
\centering
\resizebox{\hsize}{!}{
\includegraphics{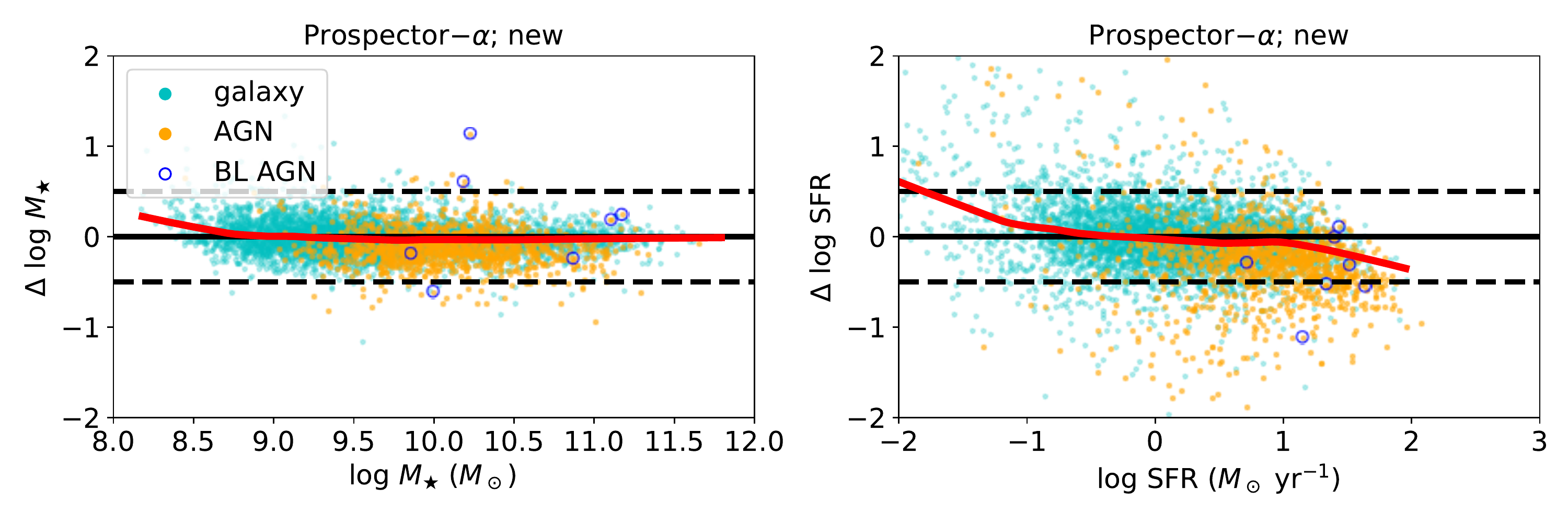}
}
\caption{Comparisons between our corrected $M_\star$ and SFR (Eq.~\ref{eq: corr_mstar_Joel} and \ref{eq: corr_sfr_Joel}) and the \texttt{Prospector-$\alpha$} values. Galaxies generally follow one-to-one relations well while there are still systematic offsets for AGNs, possibly because the \texttt{Prospector-$\alpha$} fitting attributes some AGN emission to the galaxy component.}
\label{fig_comp_correction_Leja}
\end{figure*}

Although our results are generally consistent with other measurements within $\sim0.5$~dex, there are subtle systematic differences between our results and the ground-truth \texttt{Prospector-$\alpha$} ones. There is a mass-dependent offset between our $M_\star$ and the \texttt{Prospector-$\alpha$} $M_\star$, i.e., our $M_\star$ values tend to be under-estimated for low-mass galaxies. Our SFRs are also systematically higher than the \texttt{Prospector-$\alpha$} SFRs. These issues are well explored and explained in \citet{Leja19b}. Briefly, the main reason is that \texttt{Prospector-$\alpha$} uses non-parametric SFHs while ours are less-flexible, parametric SFHs, and our SFHs tend to underestimate galaxy ages, leading to the systematic offsets. Low-mass galaxies tend to be more sensitive to the adopted SFHs. This problem is fundamental and inherent in the SED-fitting methodology, and a more flexible SFH is needed to solve this issue at the expense of much heavier computational requirements, which is impractical in our case. Therefore, we simply calculate empirical corrections to match our results with the \texttt{Prospector-$\alpha$} ones. We fit the \texttt{Prospector-$\alpha$} $\log M_\star$ and $\log\mathrm{SFR}$ as polynomial functions of $z$ and our $\log M_\star$ and $\log\mathrm{SFR}$ for galaxies. For simplicity, the polynomial degree is limited not to exceed three, and the corrections are determined to be
\begin{align}
\log M_\star^\mathrm{new}=&28.06976-7.05089x_1+0.28953x_2\nonumber\\
+&4.53179z+0.76319x_1^2+0.13833x_2^2\nonumber\\
-&0.66180z^2-0.07999x_1x_2-0.71136x_1z\nonumber\\
+&0.72624x_2z-0.02378x_1^3+0.00229x_2^3\nonumber\\
-&0.03415z^3+0.00548x_1^2x_2+0.02394x_1^2z\nonumber\\
-&0.01231x_1x_2^2-0.00369x_2^2z+0.07799x_1z^2\nonumber\\
+&0.01088x_2z^2-0.07372x_1x_2z\label{eq: corr_mstar_Joel}
\end{align}
\begin{align}
\log\mathrm{SFR}^\mathrm{new}=&-39.55280+8.78759x_1-19.54933x_2\nonumber\\
+&27.63832z-0.59271x_1^2-1.55026x_2^2\nonumber\\
-&0.18362z^2+3.57024x_1x_2-5.52671x_1z\nonumber\\
+&3.79753x_2z+0.01144x_1^3-0.09988x_2^3\nonumber\\
+&0.00888z^3-0.15492x_1^2x_2+0.27169x_1^2z\nonumber\\
+&0.12660x_1x_2^2+0.13441x_2^2z+0.01775x_1z^2\nonumber\\
-&0.10721x_2z^2-0.32932x_1x_2z,\label{eq: corr_sfr_Joel}
\end{align}
where $x_1$ and $x_2$ are our $\log M_\star$ (in $M_\odot$) and $\log\mathrm{SFR}$ (in $M_\odot~\mathrm{yr^{-1}}$), respectively. Fig.~\ref{fig_comp_correction_Leja} shows our corrected $M_\star$ and SFRs compared with the \texttt{Prospector-$\alpha$} values. Galaxies generally follow one-to-one relations, except for the high- or low-value edges, but AGNs are slightly systematically below the one-to-one lines. Given the reliability of \texttt{CIGALE} in fitting AGNs and the fact that the AGN locus and galaxy locus generally overlap well when comparing our SED results with those in \citet{Guo19} and our FIR-based SFRs (see the next paragraph), we tend to prefer our measurements for AGNs in Fig.~\ref{fig_comp_correction_Leja}. We note that the calibrations in Eqs.~\ref{eq: corr_mstar_Joel} and \ref{eq: corr_sfr_Joel} are based on sources covering a limited parameter space, i.e., populated by those above the mass-completeness limit of 3D-HST between $0.5<z<3$, where the limit as a function of $z$ is tabulated in Table~1 of \citet{Leja20}. Caution should be taken if extrapolating the corrections beyond this limited parameter space. Furthermore, we note that our uncorrected $M_\star$ and SFR are measured in a self-consistent manner, but the correction inevitably breaks the self-consistency and leads to significant interplays between $M_\star$ and SFR values, as revealed in Eqs.~\ref{eq: corr_mstar_Joel} and \ref{eq: corr_sfr_Joel}.\par

To justify further the reliability of our measurements, we compare our original SED-based SFRs (i.e., not corrected by Eq.~\ref{eq: corr_sfr_Joel}) with FIR-based SFRs. The default assumption for FIR-based SFR estimations is that nearly all the UV photons are absorbed and reemitted in IR, and FIR luminosity is known to be a good tracer of SFR (e.g., \citealt{Chen13, Yang17, Zou19, Ni20}) for galaxies with SFR $\gtrsim1~M_\odot~\mathrm{yr^{-1}}$, where dust is often abundant. FIR-based SFRs are also generally reliable for AGN hosts because AGNs usually contribute little to the FIR emission. In principle, tracing SFR by the summation of UV and IR luminosities does not require the aforementioned assumption for FIR-based SFRs and thus may provide better SFR estimations, but this both faces practical problems and is unnecessary in our case. First, this procedure is problematic for AGNs, which may strongly contaminate the UV emission; second, the UV luminosities of our FIR-detected sources are negligible. We only consider sources with SNR $\ge5$ in at least one Herschel band ($100-500~\mu\mathrm{m}$). As Eq.~1 in \citet{Leja19b} indicates, adding the UV luminosity to the SFR estimation leads to a correction of $\log(1+2.2L(1216-3000\AA)/L(8-1000\mu\mathrm{m}))$ to $\log\mathrm{SFR}$. For our Herschel-detected sources, the median correction is 0.01~dex. Even for those with SFR $<1~M_\odot~\mathrm{yr^{-1}}$, the median correction is 0.04~dex. Such a small correction from the UV emission is generally expected for Herschel-detected sources (see \citealt{Lutz14} for a review) and is also much smaller than the more significant correction in Eq.~\ref{eq: correct_fir_sfr} (see below).\par
\begin{figure*}
\centering
\resizebox{\hsize}{!}{
\includegraphics{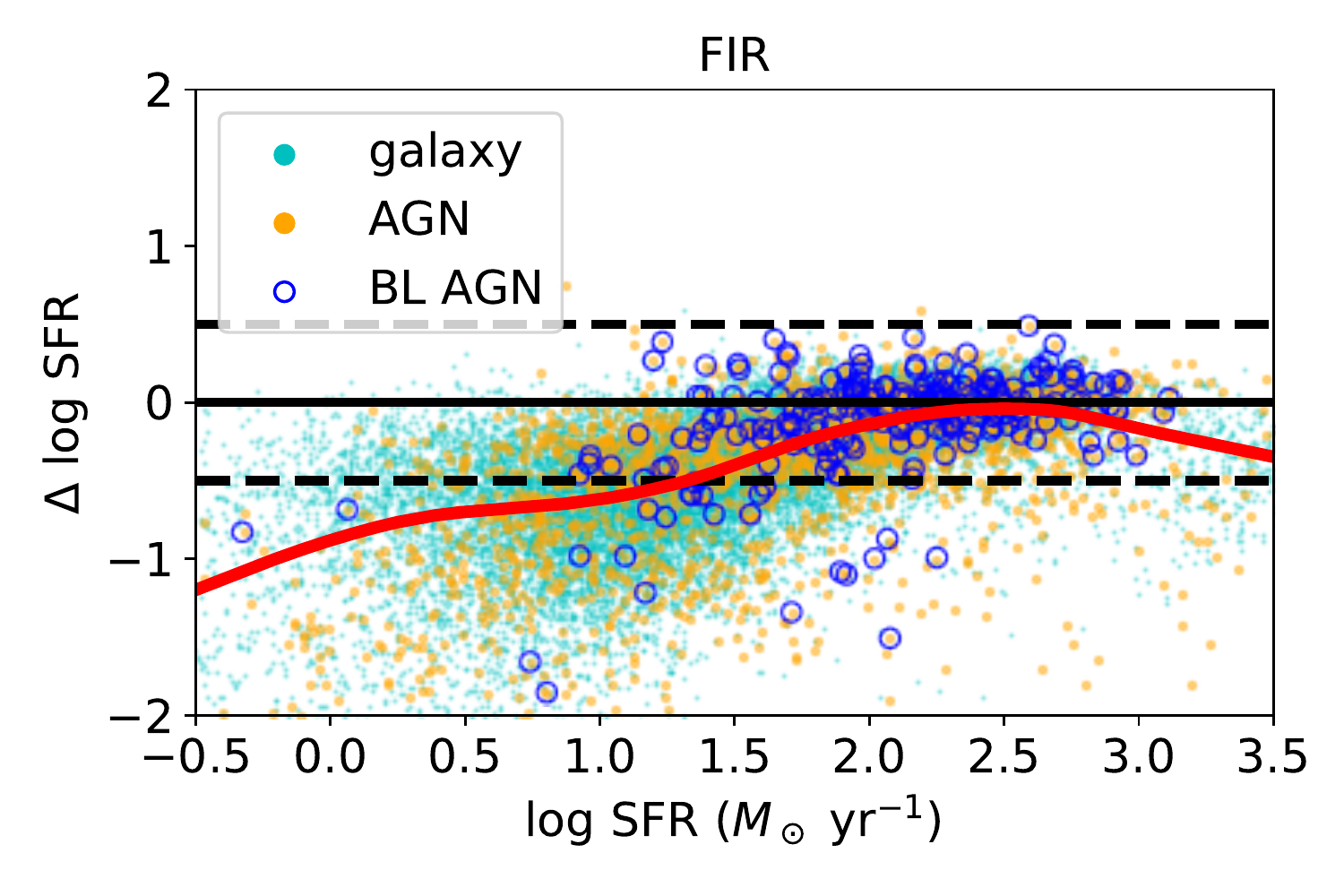}
\includegraphics{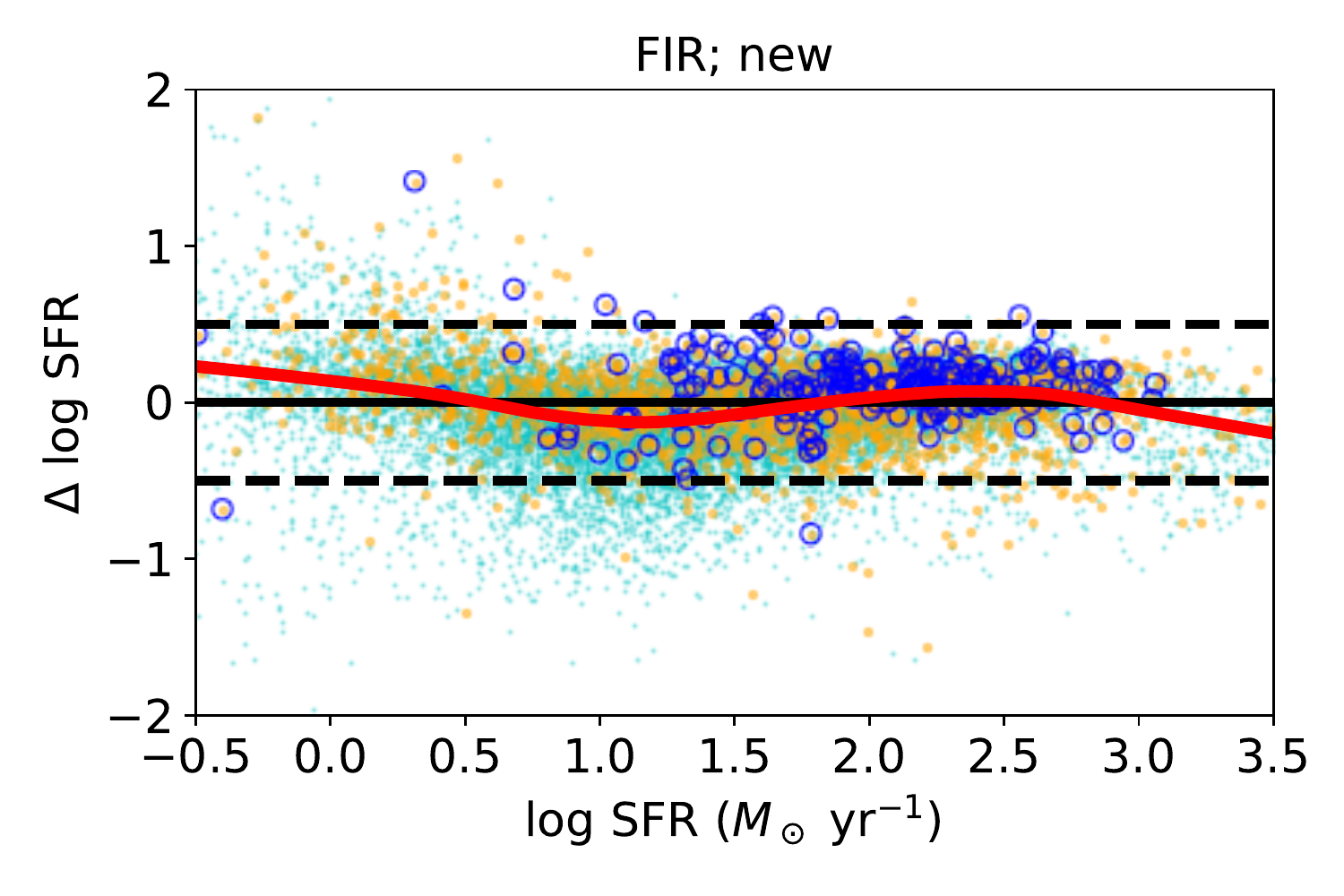}
}
\caption{Comparisons between our original SED-based SFRs and FIR-based SFRs, where the left and right panels show the original and corrected FIR-based SFRs, respectively. The original FIR-based SFRs suffer from the old-star heating bias and thus are over-estimated for low-SFR galaxies. After correcting this issue, our FIR-based and SED-based SFRs are generally consistent for both galaxies and AGNs across a wide SFR range.}
\label{fig_comp_sfr_fir}
\end{figure*}

We follow a similar method as \citet{Chen13} to measure FIR-based SFRs. Briefly, we take the observed flux from the Herschel band with SNR $\ge5$ having the longest wavelength\footnote{We found that for sources detected in multiple Herschel bands, the SFRs inferred from different bands may be different within $\sim0.6~\mathrm{dex}$. The \mbox{longest-wavelength} band is adopted to minimize possible AGN contamination.} and compare it with the redshifted IR templates in \citet{Kirkpatrick12} to estimate the total IR luminosity from $8-1000~\mu\mathrm{m}$. The luminosity is then converted to SFR by multiplying by a factor of $1.09\times10^{-10}~M_\odot~\mathrm{yr^{-1}}~L_\odot^{-1}$. The comparison is shown in the left panel of Fig.~\ref{fig_comp_sfr_fir}, and the SED-based SFRs seem to be systematically smaller than the FIR-based SFRs as the SFR decreases. This may be because the emission from old stars (i.e., stars aged $\gtrsim100~\mathrm{Myr}$) becomes increasingly important for low-SFR galaxies, and this effect is negligible and thus not considered when calibrating a linear relation between the IR luminosity and current SFR using samples of star-forming galaxies (e.g., \citealt{Leja19b}). This effect can be empirically corrected based on the relation between the old-star contribution and sSFR in \citet{Leja19b}:
\begin{align}
\mathrm{\frac{SFR_{FIR}^{new}}{SFR_{FIR}}}=10^{0.25}[-0.5\tanh(-0.8\log\mathrm{sSFR}+0.09z-8.4)],\label{eq: correct_fir_sfr}
\end{align}
where sSFR is in $\mathrm{yr}^{-1}$, and we manually multiply by a constant, $10^{0.25}$, to set the median difference in $\log\mathrm{SFR}$ to be 0. The comparison between our SED-based SFRs and corrected FIR-based SFRs is shown in the right panel of Fig.~\ref{fig_comp_sfr_fir}, which presents a better consistency than the left panel of Fig.~\ref{fig_comp_sfr_fir} and generally follows a one-to-one relation even down to small SFRs. Therefore, the old-star heating effect may be the primary cause for the deviation between our SED-based and original FIR-based SFRs for low-SFR galaxies, although the selection bias that only sources with enhanced FIR emission can be detected by Herschel when their SFRs are relatively low may also still exist. The difference between our SED-based and FIR-based SFRs for (BL) AGNs are well-confined within 0.5~dex with little systematic difference, and thus we conclude that our SFR measurements are generally more reliable for (BL) AGNs than previous works, in which the AGN contributions were not carefully considered.

\begin{figure*}
\centering
\resizebox{\hsize}{!}{
\includegraphics{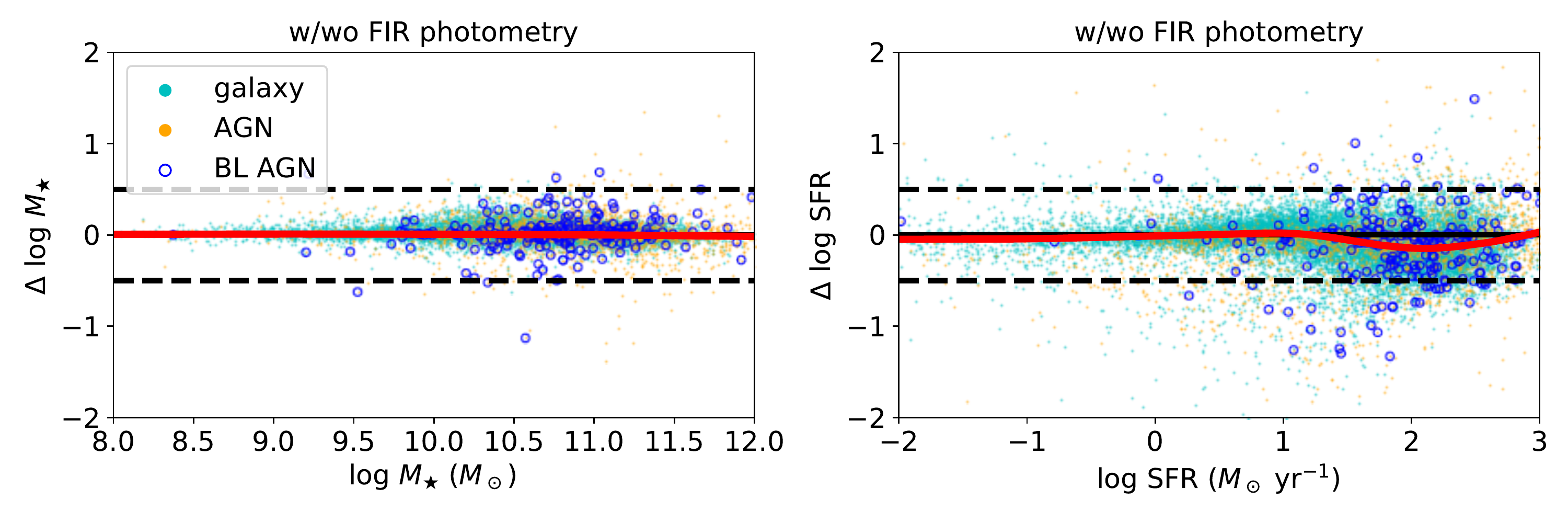}
}
\caption{Comparison between the SED-fitting results with or without FIR data for $M_\star$ (\textit{left}) and SFR (\textit{right}). The results are consistent, indicating that the fitting without FIR data does not have significant biases.}
\label{fig_comp_mstar_sfr_nofir}
\end{figure*}

Fig.~\ref{fig_comp_sfr_fir} is limited to FIR-detected sources, whose FIR photometry is included in the SED fitting, and thus it is somewhat expected that the SED-based SFRs and FIR-based SFRs will agree well. We try excluding the FIR data to see how the SED-fitting results would change, and the results are shown in Fig.~\ref{fig_comp_mstar_sfr_nofir}. Generally, there are no significant systematic differences between the results with or without FIR data for both galaxies and AGNs, and the median offsets are 0.002 and $-0.02$~dex for $M_\star$ and SFR, respectively. Therefore, the fitting without FIR data should also be reliable without significant biases (e.g., \citealt{Mountrichas21}).

\subsection{Validation of $M_\star$ and SFR Uncertainties and Nominal Depth Assessment}
\label{sec: valid_err}
We validate the $M_\star$ and SFR statistical uncertainties output by \texttt{CIGALE} in this section. \texttt{CIGALE} computes the uncertainties as the likelihood-weighted standard deviations \citep{Boquien19}, and we use the linear-space analyses in \texttt{CIGALE}. It is usually difficult to test the uncertainties directly because real $M_\star$ and SFR values are unknown. However, the \texttt{Prospector-$\alpha$} results, which are based on ultradeep multi-wavelength data, provide highly accurate measurements that largely solve this problem.\par
For each source, $s$, we denote the difference in $\log M_\star$ or $\log\mathrm{SFR}$ between our values and \texttt{Prospector-$\alpha$} values as $X_s$ and assume $X_s\sim Normal(o_s, e_{s; \texttt{CIGALE}}^2+e_{s; \texttt{Prospector}}^2)$, where $o_s$ is the expected offset between our and \texttt{Prospector-$\alpha$} measurements, and $e_{s; \texttt{CIGALE}}$ and $e_{s; \texttt{Prospector}}$ are our and \texttt{Prospector-$\alpha$} uncertainties, respectively. $o_s$ is usually not 0, as discussed in Section~\ref{sec: compare_results}; $o_s$ and $e_s$ also vary from source to source. It can be shown that the expected sample variance of $X_s$ is
\begin{align}
\mathrm{Var}(X_s)=&E(e_{s; \texttt{CIGALE}}^2)+E(e_{s; \texttt{Prospector}}^2)+\mathrm{Var}(o_s),\label{eq: valid_error}
\end{align}
where the full derivation is presented in Appendix~\ref{append: eq26}. By checking if the above equation holds, we can test if $e_{s; \texttt{CIGALE}}$ is reliable. For a meaningful comparison, the first term in Eq.~\ref{eq: valid_error} should contribute a sufficiently large portion of $\mathrm{Var}(X_s)$, and this will be checked in the following text. To mitigate the impact of outliers, we further replace the terms in Eq.~\ref{eq: valid_error} by their robust estimators -- we use $e\equiv\mathrm{NMAD}\{X_s\}$, $\mathrm{median}\{e_{s; \texttt{CIGALE}}^2\}$, and $\mathrm{median}\{e_{s; \texttt{Prospector}}^2\}$ to estimate $\sqrt{\mathrm{Var}(X_s)}$, $E(e_{s; \texttt{CIGALE}}^2)$, and $E(e_{s; \texttt{Prospector}}^2)$, respectively. We further define
\begin{align}
\hat{e}^2=\mathrm{median}\{e_{s; \texttt{CIGALE}}^2\}+\mathrm{median}\{e_{s; \texttt{Prospector}}^2\}+\mathrm{Var}(o_s).\label{eq: error_estimator}
\end{align}
The validation of Eq.~\ref{eq: valid_error} is thus to check if $e\approx\hat{e}$.\par
The difficulty arises from the fact that $o_s$ is actually unknown, and thus we cannot calculate $\mathrm{Var}(o_s)$, but we can still give a reasonable range for it. The lower limit of $\mathrm{Var}(o_s)$ is 0, and we adopt its upper limit as the variance of a uniform distribution spanning 0.5 and 1 for $\log M_\star$ and $\log\mathrm{SFR}$, respectively. The spanning range can be justified in the middle panels of Fig.~\ref{fig_comp_wcdfs_master}, where the deviations of the LOESS curves from the one-to-one relationships are confined within a 0.5 (1) dex range for $\log M_\star$ ($\log\mathrm{SFR}$). For $\log M_\star$, $\hat{e}$ is thus estimated to be within the range of $0.17-0.23$, and this range is narrow enough to fairly accurately constrain the uncertainties. The first term in Eq.~\ref{eq: error_estimator} accounts for $51\%-85\%$ of the total $\hat{e}^2$; this fraction is large, and thus $\mathrm{Var}(X_s)$ should be sensitive to $e_{s; \texttt{CIGALE}}$, though the contributions from the other two terms are not negligible. $e$ is 0.22 for $\log M_\star$, within the expected range of $\hat{e}$. For $\log\mathrm{SFR}$, $\mathrm{Var}(o_s)$ is more significant. The contribution of the first term in Eq.~\ref{eq: error_estimator} is $32\%-70\%$, and the estimated range of $\hat{e}$ is $0.27-0.39$. $e$ is 0.37, also within the expected interval. Note that the possible bias of the second term in Eq.~\ref{eq: error_estimator} is not considered, and it may also slightly change the intervals. The analyses of errors are summarized in Table~\ref{tbl_valid_err} for easy reference. Overall, $e$ is largely consistent with $\hat{e}$, and thus we conclude that our uncertainties are generally reliable. We also divide the sources into several $i$-band magnitude bins and do not detect strong dependences of the above analysis results on the magnitude.\par

\begin{table*}
\caption{Analyses of errors in Section~\ref{sec: valid_err}}
\label{tbl_valid_err}
\centering
\begin{threeparttable}
\begin{tabular}{cccccc}
\hline
\hline
& $\mathrm{median}\{e_{s; \texttt{CIGALE}}^2\}$ & $\mathrm{median}\{e_{s; \texttt{Prospector}}^2\}$ & $\mathrm{Var}(o_s)$ & $\hat{e}^2$ & $e^2$\\
& (1) & (2) & (3) & (4) & (5)\\
\hline
$\log M_\star$ & 0.026 & 0.004 & $0-0.021$ & $0.030-0.051$ & 0.047\\
\hline
log SFR & 0.049 & 0.021 & $0-0.083$ & $0.070-0.154$ & 0.139\\
\hline
\hline
\end{tabular}
\begin{tablenotes}
\item
\textit{Notes.} (1), (2), and (3) are the first, second, and third terms in Eq.~\ref{eq: error_estimator}, respectively. (4) = (1) + (2) + (3) is the total expected variance, and (5) is the measured variance. See Section~\ref{sec: valid_err} for more details.
\end{tablenotes}
\end{threeparttable}
\end{table*}

The general reliability of the statistical uncertainties of our $M_\star$ and SFR measurements ultimately arises from the reliability of the statistical uncertainties of the SEDs, which are justified indirectly and independently by the photo-$z$ uncertainties in \citet{Zou21b}. They showed that 78\% of spec-$z$s reside in the 68\% photo-$z$ intervals. 78\% roughly corresponds to $1.2\sigma$ for a normal distribution, roughly consistent with $1\sigma$. This should not be taken for granted because it is a challenging problem to measure accurately the photometric errors accounting for both the pixel correlations in single bands and cross-band systematic uncertainties; especially, within the SED context, the photometric errors should also include the uncertainties of the physical SED models. Due to all these complicated issues, it is not surprising that some previous works found that their uncertainties were underestimated (e.g., \citealt{Luo10}). In principle, single-band photometric uncertainties can be addressed with detailed analyses (e.g., \citealt{Wold19}), but the choice of the systematic uncertainties for subsequent SED analyses often lacks clear guidelines because, at least, it is challenging to quantify the effective contributions from imperfect SED models. In Nyland et al. (in preparation), the smallest error in each band is around $3\%-9\%$ of the flux, and our results suggest that this is a suitable choice when constructing SEDs in our case. We emphasize that the uncertainties generally have little impact on the returned $M_\star$ and SFR values, which mainly depend on the photometric data points instead of their errors.\par
\begin{figure}
\centering
\resizebox{\hsize}{!}{
\includegraphics{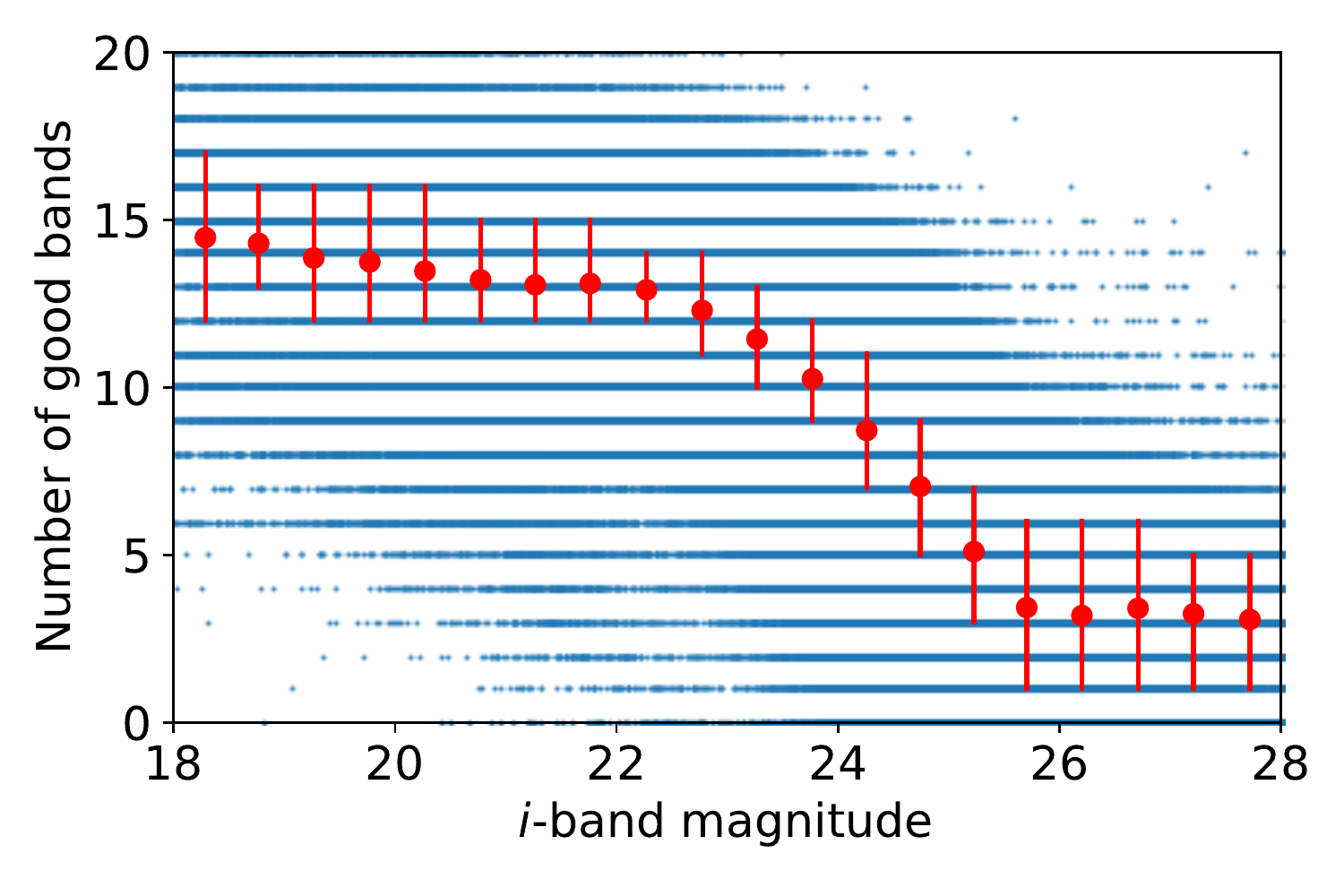}
}
\caption{The number of good bands vs. $i_\mathrm{mag}$. Each background point represents one source, and the large red points with error bars represent the median, $25^\mathrm{th}$, and $75^\mathrm{th}$ percentiles of the number of good bands in each magnitude bin. The number of good bands drops significantly around $i_\mathrm{mag}=24$.}
\label{fig_ngoodband_mag}
\end{figure}

We further estimate a nominal ``depth'' of our SEDs. We define ``good bands'' as those with ratios between their fluxes and flux errors above five and show the number of good bands of each source versus its $i_\mathrm{mag}$ in Fig.~\ref{fig_ngoodband_mag}. We use $i$ band to be consistent with the choice in \citet{Zou21b}, and this band is also sufficiently deep and red. The figure indicates that the number of good bands drops significantly below $i_\mathrm{mag}\approx24$, and thus we claim that our nominal depth is $i_\mathrm{mag}=24$. This is also supported by the fact that the nominal high-quality photo-$z$ depth is $i_\mathrm{mag}\approx24$ in \citet{Zou21b}. We found that this deterioration of SEDs when $i_\mathrm{mag}$ becomes fainter is generally contributed by all the bands between $u$ to Spitzer $4.5~\mu\mathrm{m}$, and no specific bands significantly dominate the band loss. Similarly, we repeat the analyses for the $K_s$ band, and the nominal $K_s$ depth of our SEDs is around 23. About 40\% of our sources are brighter than these magnitude depths.

\subsection{Additional Errors from Photo-$z$ Uncertainties}
\label{sec: adderr}
Photo-$z$s are only estimations of real redshifts, but we fix photo-$z$s during the SED fitting. Photo-$z$ uncertainties can also contribute to the uncertainties of the fitting results, and we probe this additional error term in this section.\par
\begin{figure*}
\centering
\resizebox{\hsize}{!}{
\includegraphics{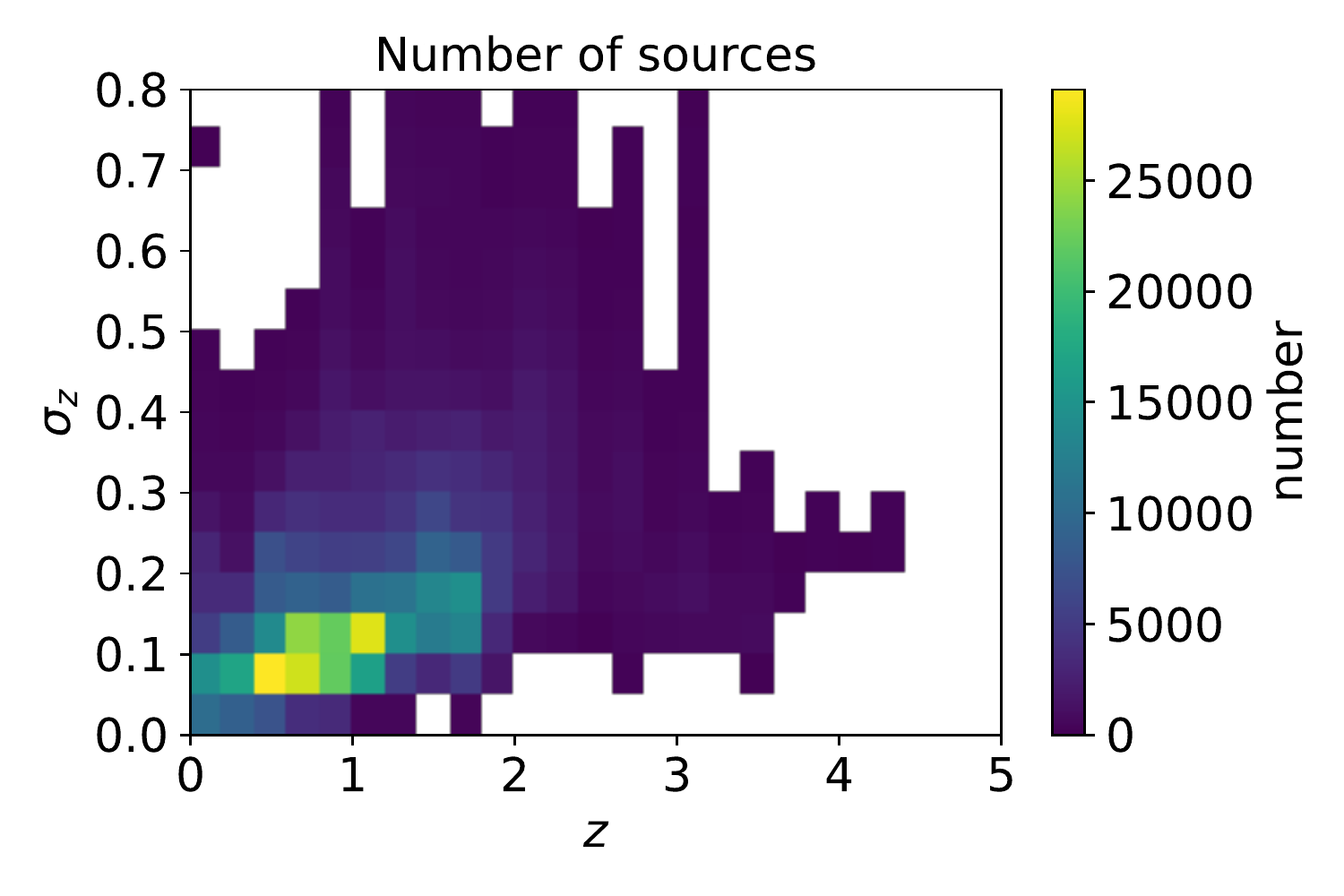}
\includegraphics{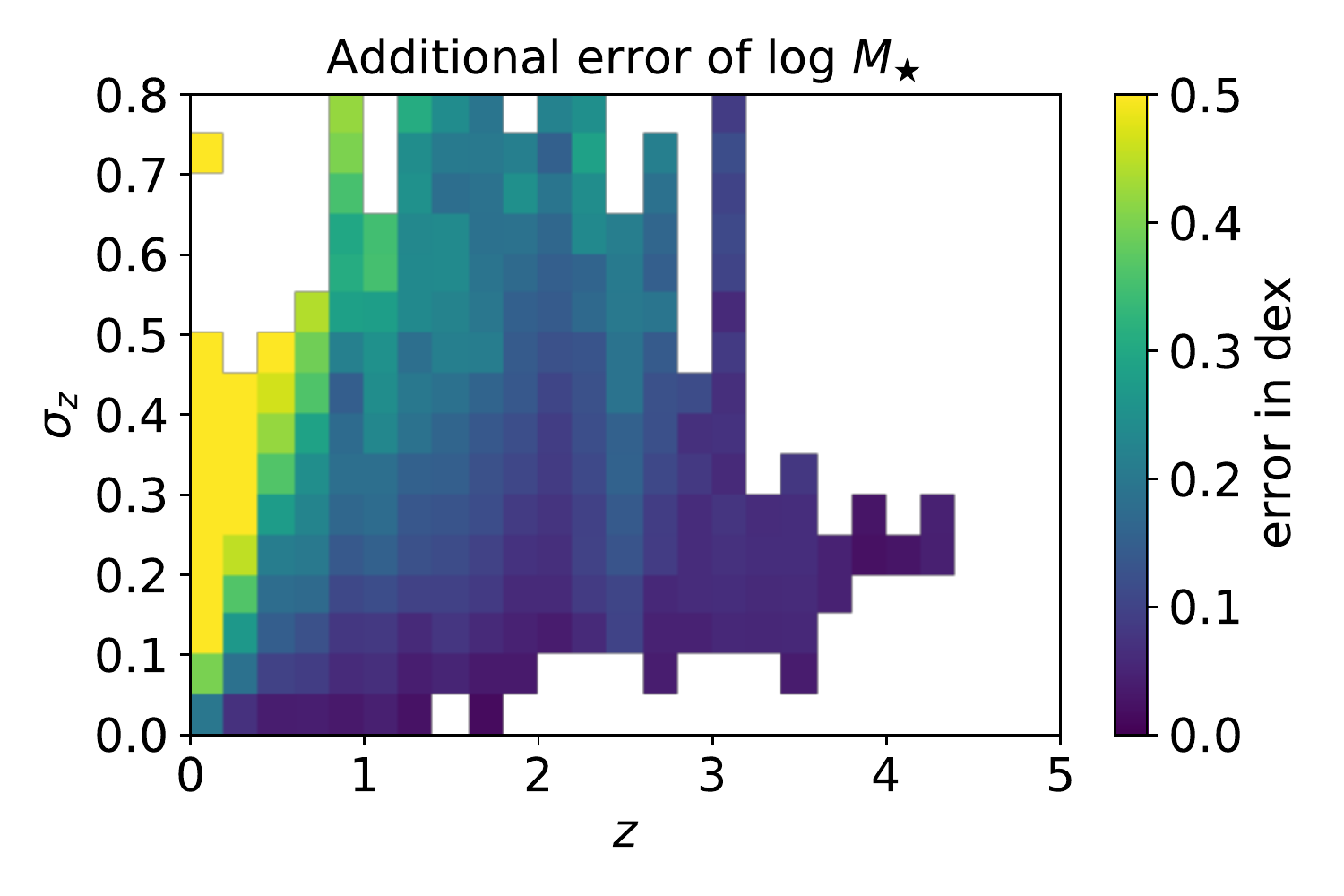}
\includegraphics{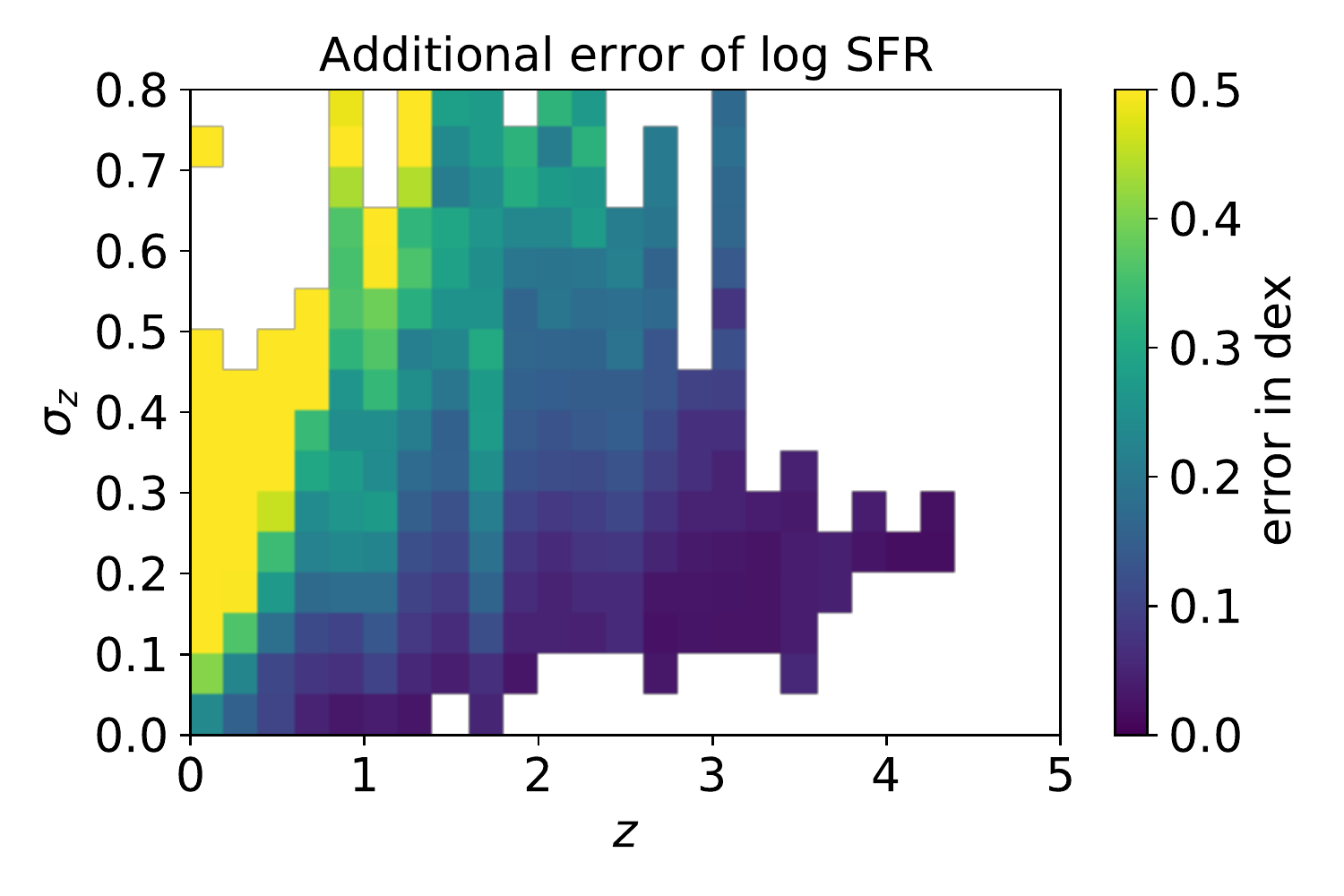}
}
\caption{\textit{Left}: The distribution of our sources in the $z-\sigma_z$ plane, where only bins with at least 200 sources are shown. \textit{Middle}: The map of the additional error from photo-$z$ uncertainties for $\log M_\star$. \textit{Right}: The map of the additional error from photo-$z$ uncertainties for $\log\mathrm{SFR}$.}
\label{fig_zperturb}
\end{figure*}

We estimate the photo-$z$ error ($\sigma_z$) as half of the difference between the 68\% photo-$z$ lower limit and upper limit in \citet{Zou21b}, who have already justified the general reliability of the photo-$z$ limits. We then divide the $z-\sigma_z$ plane into a grid with a bin size of $\Delta z=0.2$ and $\Delta\sigma_z=0.05$, and the left panel of Fig.~\ref{fig_zperturb} shows the distribution of our sources in this plane. For each source, we perturb its $z$ value following a distribution combined from two half-normal distributions -- both peak at the best-fit photo-$z$ value, and the left (right) part has a $\sigma$ value of the difference between the best-fit photo-$z$ and the 68\% lower (upper) limit. We then fit the perturbed data to obtain the differences between the resulting parameters and the unperturbed parameters for each selected source. For each parameter and each $z-\sigma_z$ bin, we take the NMAD of the differences for sources within this bin as the additional error of this parameter at this specific bin. The results for $\log\mathrm{SFR}$ and $\log M_\star$ are shown in Fig.~\ref{fig_zperturb}, and the additional errors increase with $\sigma_z$, as expected. Although the additional errors can be large ($\gtrsim0.5~\mathrm{dex}$) when $\sigma_z$ is large and $z$ is small, few sources populate these regions (Fig.~\ref{fig_zperturb}). Most sources have largely accurate photo-$z$ measurements, and the typical additional errors from photo-$z$ uncertainties are 0.14 dex and 0.11 dex for $\log\mathrm{SFR}$ and $\log M_\star$, respectively.\par

We add these errors to the \texttt{CIGALE}-output errors in quadrature for sources without spec-$z$s, and the errors in those $z-\sigma_z$ bins not covered by Fig.~\ref{fig_zperturb} are linearly extrapolated from the other bins. One caveat is that the above analyses are based on Gaussian uncertainties and thus may not fully address large photo-$z$ errors or multi-modal posteriors. Especially, one case worth noting is $z=0.01$, as mentioned in Section~\ref{sec: select_bqgal}. Such sources usually have monotonically decreasing photo-$z$ posteriors with the largest posterior values occurring at 0.01, the smallest allowed redshift value. Their $M_\star$ and SFR values are thus heavily underestimated and should not be used directly.

\subsection{Data Release}
\label{sec: release}
We release our SED-fitting results on the Zenodo repository (\url{https://zenodo.org/communities/ddfdata/}), including the catalog, auxiliary photometry products, and full best-fit decomposed SED curves for all the sources. The catalog columns are explained below. The relevant sections referred to below are mainly for \mbox{W-CDF-S}, and users should also check additional notes in Appendices~\ref{append: es1} and \ref{append: xmmlss} for ELAIS-S1 and XMM-LSS, respectively.
\begin{itemize}
\item{Column 1, \texttt{Tractor\_ID}: unique source identifier used in internal \textit{The Tractor} photometry catalogs. These identifiers are the same as those in \citet{Ni21}, \citet{Zou21a, Zou21b}, and our auxiliary photometry products.}
\item{Column 2--3, \texttt{RA, Dec}: J2000 RA and Dec in Nyland et al. (in preparation).}
\item{Column 4, \texttt{redshift}: redshift.}
\item{Column 5, \texttt{ztype}: the type of the redshift. ``\texttt{zphot}'' and ``\texttt{zspec}'' stand for photometric and spectroscopic redshift, respectively.}
\item{Column 6--7, \texttt{zphot\_lowlim} and \texttt{zphot\_upplim}: the 68\% lower and upper limit of photo-$z$. These columns are set to $-1$ for sources with spec-$z$s.}
\item{Column 8, \texttt{flag\_star}: Whether the source is a star. ``1'' and ``0'' stand for yes and no, respectively.}
\item{Column 9, \texttt{flag\_Xrayagn}: Whether the source is an \mbox{X-ray} AGN in \citet{Ni21}. ``1'' means that this source is an \mbox{X-ray} AGN, ``0'' means that this source is an \mbox{X-ray}-detected non-AGN, and ``$-1$'' means that this source is undetected in \mbox{X-rays}.}
\item{Column 10, \texttt{flag\_IRagn\_S05}: Whether this source is an IR AGN that satisfies the criteria in \citet{Stern05}. ``1'' means yes, ``0'' means that this source is detected in all four IRAC bands with SNR above three but unclassified as an AGN, and ``$-1$'' indicates other sources.}
\item{Column 11, \texttt{flag\_IRagn\_L07}: Same as Column~10, but for the criteria in \citet{Lacy07}.}
\item{Column 12, \texttt{flag\_IRagn\_D12}: Same as Column~10, but for the criteria in \citet{Donley12}.}
\item{Column 13, \texttt{flag\_reliablesedagn}: Whether this source is a reliable SED AGN that satisfies Eq.~\ref{eq: reliable_sedagn} and \ref{eq: crit_sedagn} for \mbox{W-CDF-S} and ELAIS-S1 or Eq.~\ref{eq: crit_sedagn_xmmlss} for XMM-LSS. ``1'' and ``0'' stand for yes and no, respectively.}
\item{Column 14--15, \texttt{detBIC1\_agn} and \texttt{detBIC2\_agn}: $\Delta\mathrm{BIC_1(AGN)}$ and $\Delta\mathrm{BIC_2(AGN)}$ between the normal galaxy and AGN fitting in Section~\ref{sec: select_agn}.}
\item{Column 16--17, \texttt{detBIC1\_bqgal} and \texttt{detBIC2\_bqgal}: $\Delta\mathrm{BIC_1(BQ)}$ and $\Delta\mathrm{BIC_2(BQ)}$ between the normal and BQ galaxy fitting in Section~\ref{sec: select_bqgal}.}
\item{Column 18, \texttt{redchi2\_gal}: Best-fit $\chi^2_r$ values using normal-galaxy templates.}
\item{Column 19, \texttt{redchi2\_agn}: Best-fit $\chi^2_r$ values using AGN templates. Sources other than IR AGNs, \mbox{X-ray} sources, and those with $\Delta\mathrm{BIC_1(AGN)}>2$ have NaN values.}
\item{Column 20, \texttt{redchi2\_bqgal}: Best-fit $\chi^2_r$ values using BQ-galaxy templates. Sources other than those with $\Delta\mathrm{BIC_1(BQ)}>2$ have NaN values.}
\item{Column 21, \texttt{redchi2\_best}: Adopted best-fit $\chi^2_r$ values, as described in Section~\ref{sec: bestsedfittingresults}.}
\item{Column 22--29, \texttt{Mstar\_gal}, \texttt{Mstar\_gal\_err}, \texttt{Mstar\_agn}, \texttt{Mstar\_agn\_err}, \texttt{Mstar\_bqgal}, \texttt{Mstar\_bqgal\_err}, \texttt{Mstar\_best}, and \texttt{Mstar\_best\_err}: $M_\star$ and uncertainties in $M_\odot$, with the additional errors in Section~\ref{sec: adderr} added. The suffix (``\texttt{gal}'', ``\texttt{agn}'', ``\texttt{bqgal}'', or ``\texttt{best}'') follows the same rule in Columns~18--21.}
\item{Column 30, \texttt{logMstar\_err\_from\_zphot}: The additional error of $\log M_\star$ from photo-$z$ uncertainties in Section~\ref{sec: adderr}. This column is set to 0 for sources with spec-$z$s.}
\item{Column 31, \texttt{logMstar\_new}: Corrected $\log M_\star$ derived from Eq.~\ref{eq: corr_mstar_Joel} for \mbox{W-CDF-S and ELAIS-S1} or Eq.~\ref{eq: corr_mstar_Joel_xmmlss} for XMM-LSS. This column should be used with appropriate consideration due to the related caveats discussed in Section~\ref{sec: compare_results}.}
\item{Column 32--41, \texttt{SFR\_gal}, \texttt{SFR\_gal\_err}, \texttt{SFR\_agn}, \texttt{SFR\_agn\_err}, \texttt{SFR\_bqgal}, \texttt{SFR\_bqgal\_err}, \texttt{SFR\_best}, \texttt{SFR\_best\_err}, \texttt{logSFR\_err\_from\_zphot}, and \texttt{logSFR\_new}: Similar to Columns~22--31, but for SFR.}
\item{Column 42, \texttt{SFR\_FIR}: Original, uncorrected FIR-based SFRs.}
\item{Column 43, \texttt{SFR\_FIR\_new}: Corrected FIR-based SFRs based on Eq.~\ref{eq: correct_fir_sfr}.}
\item{Column 44--45, \texttt{V\_J\_gal} and \texttt{V\_J\_gal\_err}: Rest-frame $V-J$ colors and uncertainties using the galaxy templates.}
\item{Column 46--47, \texttt{U\_V\_gal} and \texttt{U\_V\_gal\_err}: Rest-frame $U-V$ colors and uncertainties using the galaxy templates.}
\item{Column 48--49, \texttt{FUV\_V\_gal} and \texttt{FUV\_V\_gal\_err}: Rest-frame $FUV-V$ colors and uncertainties using the galaxy templates.}
\item{Column 50--51, \texttt{fracagn} and \texttt{fracagn\_err}: AGN fractions and uncertainties. Sources other than IR AGNs, \mbox{X-ray} sources, and those with $\Delta\mathrm{BIC_1(AGN)}>2$ have $\texttt{fracagn}=0$ and $\texttt{fracagn\_err}=-1$. We reiterate that this parameter can hardly be constrained well by the available data; see Section~\ref{sec: sedagn}.}
\item{Column 52, \texttt{logL\_6um\_AGN}: Decomposed best-fit AGN rest-frame $6~\mu\mathrm{m}$ luminosity.}
\item{Column 53--54, \texttt{rSFR} and \texttt{rSFR\_err}: $r_\mathrm{SFR}$ and uncertainties in Eq.~\ref{eq: delayedsfhbq}. Sources other than those with $\Delta\mathrm{BIC_1(BQ)}>2$ have $\texttt{rSFR}=1$ and $\texttt{rSFR\_err}=-1$.}
\item{Column 55, \texttt{ngoodband}: Number of bands with $\mathrm{SNR>5}$, as defined in Section~\ref{sec: valid_err}.}
\end{itemize}
$\Delta\mathrm{BIC}$ values are included in our catalog, but we do not recommend directly linking them to statistical probabilities because the real physical case is much more complicated than the underlying assumptions of $\Delta\mathrm{BIC}$. Instead, detailed calibrations are usually needed. We refer readers to Section~\ref{sec: sedagn} and \citet{Ciesla18} for discussions of using $\Delta\mathrm{BIC}$ and other parameters to reliably select SED AGNs and rapidly quenching galaxies, respectively. The $M_\star$ and SFR measurements may become unreliable and/or have large errors when the photometric quality decreases, and thus we record the number of bands with $\mathrm{SNR>5}$ as a basic quality indicator worth considering.\par
Our best-fit decomposed SED curves are also released to facilitate future studies of our sources. Their individual SEDs are merged into several large Hierarchical Data Format (HDF5) files, in which the \texttt{group} name is the same as the \texttt{Tractor\_ID} for each source, and the \texttt{dataset}s under each \texttt{group} record rasterized wavelengths or specific luminosities of all the components. We release the SEDs in the HDF5 format instead of the traditional FITS format because the HDF5 format has a better I/O performance (e.g., \citealt{Price15}), which is important in our case as millions of SEDs are involved. The files are downsized by resampling the SEDs to a sparser wavelength grid. We adopt the flux-conservation algorithm in the \texttt{SpectRes} package \citep{Carnall17} to do the resampling. We rewrote its code in \texttt{Julia} and increased the speed by a factor of $>100$. The resolution of the new grid is around eight times better than those of broad photometric bands at the corresponding wavelengths, and thus the downsizing procedure does not affect broad-band characterizations.\par
Other intermediate data can be shared upon reasonable request to the authors.

\section{Summary and Future Work}
\label{sec: summary}
In this work, we have derived and analyzed the \mbox{X-ray} to FIR SEDs of nearly three million sources in \mbox{W-CDF-S}, ELAIS-S1, and XMM-LSS. The main text focuses on \mbox{W-CDF-S} as a representative example, and the results for ELAIS-S1 and XMM-LSS are presented in Appendices~\ref{append: es1} and \ref{append: xmmlss}, respectively. Appendix~\ref{append: compare_fields} further makes a check that there are no significant systematic differences in our SED-fitting results among different fields. The main results are summarized below.
\begin{itemize}
\item{We collect and reduce the data from \mbox{X-ray} to FIR. The intrinsic \mbox{X-ray} luminosities are estimated using a Bayesian approach. We also generate flux upper-limit maps in the \mbox{X-ray} and FIR to feed into the SED fitting. See Section~\ref{sec: data}.}
\item{We select AGNs or AGN candidates through \mbox{X-ray}, MIR, and SED methods and compare the selection results of these methods. By calibrating the SED method using the deep Chandra data in the \mbox{CDF-S}, we find that the SED method based on the existing data can hardly select more AGNs missed by other methods if we require high purity. The SED method can thus only select AGN candidates but may not be able to reliably justify whether a source is an AGN or not. See Section~\ref{sec: select_agn}.}
\item{We select BQ-galaxy candidates that may be undergoing recent rapid changes in their SFRs. See Section~\ref{sec: select_bqgal}.}
\item{We provide a catalog recording the source properties (e.g., SFR and $M_\star$) and briefly analyze them, including quiescent-galaxy colors, AGN MIR-X-ray relations, and comparisons between our measurements and others. Especially, we assess and calibrate our measurements by comparing them with the reference \texttt{Prospector-$\alpha$} results for small parts of the \mbox{W-CDF-S} and XMM-LSS fields. Empirical corrections of our SFR (SED-based and FIR-based) and $M_\star$ are given. The detailed decomposed SEDs are also publicly available along with the catalog. See Section~\ref{sec: sedfitting}.}
\end{itemize}
Overall, our data products provide a valuable resource for future extragalactic research in these LSST DDFs. New datasets are also constantly being released in these fields. For example, at the time of writing this article, slightly deeper Spitzer images from the Cosmic Dawn Survey than what we are using were released in \mbox{W-CDF-S} and XMM-LSS \citep{Moneti21}. The Hawaii Two-0 Survey (H20) will soon provide deep HSC images in \mbox{W-CDF-S} comparable to the LSST depth. Our \mbox{W-CDF-S} field has been selected as the Euclid Deep Field-Fornax, which will be deeply observed by Euclid in the NIR. The upcoming LSST DDF observations will also provide deeper optical data with hundreds of observation epochs. Re-analyses of the new data, including measuring forced photometry and subsequently conducting SED fitting, will take much more effort and time. We leave the utilization of the new data to future updates of our catalogs.\par
There are also many ongoing or forthcoming spectroscopic surveys in our fields, and we list them here to the best of our knowledge: CSI (The Carnegie-Spitzer-IMACS Survey; \citealt{Kelson14}), DESI (The Dark Energy Spectroscopic Instrument; \citealt{Levi19}), DEVILS (The Deep Extragalactic Visible Legacy Survey; \citealt{Davies18}), H20, PFS (The Subaru Prime Focus Spectrograph; \citealt{Takada14}), SDSS-V BHM (Black Hole Mapper; \citealt{Kollmeier17}), VLT MOONS (The Multi-Object Optical and Near-infrared Spectrograph for the Very Large Telescope; \citealt{Maiolino20}), 4MOST WAVES (Wide-Area VISTA Extragalactic Survey with the 4-metre Multi-Object Spectroscopic Telescope; \citealt{Driver19}). Our work can help these projects select targets for spectroscopic observations, and they will further provide better redshift measurements and additional data for source characterization. Especially, with these upcoming spectra, as we briefly discussed in Sections~\ref{sec: bqagn} and \ref{sec: xraydatapoint}, co-analyses of spectra and photometry can provide further insights about our sources, and \citet{Villa-Velez21} is a recent example in COSMOS.\par
Meanwhile, our catalogs will have rich legacy value and enable many scientific projects on different topics. To name just a few possibilities, our catalog helps in characterizing rapidly quenching or bursting galaxies; the links between AGNs and their host-galaxy properties can be probed; the cosmic growth of SMBHs and galaxies can be constructed; many high-redshift active dwarf galaxies with intermediate-mass black hole candidates can be selected; and hosts of transients (e.g., supernovae and TDEs) found in these DDFs can be analyzed. A notable example is that our catalog will be directly helpful for the LADUMA (Looking At the Distance Universe with the MeerKAT Array) survey \citep{Blyth16}, which will measure the amount of neutral atomic hydrogen; together with our $M_\star$ and SFR measurements, one can measure how long galactic star formation can continue in the future and the relative importance of the future star formation compared to the past star formation (characterized by $M_\star$). Overall, all these studies will greatly benefit from our large sample size. For instance, as far as we know, our catalog includes the largest sample of medium-depth ($\approx50$~ks exposure) \mbox{X-ray} AGNs and should thus be superb for AGN studies that were previously limited by the sample size. Besides our SED-fitting catalog, our compilation of photometric and redshift data in Section~\ref{sec: data} is also valuable, and users can perform further analyses with these depending upon their needs.\par
Our analyses could further be extended to COSMOS and EDF-S, which would provide the community with self-consistent and easy-to-access catalogs covering all the LSST DDFs. Analyses in COSMOS would also provide opportunities for extensive comparisons and calibrations with many previous works, and those in EDF-S, when the data are ripe, will fill the vacancy of systematic catalogs of source SEDs and properties in EDF-S.

\acknowledgments
We thank the referee for a thorough and constructive review, which has greatly improved this article. We thank Franz E. Bauer, Zhenyuan Wang, and Junyu Zhang for helpful discussions. FZ, WNB, WY, and SZ acknowledge support from NASA grant 80NSSC19K0961, NSF grant AST-2106990, Penn State ACIS Instrument Team Contract SV4-74018 (issued by the Chandra X-ray Center, which is operated by the Smithsonian Astrophysical Observatory for and on behalf of NASA under contract NAS8-03060), and the V.M. Willaman Endowment. We acknowledge support from the LSST Corporation through an Enabling Science grant. QN acknowledges support from a UKRI Future Leaders Fellowship (grant code: MR/T020989/1). GY acknowledges support from the George P. and Cynthia Woods Mitchell Institute for Fundamental Physics and Astronomy at Texas A\&M University. BL acknowledges support from the National Natural Science Foundation of China grant 11991053. Basic research in radio astronomy at the U.S. Naval Research Laboratory is supported by 6.1 Base Funding. KN acknowledges support from the NRL Karles Fellow program. YQX acknowledges support from NSFC-12025303 and 11890693, the CAS Frontier Science Key Research Program (QYZDJ-SSW-SLH006), the K.C. Wong Education Foundation, and the science research grants from the China Manned Space Project with NO. CMS-CSST-2021-A06. The Chandra ACIS Team Guaranteed Time Observations (GTO) utilized were selected by the ACIS Instrument Principal Investigator, Gordon P. Garmire, currently of the Huntingdon Institute for X-ray Astronomy, LLC, which is under contract to the Smithsonian Astrophysical Observatory via Contract SV2-82024.

\appendix
\section{Derivation of Equation~1}
\label{append: eq1}
Eq.~\ref{eq_def_fx} or related formulae have been derived in previous literature (e.g., \citealt{Schmidt86}), and we present a brief derivation here for easy reference.\par
An unabsorbed power-law spectrum with a photon index of $\Gamma$ is defined as $C_E(E)=KE^{-\Gamma}$, where $C_E$ is the count rate per unit area per unit energy range, $K$ is the normalization, and $E$ is the observed-frame photon energy. The specific flux is thus $f_E(E)=EC_E(E)=KE^{1-\Gamma}$. Without loss of generality, we assume $\Gamma\neq2$. The intrinsic \mbox{X-ray} flux between the observed-frame energy range, $E_\mathrm{low}-E_\mathrm{high}$, is
\begin{align}
f_\mathrm{X}^\mathrm{int}=\int_{E_\mathrm{low}}^{E_\mathrm{high}}f_E(E)dE=\frac{K}{2-\Gamma}(E_\mathrm{high}^{2-\Gamma}-E_\mathrm{low}^{2-\Gamma}).\label{eq: eq1_fxint}
\end{align}
The specific luminosity at rest-frame $E_{(r)}$ is
\begin{align}
L_{E_{(r)}}(E_{(r)})=\frac{4\pi D_L^2}{1+z}f_E\left(\frac{E_{(r)}}{1+z}\right)=\frac{4\pi D_L^2}{(1+z)^{2-\Gamma}}KE_{(r)}^{1-\Gamma}.
\end{align}
Therefore, the intrinsic \mbox{X-ray} luminosity between rest-frame $2-10~\mathrm{keV}$ is
\begin{align}
L_\mathrm{X}=\int_2^{10}L_{E_{(r)}}(E_{(r)})dE_{(r)}=\frac{4\pi D_L^2}{(1+z)^{2-\Gamma}}\frac{K}{2-\Gamma}(10^{2-\Gamma}-2^{2-\Gamma}).\label{eq: eq1_lx}
\end{align}
Based on Eqs.~\ref{eq: eq1_fxint} and \ref{eq: eq1_lx}, we obtain
\begin{align}
f_\mathrm{X}^\mathrm{int}=\frac{L_\mathrm{X}}{4\pi D_L^2}(1+z)^{2-\Gamma}\frac{E_\mathrm{high}^{2-\Gamma}-E_\mathrm{low}^{2-\Gamma}}{10^{2-\Gamma}-2^{2-\Gamma}}.\label{eq: eq1_fxint_result}
\end{align}
When $\Gamma=2$,
\begin{align}
f_\mathrm{X}^\mathrm{int}(\Gamma=2)=\lim_{\Gamma\to2}f_\mathrm{X}^\mathrm{int}=\frac{L_\mathrm{X}}{4\pi D_L^2}\frac{\ln\frac{E_\mathrm{high}}{E_\mathrm{low}}}{\ln5}.\label{eq: eq1_fxint_result_gamma2}
\end{align}
We further denote $\eta=f_\mathrm{X}/f_\mathrm{X}^\mathrm{int}$ to account for the absorption; combining Eqs.~\ref{eq: eq1_fxint_result} and \ref{eq: eq1_fxint_result_gamma2} returns Eq.~\ref{eq_def_fx}.\par

\section{X-Ray Detection Function}
\label{append: xray_detection}
\restartappendixnumbering
The \mbox{X-ray} band-merged detection function is
\begin{align}
D=1-\mathfrak{Prob}\{\mathbb{POI}(M(\mathrm{SB})+B(\mathrm{SB}))\leq N_\mathrm{thres}(\mathrm{SB}),\nonumber\\
\mathbb{POI}(M(\mathrm{HB})+B(\mathrm{HB}))\leq N_\mathrm{thres}(\mathrm{HB}),\nonumber\\
\mathbb{POI}(M(\mathrm{FB})+B(\mathrm{FB}))\leq N_\mathrm{thres}(\mathrm{FB})\}.
\end{align}
We assume the SB and HB counts are independent for simplicity, and FB counts are the sum of SB and HB counts. Then, the above formula can be reduced to the following problem: calculate $D\equiv\mathfrak{Prob}(X_1\leq C_1, X_2\leq C_2, X_1+X_2\leq C)$ if $X_i\sim\mathbb{POI}(\lambda_i)$ ($i=1, 2$), and $X_1$ and $X_2$ are independent. We further assume $C_1\leq C_2$ without loss of generality.\par
If $C\leq C_2$,
\begin{align}
D&=\sum_{n=0}^{C_1}\mathfrak{Prob}\{X_1=n\}\mathfrak{Prob}\{X_2\leq C_2, X_2\leq C-n\}\\
&=\sum_{n=0}^{\min\{C, C_1\}}\mathfrak{Prob}\{X_1=n\}\mathfrak{Prob}\{X_2\leq C-n\}\\
&=\sum_{n=0}^{\min\{C, C_1\}}\mathfrak{Prob}\{\mathbb{POI}(\lambda_1)=n\}\mathcal{Q}_\mathrm{IG}(C-n+1, \lambda_2).
\end{align}
Similarly, we can derive $D$ when $C>C_2$, and the results are summarized below.
\begin{align}
D=
\begin{cases}
\sum_{n=0}^{\min\{C, C_1\}}\mathfrak{Prob}\{\mathbb{POI}(\lambda_1)=n\}\mathcal{Q}_\mathrm{IG}(C-n+1, \lambda_2),~C\leq C_2\\
\mathcal{Q}_\mathrm{IG}(C-C_2+1, \lambda_1)\mathcal{Q}_\mathrm{IG}(C_2+1, \lambda_2)+\\
~~~~~~~~~~\sum_{n=C-C_2+1}^{C_1}\mathfrak{Prob}\{\mathbb{POI}(\lambda_1)=n\}\mathcal{Q}_\mathrm{IG}(C-n+1, \lambda_2),\\
~~~~~~~~~~~~~~~~~~~~~~~~~~~~~~~~~~~~~~~~~~~~~~~~~~C_2<C<C_1+C_2\\
\mathcal{Q}_\mathrm{IG}(C_1+1, \lambda_1)\mathcal{Q}_\mathrm{IG}(C_2+1, \lambda_2),~C\ge C_1+C_2
\end{cases}
\end{align}
In our case, $C=N_\mathrm{thres}(\mathrm{FB})$, $C_i$ ($i=1, 2$) is $N_\mathrm{thres}(\mathrm{SB})$ or $N_\mathrm{thres}(\mathrm{HB})$, and $\lambda_i$ is $M(\mathrm{SB})+B(\mathrm{SB})$ or $M(\mathrm{HB})+B(\mathrm{HB})$. If $N_\mathrm{thres}(\mathrm{SB})\leq N_\mathrm{thres}(\mathrm{HB})$, then the subscript (1, 2) denotes (SB, HB); otherwise, (1, 2) means (HB, SB).

\section{Derivation of Equation 26}
\label{append: eq26}
\restartappendixnumbering
Considering a general problem that a random variable $X$ follows a distribution determined by two parameters, $\mu$ and $\sigma$, which are also random variables, we can use the law of total variance to calculate $\mathrm{Var}(X)$:
\begin{align}
\mathrm{Var}(X)=&\mathrm{Var}_\mu(E_X(X\mid\mu))+E_\mu(\mathrm{Var}_X(X\mid\mu))\nonumber\\
=&\mathrm{Var}_\mu(E_X(X\mid\mu))+E_\mu(\mathrm{Var}_\sigma(E_X(X\mid\mu,\sigma)))\nonumber\\
&+E_\mu(E_\sigma(\mathrm{Var}_X(X\mid\mu,\sigma))),\label{eq: append_C}
\end{align}
where the second equation expands $\mathrm{Var}_X(X\mid\mu)$ by $\sigma$ following the same way of expanding $\mathrm{Var}(X)$ by $\mu$ in the first equation. We explicitly write the objects to which the ``$E$'' and ``$\mathrm{Var}$'' operators are applied as these operators' subscripts, without which appropriate conditions should be added; for example, $\mathrm{Var}_\sigma(E_X(X\mid\mu,\sigma))$ is the same as $\mathrm{Var}(E(X\mid\mu,\sigma)\mid\mu)$.\par
If $X\sim Normal(\mu, \sigma^2)$, Eq.~\ref{eq: append_C} returns $\mathrm{Var}(X)=E(\sigma^2)+\mathrm{Var}(\mu)$. In Section~\ref{sec: valid_err}, $\mu=o_s$, and $\sigma^2=e_{s; \texttt{CIGALE}}^2+e_{s; \texttt{Prospector}}^2$. Combining these together returns Eq.~\ref{eq: valid_error}.

\section{SEDs in ELAIS-S1}
\label{append: es1}
\restartappendixnumbering
We apply the same methods used in \mbox{W-CDF-S} to ELAIS-S1. Generally, the results in ELAIS-S1 are similar to those for \mbox{W-CDF-S}, and we only highlight the most important aspects here.\par
The \mbox{X-ray} data are from \citet{Ni21}, the $0.36-4.5~\mu\mathrm{m}$ data are from \textit{The Tractor} catalog in \citet{Zou21a}, and we also add the GALEX data and the photometry between $5.8-500~\mu\mathrm{m}$, as detailed in Section~\ref{sec: data}. The $0.36-4.5~\mu\mathrm{m}$ photometry includes VOICE $u$ \citep{Vaccari16}, ESIS $BVR$ \citep{Berta06, Vaccari16}, DES DR2 $grizY$ \citep{Abbott21}, VIDEO $ZYJHK_s$ \citep{Jarvis13}, and DeepDrill IRAC 3.6 and 4.5~$\mu\mathrm{m}$ \citep{Lacy21}; also see Table~\ref{tbl_fieldinfo}. The redshifts are from \citet{Zou21b}. The general data quality in ELAIS-S1 is slightly worse than in \mbox{W-CDF-S} for VIDEO-detected sources. Especially, the \mbox{X-ray} and MIR depths are slightly shallower, and the spectroscopic coverage is around one magnitude brighter than for \mbox{W-CDF-S}. However, the overall differences in the depths are only minor for optical-to-NIR SEDs, and Appendix~B in \citet{Zou21b} shows that the high-quality photo-$z$s, derived based on the $0.36-4.5~\mu\mathrm{m}$ SEDs, reach similar depths (differing by $\approx0.2$~mag) between \mbox{W-CDF-S} and ELAIS-S1.\par
The source classifications are the same as for \mbox{W-CDF-S}, except for a small difference in selecting stars. We use HSC optical morphological selection in \mbox{W-CDF-S} (Section~\ref{sec: select_star}); particularly, extended sources are excluded from being classified as SED-selected stars. Similarly, we use DES DR2 morphology to help select stars in ELAIS-S1, but DES DR2 in ELAIS-S1 is not as deep as HSC in \mbox{W-CDF-S}, and thus it cannot efficiently remove extended sources that may be misclassified as stars through SED fitting. We thus apply an empirical color-color cut to prevent the SED selection from misclassifing too many galaxies as stars -- stars selected through SED fitting are required to satisfy $Z-K_s\leq0.4(B-Z)-0.65$. Such color-color cuts are necessary when the morphological information is limited. For example, similar cuts were used when selecting stars in COSMOS \citep{Laigle16}, and we would also need similar cuts in \mbox{W-CDF-S} if we did not utilize the HSC morphology. However, this color-color cut may not be as efficient as deep optical morphology in cleaning SED-selected stars, and thus the contamination from galaxies to ELAIS-S1 stars may be slightly larger than for \mbox{W-CDF-S}. The deep imaging from the upcoming LSST will help the star-galaxy separation.\par
\begin{figure*}
\centering
\resizebox{\hsize}{!}{
\includegraphics{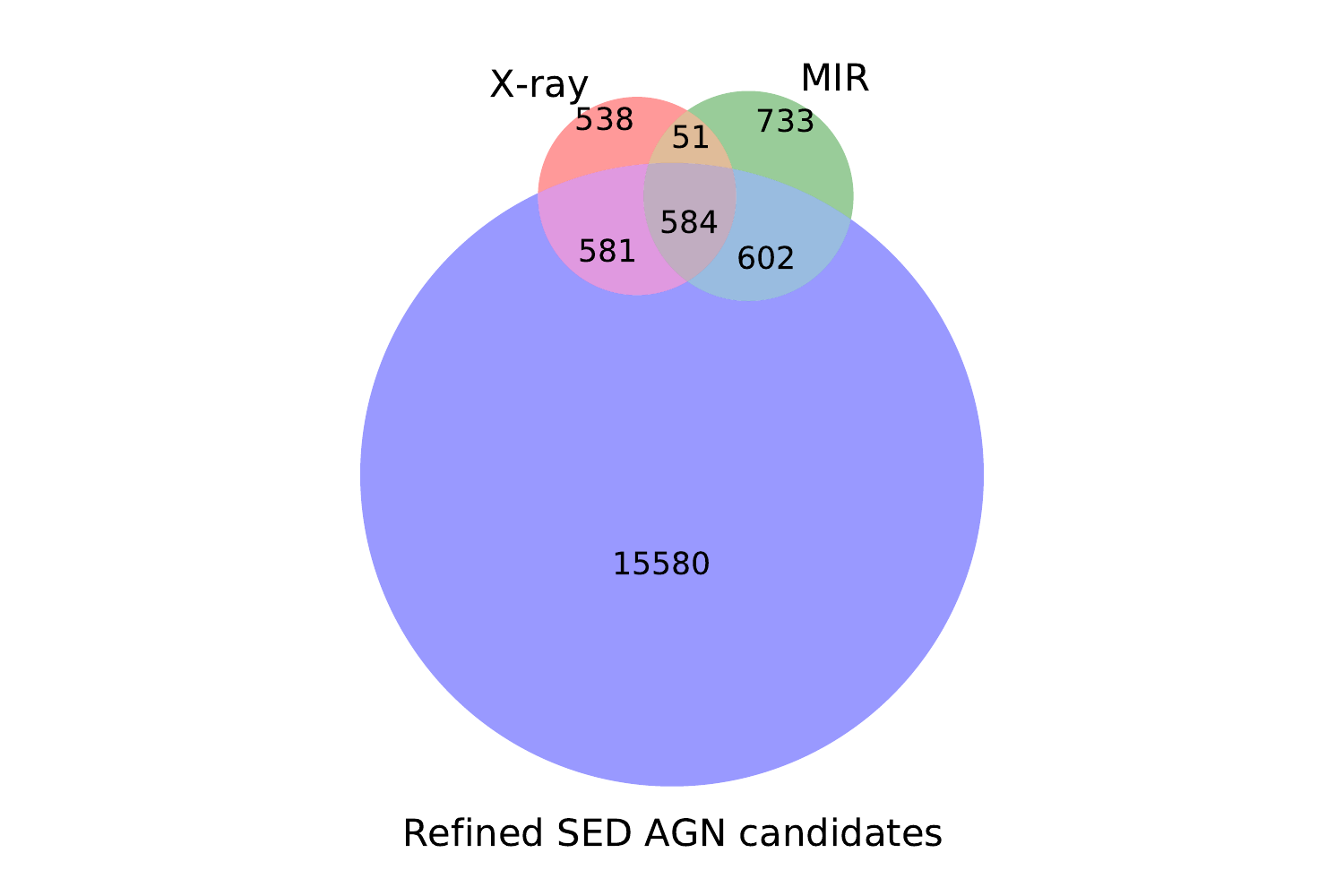}
\includegraphics{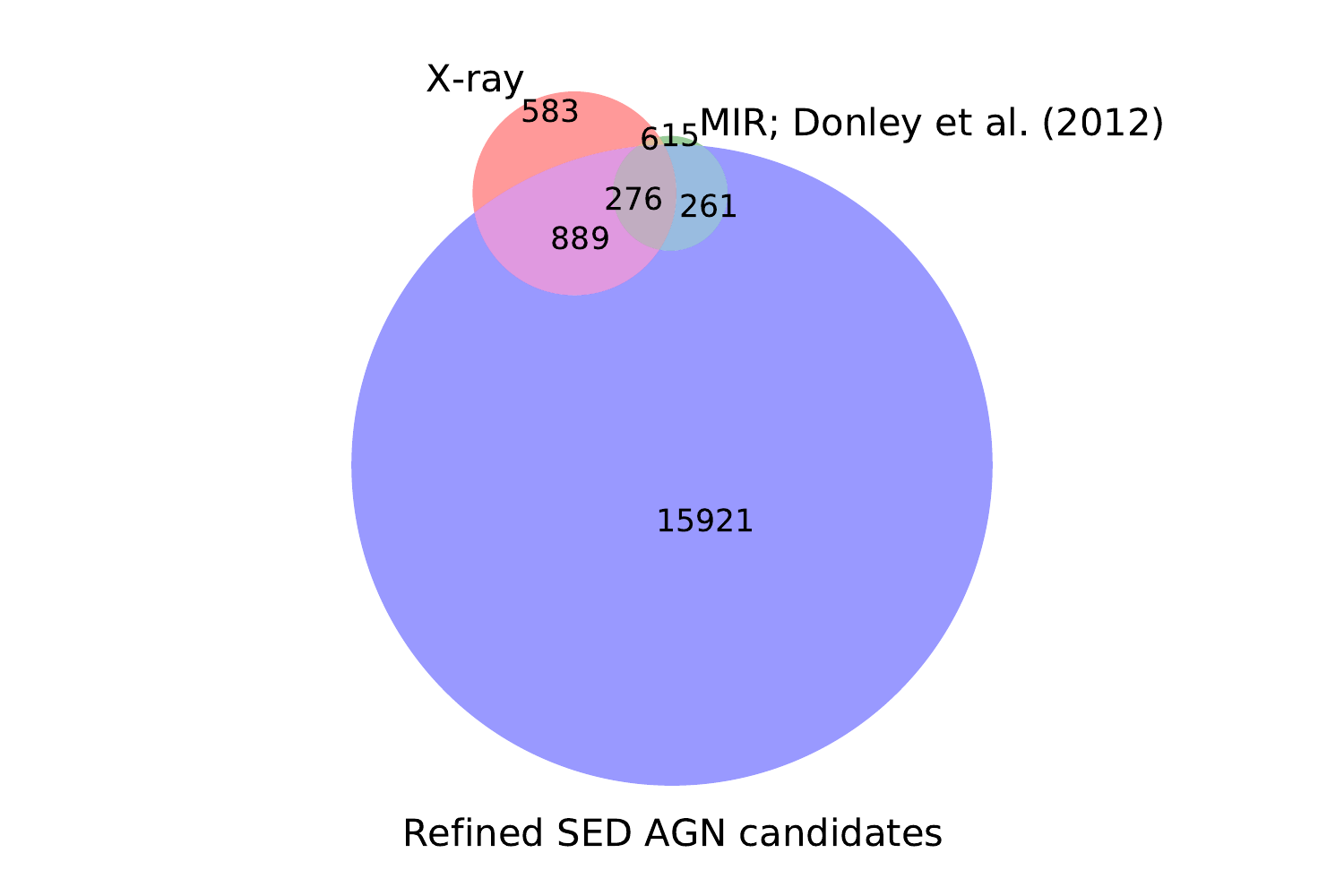}
\includegraphics{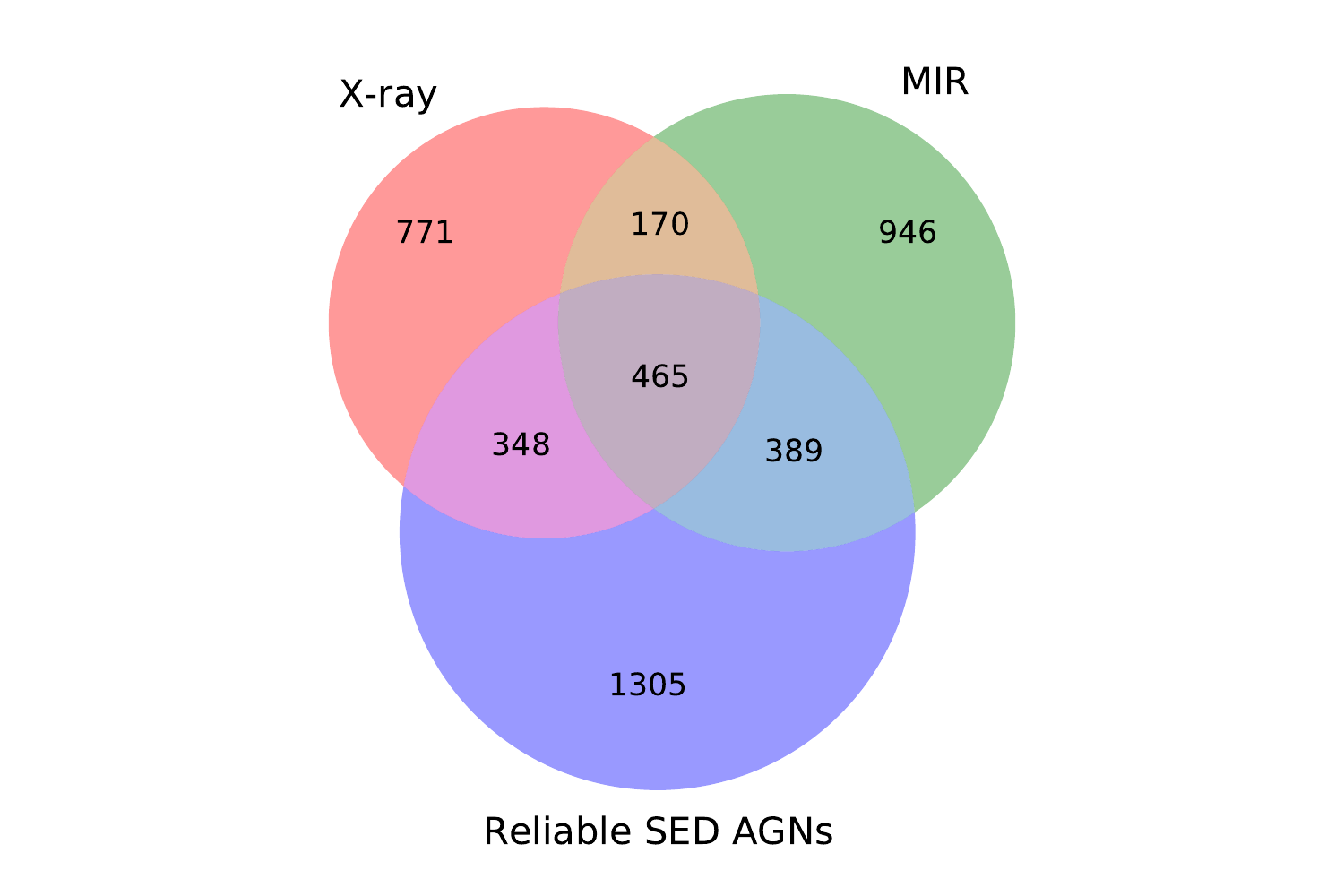}
}
\caption{Venn diagrams comparing different AGN-selection results in ELAIS-S1. \textit{Left}: Comparison among \mbox{X-ray} AGNs, MIR AGNs, and the refined SED AGN candidates. \textit{Middle}: The MIR AGNs are limited to those satisfying the criterion in \citet{Donley12}. \textit{Right}: The SED AGNs are only limited to those satisfying the reliable-AGN criterion in \mbox{W-CDF-S} (see Section~\ref{sec: sedagn}). The left and middle panels correspond to Fig.~\ref{fig_venn_agn}, and the right panel corresponds to the right panel of Fig.~\ref{fig_venn_reliable_agn_cdfs_L17}.}
\label{fig_venn_agn_es1}
\end{figure*}

The Venn diagrams comparing different AGN selections in ELAIS-S1 are presented in Fig.~\ref{fig_venn_agn_es1}. 66\% of \mbox{X-ray} AGNs and 96\% of MIR AGNs based on the criterion in \citet{Donley12} are also selected as refined SED AGN candidates. Unlike for \mbox{W-CDF-S}, we cannot select a complete and pure AGN sample based on ultradeep \mbox{X-ray} coverage in ELAIS-S1, and thus we cannot calibrate its AGN selection following the method in Section~\ref{sec: sedagn}. Therefore, we simply apply the calibration results in \mbox{W-CDF-S} to ELAIS-S1 to select reliable SED AGNs, and 48\% of the reliable SED AGNs are also identified by the \mbox{X-ray} or MIR selections. If we adopt the calibration results in XMM-LSS (Appendix~\ref{append: xmmlss}), then the fraction of reliable SED AGNs being identified by other methods becomes 53\%. We adopt the \mbox{W-CDF-S} calibration in ELAIS-S1 because it is based on a deeper \mbox{X-ray} survey, but there is generally no strong preference for the \mbox{W-CDF-S} calibration over the XMM-LSS one. The fraction of reliable SED AGNs identified by \mbox{X-ray} or MIR is slightly smaller in ELAIS-S1 than for \mbox{W-CDF-S} because of both the variation of the data between the two fields and the shallower MIR depth in ELAIS-S1, but the difference is only moderate and all the relevant qualitative results in \mbox{W-CDF-S} still hold for ELAIS-S1.\par

\begin{figure*}
\centering
\resizebox{\hsize}{!}{
\includegraphics{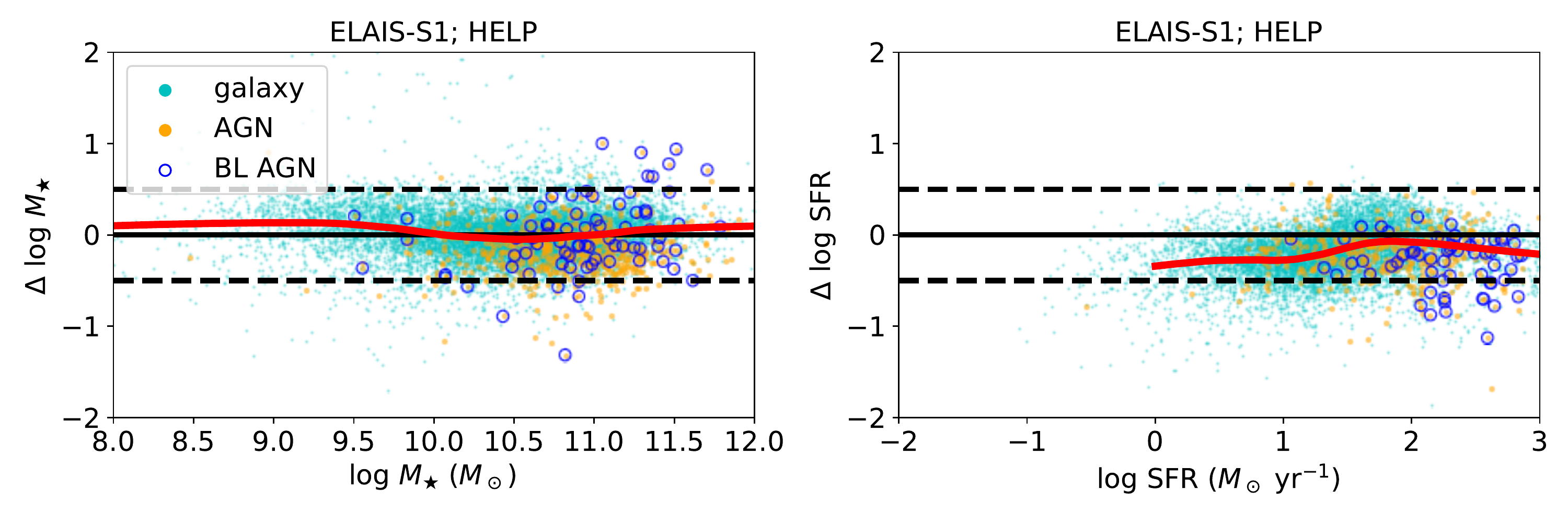}
}
\caption{Comparisons of our $M_\star$ (\textit{left}) and SFR (\textit{right}) measurements with the HELP values. They are the same as the left panels of Fig.~\ref{fig_comp_wcdfs_master}, but for ELAIS-S1.}
\label{fig_comp_mstar_sfr_literature_es1}
\end{figure*}

\begin{figure*}
\centering
\resizebox{\hsize}{!}{
\includegraphics{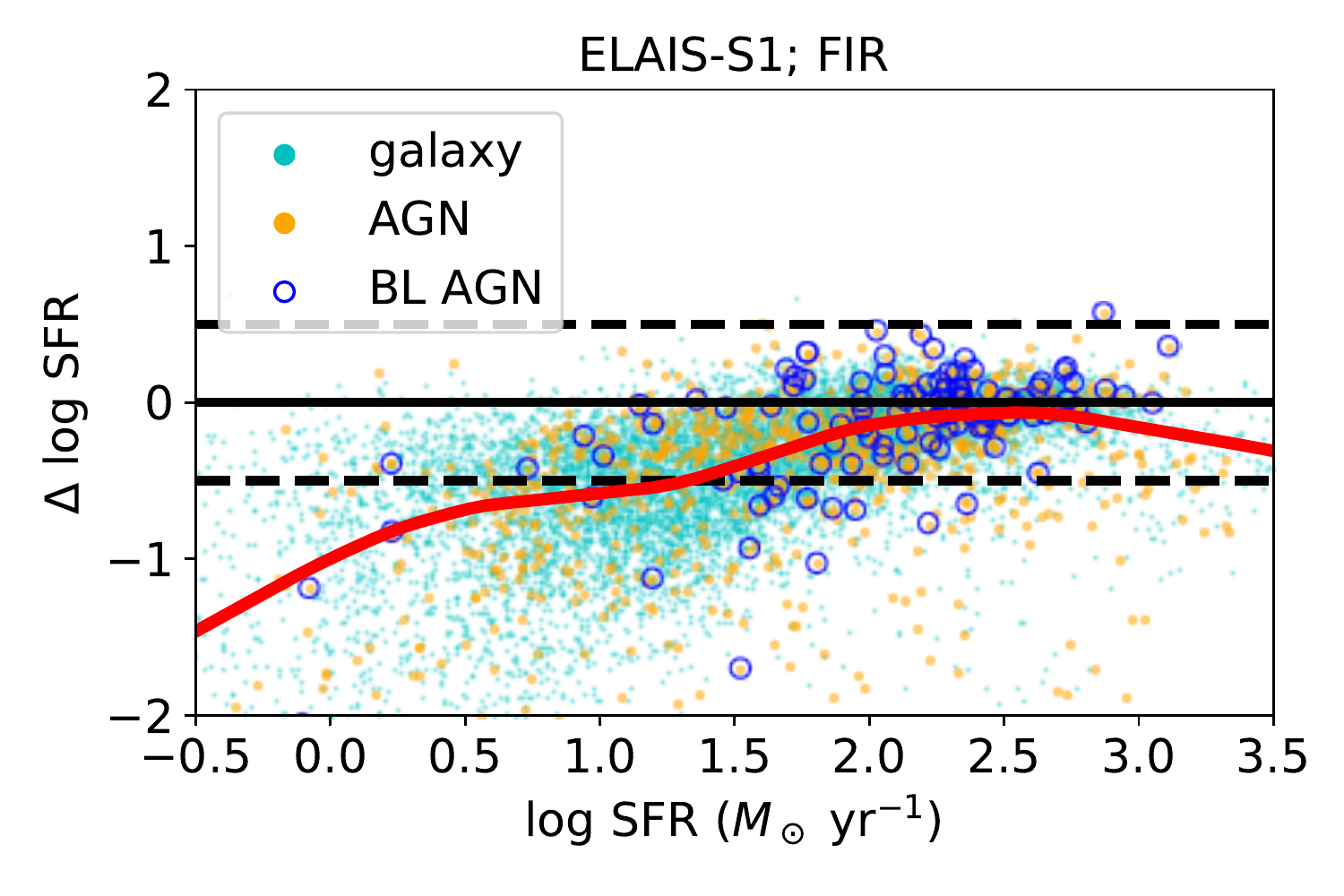}
\includegraphics{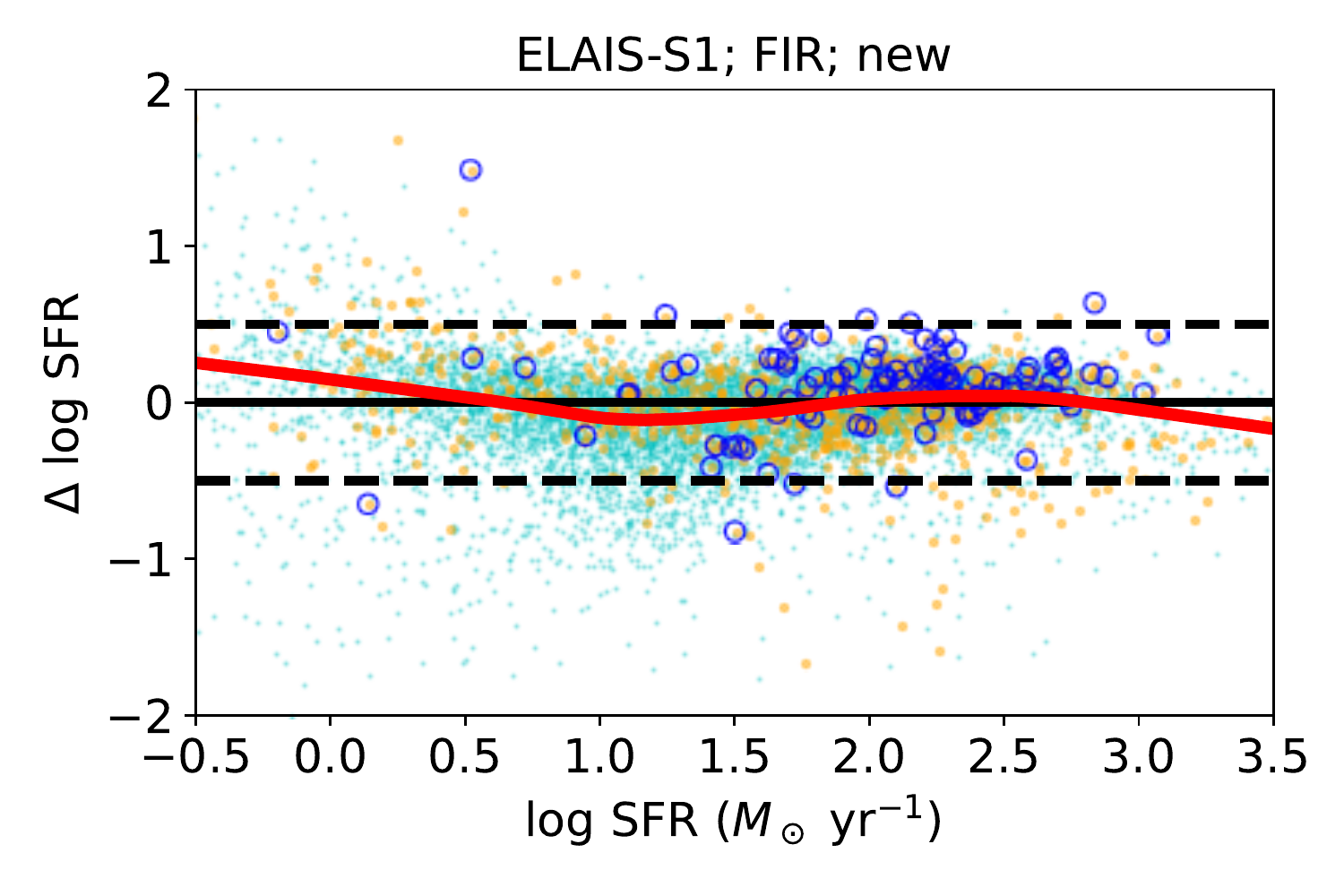}
}
\caption{Comparisons between our SED-based and FIR-based SFRs in ELAIS-S1. This figure corresponds to Fig.~\ref{fig_comp_sfr_fir}.}
\label{fig_comp_sfr_fir_es1}
\end{figure*}

There are 56850, 746634, 18454, and 4304 sources whose best results are based on star, normal-galaxy, AGN, and BQ-galaxy fits, respectively. We compare our $M_\star$ and SFRs with HELP values in Fig.~\ref{fig_comp_mstar_sfr_literature_es1}, and they generally agree well. Just as for the discussion in Section~\ref{sec: compare_results}, only sources with $\mathrm{sSFR>10^{-9.8}~yr^{-1}}$ are shown when comparing the SFR. Also note that our catalog provides $M_\star$ and SFR measurements for over 30 times more sources than the HELP catalog in our footprint. The comparison between our SED-based and FIR-based SFRs is shown in Fig.~\ref{fig_comp_sfr_fir_es1}. They follow similar patterns to those in \mbox{W-CDF-S}, i.e., the old-star heating bias for the FIR-based SFR measurement also exists and can be empirically corrected using Eq.~\ref{eq: correct_fir_sfr}. The corrected FIR-based SFRs are consistent with the SED-based SFRs, as justified in the right panel of Fig.~\ref{fig_comp_sfr_fir_es1}. Also, Eqs.~\ref{eq: corr_mstar_Joel} and \ref{eq: corr_sfr_Joel} (or similarly, Eqs.~\ref{eq: corr_mstar_Joel_xmmlss} and \ref{eq: corr_sfr_Joel_xmmlss}) should roughly hold as we face the same issue that our SFHs are not sufficiently flexible in ELAIS-S1. Nevertheless, caution should be taken when applying the equations to ELAIS-S1 because we do not have reference \texttt{Prospector-$\alpha$} results in ELAIS-S1 to calibrate the coefficients in the equations.

\section{SEDs in XMM-LSS}
\label{append: xmmlss}
\restartappendixnumbering
The same procedures are applied to XMM-LSS as well, and only the most important aspects are highlighted in this appendix.\par
The \mbox{X-ray} data are from \citet{Chen18}, the \mbox{$0.36-4.5~\mu\mathrm{m}$} data are from \citet{Hudelot12} and Nyland et al. (in preparation), and the GALEX and $5.8-500~\mu\mathrm{m}$ photometry are included as well. Nyland et al. (in preparation) provide \textit{The Tractor} photometry for CFHTLS $u^*$, HSC $grizy$ \citep{Aihara18}, VIDEO $ZYJHK_s$ \citep{Jarvis13}, and DeepDrill IRAC 3.6 and 4.5~$\mu\mathrm{m}$ \citep{Lacy21} in the whole XMM-LSS field (also see \citealt{Nyland17}), but these data do not include CFHTLS $griz$\footnote{There are two CFHTLS $i$-band filters because the old one was damaged halfway during the CFHT survey, and they are treated separately in our SED fitting.} photometry, and thus we supplement with the CFHTLS $griz$ photometry from \citet{Hudelot12}; also see Table~\ref{tbl_fieldinfo}. We manually add $3\%$ flux errors to the uncertainties in \citet{Hudelot12}, as done in \citet{Nyland17}. The redshifts are from Appendix~C of \citet{Chen18}.\par
\begin{figure*}
\centering
\resizebox{\hsize}{!}{
\includegraphics{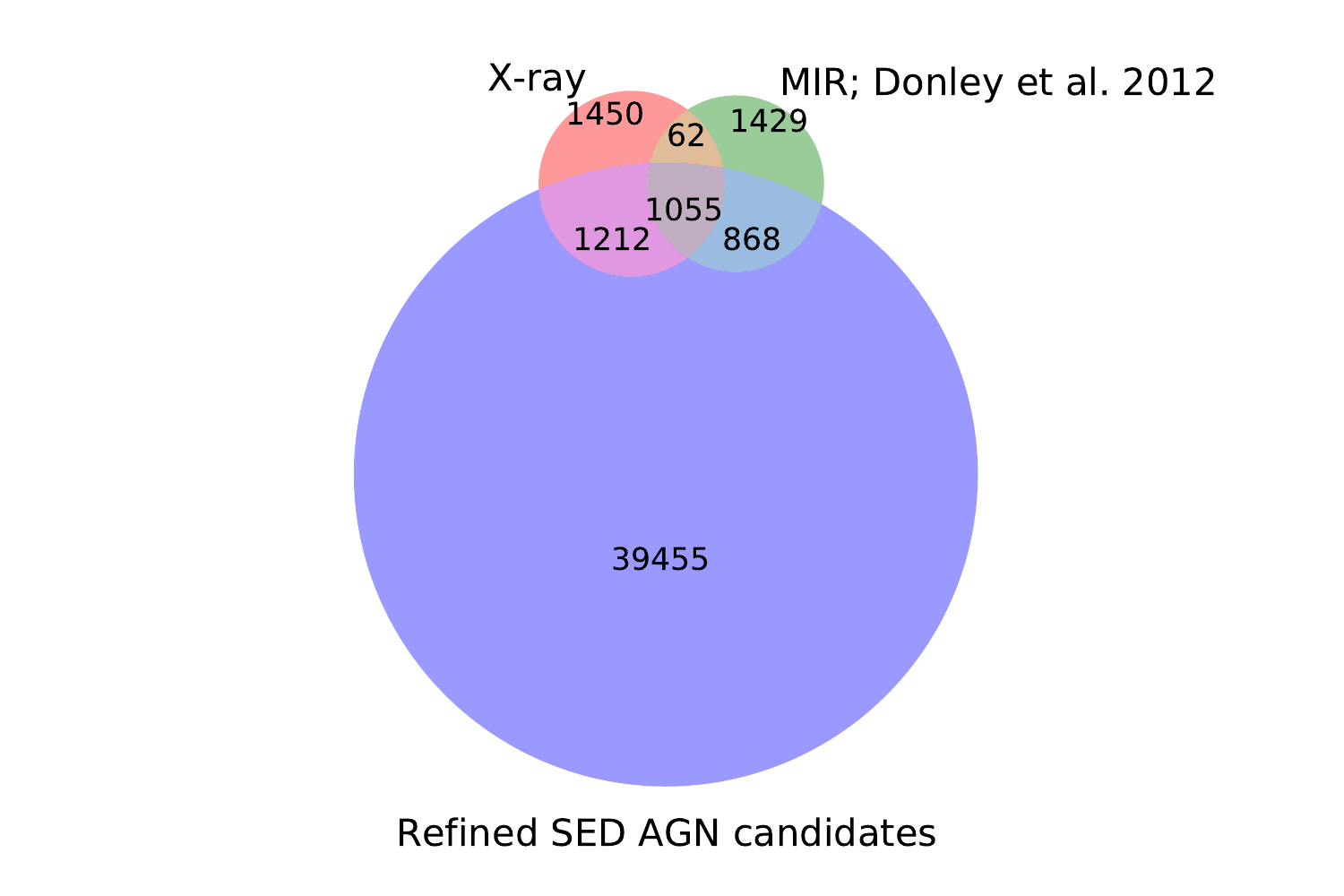}
\includegraphics{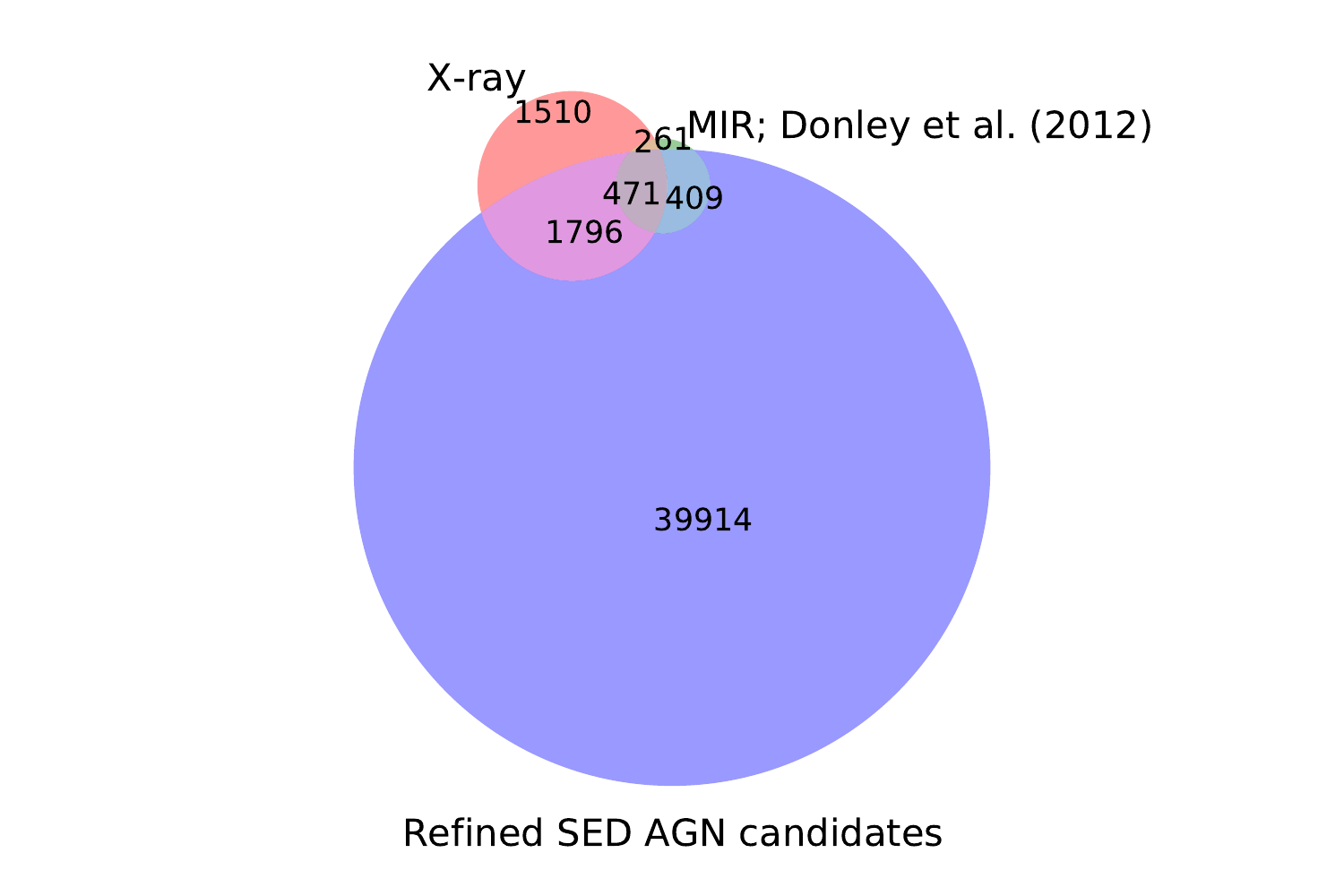}
\includegraphics{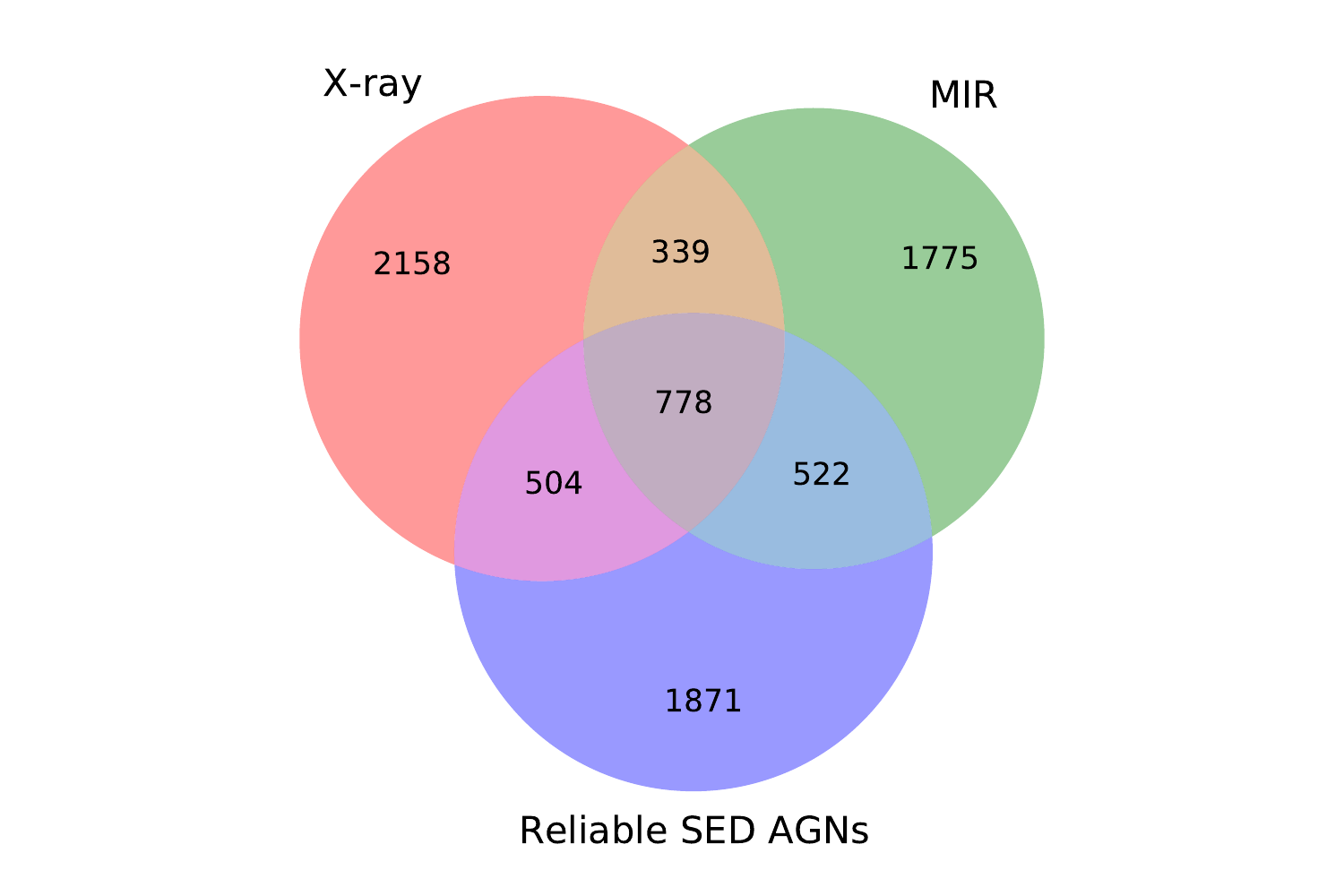}
}
\caption{Same as Fig.~\ref{fig_venn_agn_es1}, but for XMM-LSS, and the reliable-AGN criterion is Eq.~\ref{eq: crit_sedagn_xmmlss}.}
\label{fig_venn_agn_xmmlss}
\end{figure*}

Fig.~\ref{fig_venn_agn_xmmlss} shows the Venn diagrams comparing different AGN selections in XMM-LSS. 60\% of \mbox{X-ray} and 93\% of MIR AGNs based on the criterion in \citet{Donley12} are selected as refined SED AGN candidates. XMM-LSS overlaps with the X-UDS survey, in which deeper Chandra observations are available \citep{Kocevski18}, and we use X-UDS to calibrate our \mbox{X-ray} selection in XMM-LSS. We focus on the central $0.13~\mathrm{deg^2}$ area of X-UDS with high Chandra exposure and follow the same approach in Section~\ref{sec: sedagn} to calibrate the parameters in Eq.~\ref{eq: reliable_sedagn}. The only difference is that a smaller threshold is adopted for the detection rate because X-UDS is shallower than \mbox{CDF-S}. Among the 34 reliable SED AGNs in \mbox{CDF-S}, only 18 are expected to be detectable at the depth of X-UDS. We thus only require that at least $18/34=53\%$ of sources satisfying Eq.~\ref{eq: reliable_sedagn} are detected in X-UDS. This returns the following parameters:
\begin{align}
i_\mathrm{break}=23.3,~\delta_1=8,~\mathrm{and}~\delta_2=66.\label{eq: crit_sedagn_xmmlss}
\end{align}
The result after applying Eq.~\ref{eq: crit_sedagn_xmmlss} is displayed in the right panel of Fig.~\ref{fig_venn_agn_xmmlss}, and 49\% of the reliable SED AGNs are also identified by \mbox{X-ray} or MIR.\par

There are 49230, 1136304, 41568, and 20852 sources with best results from star, normal-galaxy, AGN, and BQ-galaxy fits, respectively. 3D-HST UDS, which has \texttt{Prospector-$\alpha$} measurements, overlaps with XMM-LSS, and we derive empirical corrections to match our results to the \texttt{Prospector-$\alpha$} values, just as for \mbox{W-CDF-S} (Section~\ref{sec: compare_results}):
\begin{align}
\log M_\star^\mathrm{new}=&19.17406-4.34508x_1-0.97304x_2\nonumber\\
+&4.46941z+0.49474x_1^2+0.29096x_2^2\nonumber\\
-&0.09364z^2+0.24277x_1x_2-0.81604x_1z\nonumber\\
+&0.29206x_2z-0.01504x_1^3+0.01873x_2^3\nonumber\\
-&0.01633z^3-0.01304x_1^2x_2+0.03392x_1^2z\nonumber\\
-&0.02143x_1x_2^2-0.00491x_2^2z+0.02152x_1z^2\nonumber\\
-&0.02532x_2z^2-0.03346x_1x_2z\label{eq: corr_mstar_Joel_xmmlss}
\end{align}
\begin{align}
\log\mathrm{SFR}^\mathrm{new}=&-19.56269+4.31679x_1-5.77890x_2\nonumber\\
+&11.12319z-0.28805x_1^2-0.54081x_2^2\nonumber\\
+&0.50187z^2+1.12262x_1x_2-2.37458x_1z\nonumber\\
+&0.91865x_2z+0.00570x_1^3-0.04670x_2^3\nonumber\\
-&0.02217z^3-0.04754x_1^2x_2+0.12051x_1^2z\nonumber\\
+&0.04291x_1x_2^2+0.08807x_2^2z-0.02699x_1z^2\nonumber\\
-&0.09507x_2z^2-0.07334x_1x_2z,\label{eq: corr_sfr_Joel_xmmlss}
\end{align}
where $x_1$ and $x_2$ are our $\log M_\star$ (in $M_\odot$) and $\log\mathrm{SFR}$ (in $M_\odot~\mathrm{yr^{-1}}$), respectively.\par
\begin{figure*}
\centering
\resizebox{\hsize}{!}{
\includegraphics{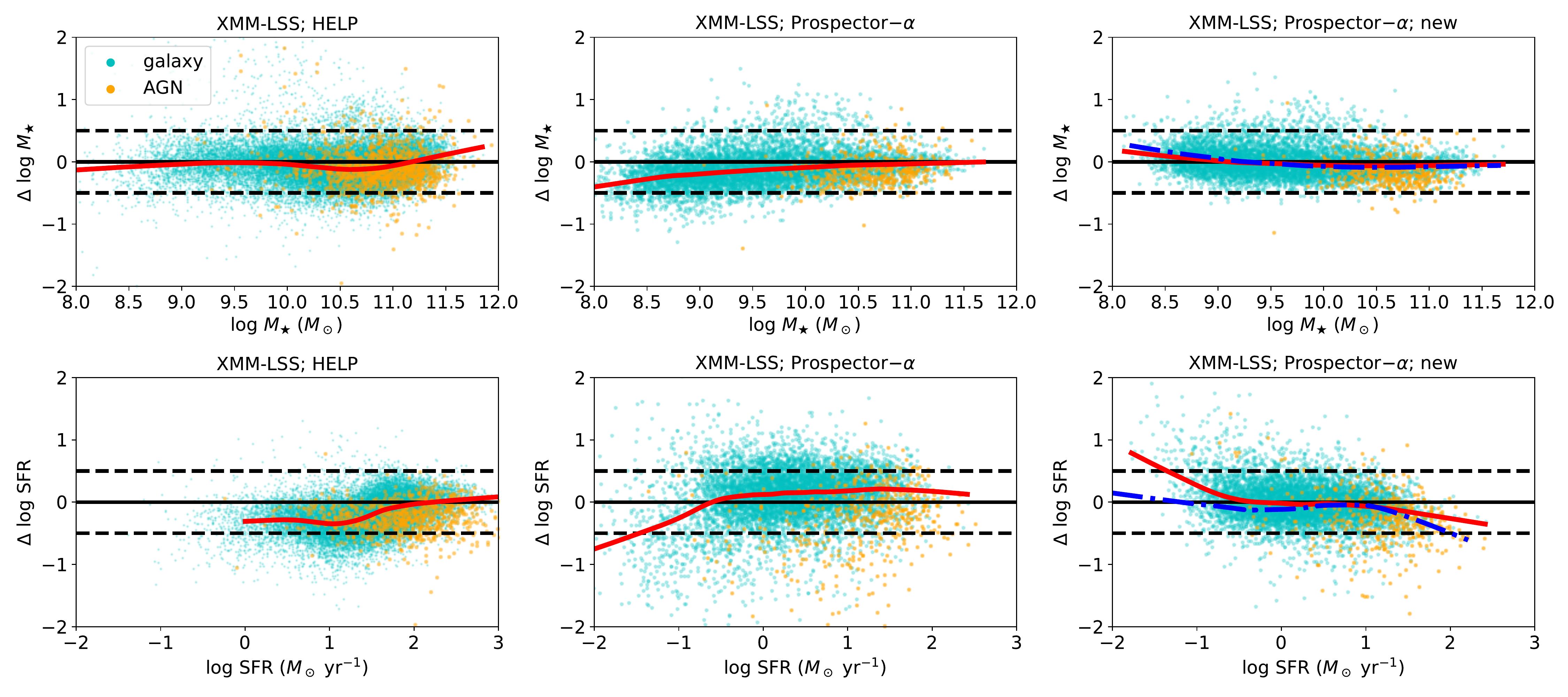}
}
\caption{Comparisons of our $M_\star$ (\textit{top}) and SFR (\textit{bottom}) measurements in XMM-LSS with the HELP values (\textit{left}), the \texttt{Prospector-$\alpha$} values (\textit{middle}; before corrections), and the \texttt{Prospector-$\alpha$} values after applying the empirical corrections in Eqs.~\ref{eq: corr_mstar_Joel_xmmlss} and \ref{eq: corr_sfr_Joel_xmmlss}. The left and middle panels are similar to those in Fig.~\ref{fig_comp_wcdfs_master}, but for XMM-LSS, and the right panels correspond to Fig.~\ref{fig_comp_correction_Leja}. The blue dash-dotted line in the right panel is the resulting LOESS curve if we apply the \mbox{W-CDF-S} corrections (Eqs.~\ref{eq: corr_mstar_Joel} and \ref{eq: corr_sfr_Joel}), which is generally consistent with the XMM-LSS corrections, as represented by the red solid line.}
\label{fig_comp_xmmlss_master}
\end{figure*}
\begin{figure*}
\centering
\resizebox{\hsize}{!}{
\includegraphics{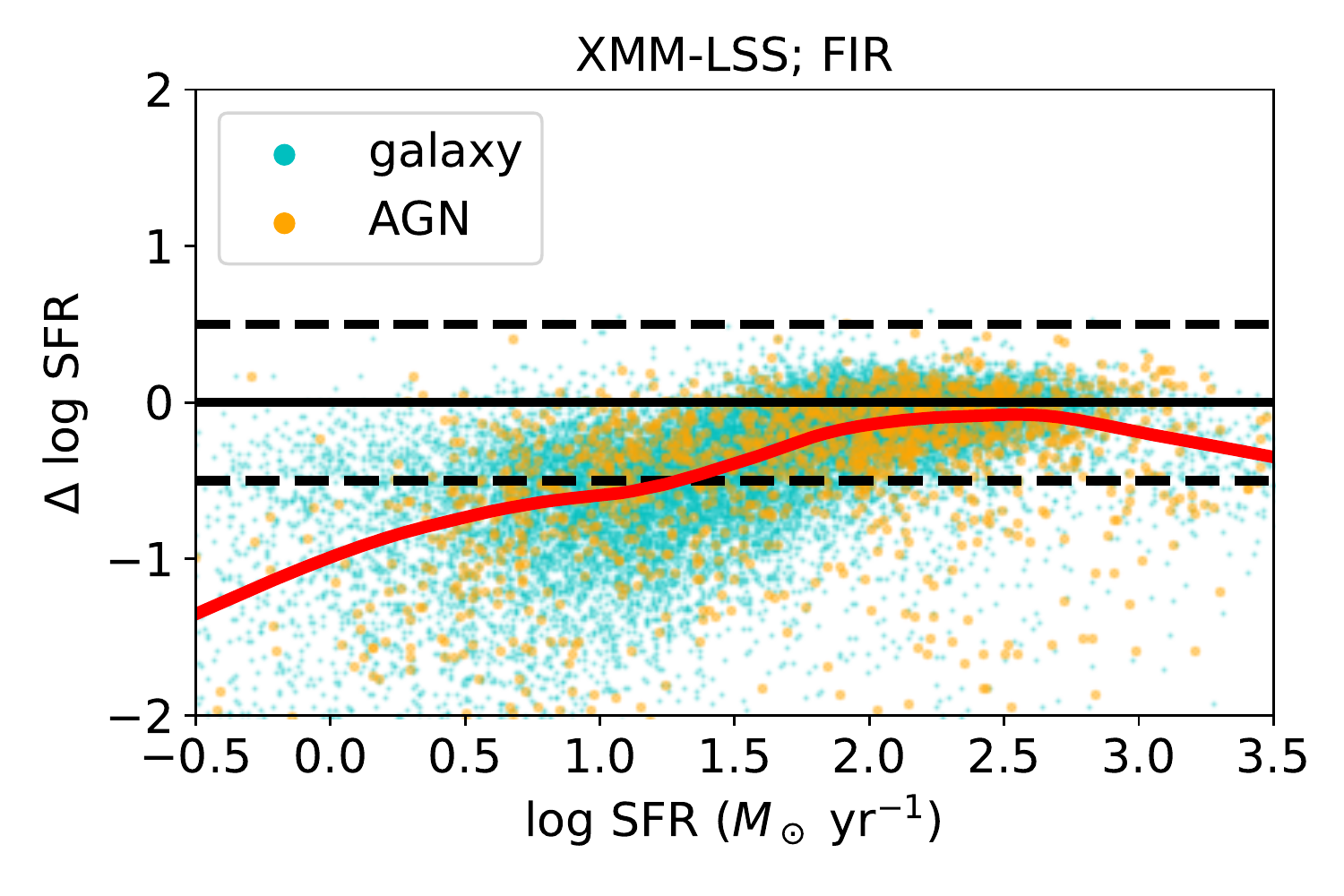}
\includegraphics{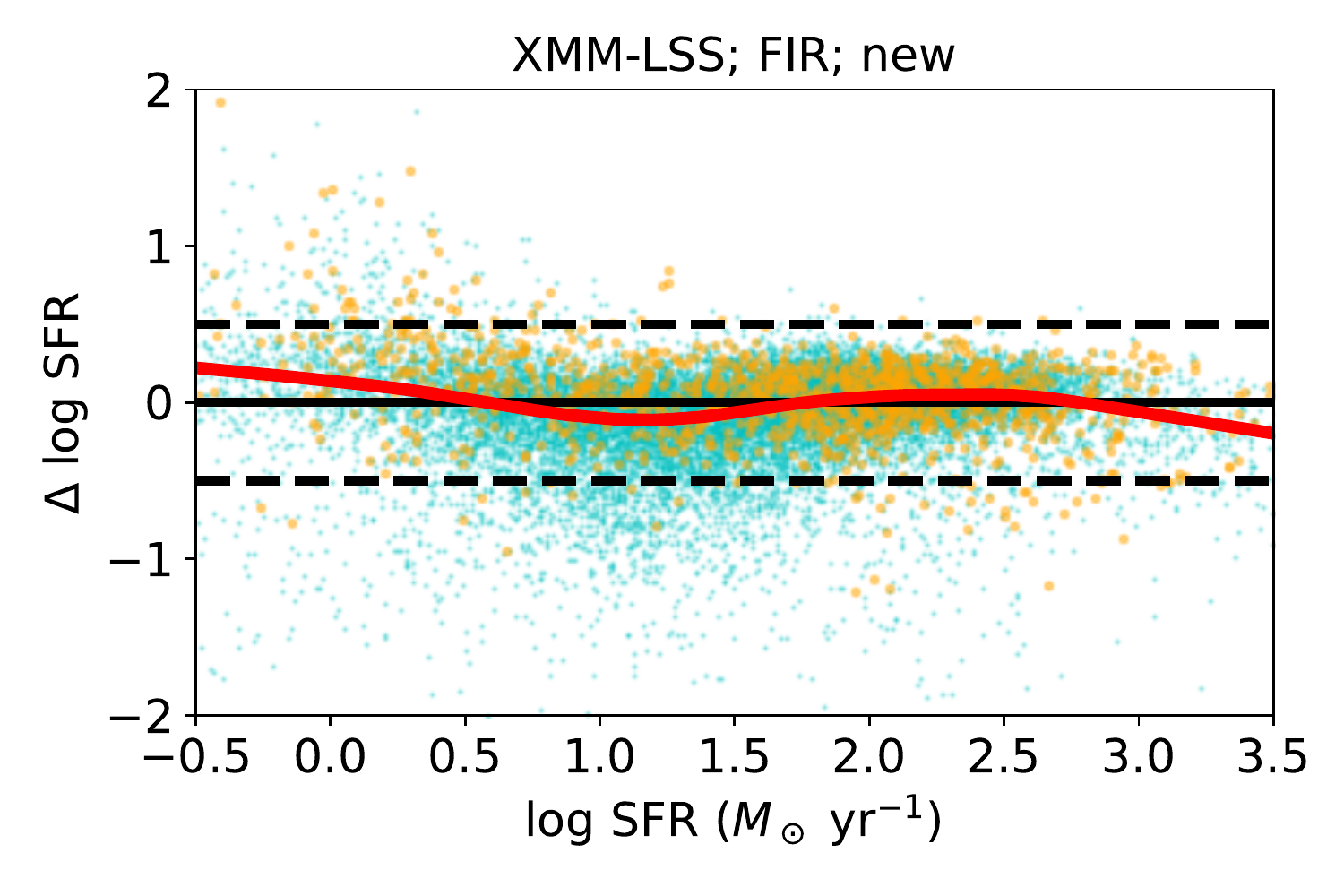}
}
\caption{Comparisons between our SED-based and FIR-based SFRs in XMM-LSS. This figure corresponds to Fig.~\ref{fig_comp_sfr_fir}.}
\label{fig_comp_sfr_fir_xmmlss}
\end{figure*}

Fig.~\ref{fig_comp_xmmlss_master} compares our $M_\star$ (top panels) and SFRs (bottom panels) with HELP values (left panels), \texttt{Prospector-$\alpha$} values before corrections (middle panels), and \texttt{Prospector-$\alpha$} values after corrections (right panels). They are generally consistent and also show similar patterns to those in \mbox{W-CDF-S} and ELAIS-S1. Especially, the empirical corrections in \mbox{W-CDF-S} (Eqs.~\ref{eq: corr_mstar_Joel} and \ref{eq: corr_sfr_Joel}) are roughly the same as the corrections in XMM-LSS, supporting the similarities between the two fields. Fig.~\ref{fig_comp_sfr_fir_xmmlss} compares our SED-based and FIR-based SFRs. The FIR-based SFRs are also corrected using Eq.~\ref{eq: correct_fir_sfr}, but with a slightly different constant factor of $10^{0.24}$ to correct the median offset.\par

We follow the same procedure in Section~\ref{sec: valid_err} to examine the statistical uncertainties in XMM-LSS by comparison with \texttt{Prospector-$\alpha$} measurements. The results are summarized in Table~\ref{tbl_valid_err_xmmlss}, showing that the measured dispersions are close to expectations.

\begin{table*}
\caption{Analyses of errors in XMM-LSS}
\label{tbl_valid_err_xmmlss}
\centering
\begin{threeparttable}
\begin{tabular}{cccccc}
\hline
\hline
& $\mathrm{median}\{e_{s; \texttt{CIGALE}}^2\}$ & $\mathrm{median}\{e_{s; \texttt{Prospector}}^2\}$ & $\mathrm{Var}(o_s)$ & $\hat{e}^2$ & $e^2$\\
\hline
$\log M_\star$ & 0.021 & 0.009 & $0-0.021$ & $0.030-0.050$ & 0.054\\
\hline
log SFR & 0.035 & 0.035 & $0-0.083$ & $0.070-0.154$ & 0.129\\
\hline
\hline
\end{tabular}
\begin{tablenotes}
\item
\textit{Notes.} This table corresponds to Table~\ref{tbl_valid_err}, but for XMM-LSS. See Section~\ref{sec: valid_err} for more details.
\end{tablenotes}
\end{threeparttable}
\end{table*}

\section{Comparisons among Different Fields}
\label{append: compare_fields}
\begin{figure*}
\centering
\resizebox{\hsize}{!}{\includegraphics{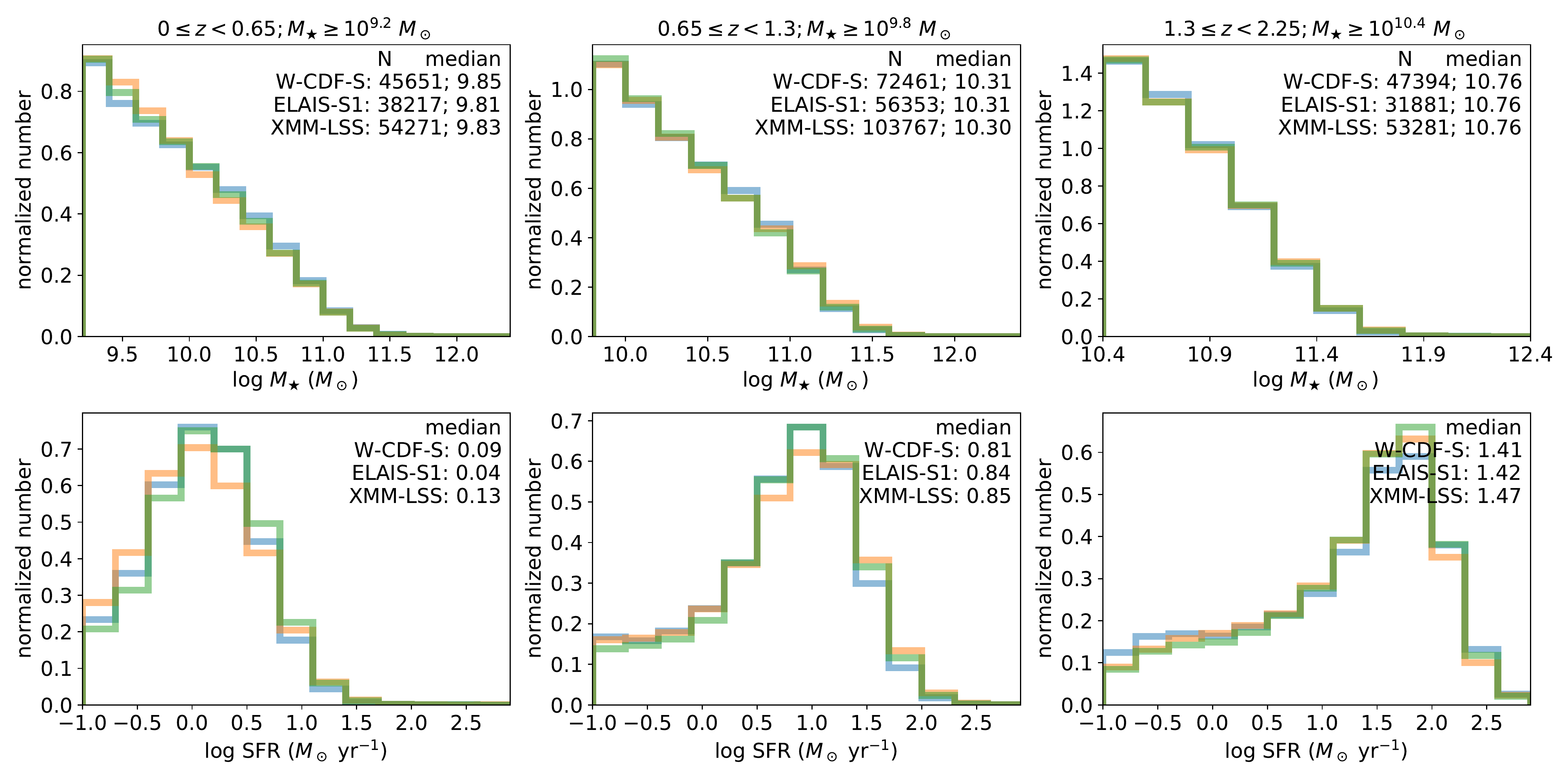}}
\caption{Comparisons of $M_\star$ and SFR distributions among our three fields. Sources are separated into three redshift bins with cuts listed in the panel titles. The upper-right corner of each panel lists the median value of each distribution, and the numbers of sources plotted are shown in the same locations of the upper panels. The distributions are generally consistent among different fields.}
\label{fig_compare_fields}
\end{figure*}

\restartappendixnumbering
The general reliability of our $M_\star$ and SFR measurements has been justified previously, and we further do a check that there are no significant systematic differences in SED-fitting results among different fields by comparing their $M_\star$ and SFR distributions. Detailed comparisons involve the galaxy stellar mass function, and complex selection effects should be considered. This is left to the future, and we simply focus on bright sources to avoid the selection effects, which mostly affect faint sources. We show the $M_\star$ and SFR distributions in three redshift bins in Fig.~\ref{fig_compare_fields}, where we apply empirical cuts. The $M_\star$ cuts are taken to be 0.5~dex above the mass limits for galaxies in the COSMOS2015 catalog \citep{Laigle16}, and we also require $\mathrm{SFR}>0.1~M_\odot~\mathrm{yr^{-1}}$ because low-SFR galaxies are more sensitive to selection effects. Our sources should be largely complete above the cuts. The distributions are generally consistent among different fields, supporting the self-consistency of our results.\par

\begin{figure*}
\centering
\resizebox{\hsize}{!}{\includegraphics{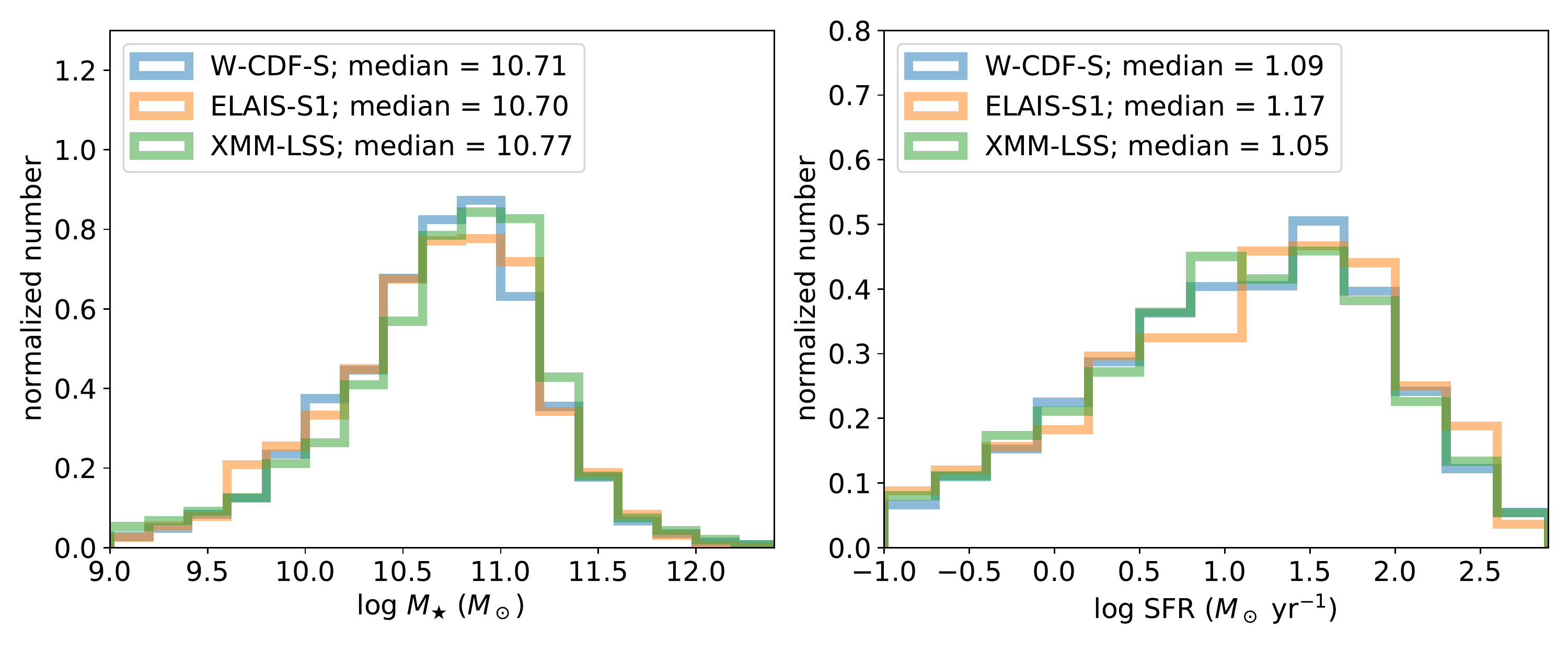}}
\caption{Comparisons of $M_\star$ and SFR distributions among our three fields for \mbox{X-ray} AGNs. The legends list the median values of the distributions.}
\label{fig_compare_fields_xrayagn}
\end{figure*}

The comparisons for AGNs are subject to the AGN selection, which is even harder to quantify. For simplicity, we compare \mbox{X-ray} AGNs in each field, and the \mbox{X-ray} depths are roughly the same for the three fields within a factor of two. The comparisons are shown in Fig.~\ref{fig_compare_fields_xrayagn}, which also do not show significant differences.

\bibliography{citations}

\end{document}